\documentclass[12pt]{iopart}

\pdfoutput=1

\usepackage{graphicx,sidecap}
\usepackage{bm}
\usepackage{braket}
\usepackage{amssymb}
\usepackage{mathrsfs}
\usepackage{amsmath}
\usepackage{xcolor}
\usepackage{hyperref}
\usepackage{enumerate}
\usepackage[toc,title]{appendix}
\usepackage{verbatim}
\usepackage[margin=30pt, bf, font=footnotesize, center, justification=justified]{caption}[2004/07/16]

\hypersetup{colorlinks,bookmarksopen,bookmarksnumbered,
citecolor=red,
linkcolor=blue,
pdfstartview=green,
urlcolor=purple}

\catcode`@=11 
\renewcommand\tableofcontents{%
  \section*{\contentsname}%
  \@starttoc{toc}%
}
\catcode`@=12

\def\be{\begin{equation}}
\def\ee{\end{equation}}

\def\bea{\begin{eqnarray}}
\def\eea{\end{eqnarray}}

\def\Tr{{\rm Tr}}
\def\iu{\textrm{i}}

\begin{document}

\title[A contour for the entanglement entropies in harmonic lattices]
{A contour for the entanglement entropies \\ in harmonic lattices}

\vspace{.5cm}

\author{Andrea Coser$^{1,2}$, Cristiano De Nobili$^3$ and Erik Tonni$^3$}
\address{$^1$\,Departamento de An\'alisis Matem\'atico, Universidad Complutense de Madrid, 28040 Madrid, Spain.}
\address{$^2$\,ICMAT, C/ Nicol\'as Cabrera, Campus de Cantoblanco, 28049 Madrid, Spain.}
\address{$^3$\,SISSA and INFN, via Bonomea 265, 34136 Trieste, Italy.}

\vspace{.5cm}

\begin{abstract}
We construct a contour function for the entanglement entropies in generic harmonic lattices.
In one spatial dimension, numerical analysis are performed by considering harmonic chains with either periodic or Dirichlet boundary conditions.
In the massless regime and for some configurations where the subsystem is a single interval, 
the numerical results for the contour function are compared to the inverse of the local weight function which multiplies the energy-momentum tensor in the corresponding entanglement hamiltonian, found through conformal field theory methods, and a good agreement is observed. 
A numerical analysis of the contour function for the entanglement entropy is performed also in a massless harmonic chain 
for a subsystem made by two disjoint intervals. 
\end{abstract}

\maketitle

\tableofcontents

\newpage

\section{Introduction}
\label{sec:intro}

Entanglement in many-body quantum systems has attracted a lot of research
during the last decade (see \cite{ccd-09-reviews} for a collection of reviews).

Consider a quantum system whose Hilbert space is bipartite, namely $\mathcal{H} = \mathcal{H}_A \otimes \mathcal{H}_B$.
Denoting by $\rho$ the density matrix which characterises the state of the whole system, $A$'s reduced density matrix $\rho_A$ is obtained by taking the partial trace of $\rho$ over $\mathcal{H}_B$.
The entanglement entropy  is defined as the Von Neumann entropy of $\rho_A$, i.e.
\be
\label{SA def}
S_A = -\, \Tr (\rho_A \log \rho_A)\,,
\ee
where the normalization condition  $\Tr \rho_A =1$ has been imposed.

Other important quantities providing useful information about the entanglement of the bipartition of the system in the state $\rho$
are the R\'enyi entropies
\be
\label{Renyi entropies def}
S_A^{(n)} = \frac{1}{1-n} \, \log \Tr \rho_A^n \,,
\ee
which are parameterised by the integer $n\geqslant 2$.
The entanglement entropy (\ref{SA def}) can be computed from the R\'enyi entropies (\ref{Renyi entropies def}) 
by performing the following analytic continuation 
\be
\label{S_A replica limit}
S_A = - \lim_{n \rightarrow 1} \partial_n \Tr \rho_A^n = \lim_{n \rightarrow 1} S_A^{(n)}\,,
\ee
which is also known as the replica limit for the entanglement entropy. 
We refer to $S_A^{(n)}$ with $n\geqslant 1$ as the entanglement entropies, assuming that $S_A^{(1)}\equiv S_A$ is the entanglement entropy (\ref{SA def}) obtained through the replica limit (\ref{S_A replica limit}).

The entanglement entropies measure the bipartite entanglement associated to the decomposition $\mathcal{H} = \mathcal{H}_A \otimes \mathcal{H}_B$ when the whole system is in a pure state. 
In this manuscript we consider bipartitions of the Hilbert space associated to spatial bipartitions of the entire space,
where $A$ is a spatial region and $B$ its complement. 

Entanglement entropies have been largely studied in many quantum systems both in lattice models \cite{ee-lattice-boson, peschel-03-modham, br-04, ee-lattice-fermion,ep-rev, ee-lattice-rev} and in quantum field theories (QFTs).
As for the latter class of models, important goals have been achieved in conformal field theories (CFTs) \cite{wilczek, cc-04, cc-rev}, free quantum field theories \cite{ch-rev} and holography \cite{RT, RT-review-hee}.

A relatively simple spatial bipartition is obtained when $A$ is a simply connected domain.
Nonetheless, it is very interesting to study also configurations where $A$ is made by disjoint regions. 
In the case of two disjoint domains $A_1$ and $A_2$, we have $A = A_1 \cup A_2$ and an important quantity to introduce is the following combination of entanglement entropies
\be
\label{MI renyi def}
I^{(n)}_{A_1, A_2} = S^{(n)}_{A_1} + S^{(n)}_{A_2} - S^{(n)}_{A_1 \cup A_2} \,,
\ee 
where the $n=1$ case corresponds to the mutual information
$I_{A_1, A_2} \equiv S_{A_1} + S_{A_2} - S_{A_1 \cup A_2} =  \lim_{n \to 1} I_{A_1,A_2}^{(n)} $, which can be obtained by taking the replica limit (\ref{S_A replica limit}) of the various terms. 
The mutual information measures the total amount of correlations between the two disjoint regions \cite{MI properties}.
The quantity (\ref{MI renyi def}) has been studied both in one \cite{furukawa-09, cct-09, 2 disjoint intervals, casini-2int} and in higher spatial dimensions \cite{ch-rev, cardy-mi-higher}.

In principle, the reduced density matrix $\rho_A$ contains more information about the entanglement between $A$ and its complement $B$ than the entanglement entropies $S_A^{(n)}$.
Since $\rho_A$ is a positive semi-definite and hermitian operator, it can be written as
\be
\label{ent ham def}
\rho_A = e^{- 2\pi K_A}\,,
\ee
where $K_A$ is known as the entanglement hamiltonian (or modular hamiltonian).

The entanglement hamiltonians for free fermions and free bosons on the lattice have been studied in \cite{peschel-03-modham} 
(see \cite{ep-rev, ch-rev} for reviews).
In QFTs the entanglement hamiltonians are generically non local operators \cite{haag}.
Nonetheless, there are interesting special configurations where $K_A$ can be expressed in terms of integrals over $A$ of local operators multiplied by proper local weight functions. 
The most important case belonging to this class is a Lorentz invariant QFT in its ground state with $A$ given by the half space $x_1 > 0$.
For this example Bisognano and Wichmann found that the entanglement hamiltonian is the generator of the Lorentz boosts in the $x_1$ direction \cite{bw}. 

Considering the special class of QFTs given by CFTs, the crucial result by Bisognano and Wichmann and the conformal symmetry allow to write the entanglement hamiltonians corresponding to other interesting configurations as the integral over the domain $A$ of the component $T_{00}$ of the energy-momentum tensor multiplied by a suitable local weight function 
\cite{chm, klich-13, ct-16}.
For free fermions in one spatial dimension, also the entanglement hamiltonian of disjoint intervals has been found \cite{casini-2int}.

In this manuscript we study functions $s_A^{(n)}: A \to \mathbb{R}$ which assign a real number to every site in the spatial domain $A$ such that
\be
\label{E_A sum_i}
S_A^{(n)} = \sum_{i \,\in\,A} s_A^{(n)}(i)\,,
\ee
where the subsystem $A$ is made by either a connected region or disjoint regions.
Since this property naturally leads to interpret $s_A^{(n)}(i)$ as a density for the entanglement entropies \cite{ent-density}, it is natural to impose also the following positivity condition
\be
\label{positiv}
s_A^{(n)}(i) \,\geqslant \,0\,,
\qquad
\forall \, i \in A\,.
\ee

The aim is to construct these functions $s_A^{(n)}(i)$ from the methods employed to compute the entanglement entropies, in order to identify the contribution of the $i$-th site to the entanglement entropies.
These functions have to fulfil  (\ref{E_A sum_i}), (\ref{positiv}) and possibly other proper requirements and they are expected to provide information about the spatial structure within $A$ of the entanglement between $A$ and $B$.
The same question can be formulated for QFTs in the continuum, where the discrete sum in (\ref{E_A sum_i}) becomes an integral over the region $A$ and $s_A^{(n)}$ is a function of the position $x\in A$.

A function fulfilling the constraints (\ref{E_A sum_i}) and (\ref{positiv}) for free fermions on the lattice has been studied by Chen and Vidal \cite{chen-vidal}.
As for the harmonic lattices, a proposal which satisfies (\ref{E_A sum_i}) is based on the results obtained by Botero and Reznik \cite{br-04} and it has been studied in \cite{frerot-roschilde}.
Following \cite{chen-vidal}, throughout this manuscript the function $s_A^{(n)}(i)$ will be called contour function for the entanglement entropies.

Many functions satisfying (\ref{E_A sum_i}) and (\ref{positiv}) could be constructed and a complete list of properties characterising the contour function for the entanglement entropies in a unique way is not available.
In \cite{chen-vidal}, the authors made a step in this direction by proposing three further reqirements beside (\ref{E_A sum_i}) and (\ref{positiv}) involving the contour functions for the entanglement entropies. 
They are given by: 
(a) a constraint implementing the consistency with any spatial symmetry of the subsystem; 
(b) the requirement that the contour integrated over any subregion $G \subseteq A$ must be invariant under local unitary transformations in that subregion;
(c) a bound meaning that the contour integrated over any subregion $G \subseteq A$ must be smaller or equal than the entanglement of any factor space of $\mathcal{H}_A$ which contains the Hilbert space $\mathcal{H}_G$ of that subregion.
The authors of \cite{chen-vidal} constructed a contour function for the entanglement entropy of free fermions on the lattice which fulfils these five requirements.
These properties could be helpful in order to characterise the contour function within the set made by the densities of entanglement, which can be 
naturally defined as the functions satisfying the properties (\ref{E_A sum_i}) and (\ref{positiv}).
For instance, a natural function to consider is the flat contour $s_A^{(n)}(i)=S_A^{(n)}/ |A|$, which assigns the same value to all the sites in $A$, being $|A|$ the total number of sites contained in $A$. 
This function fulfils the constraints (\ref{E_A sum_i}) and (\ref{positiv}), but it does not provide any useful information about the spatial structure of the entanglement within $A$.
As for the contour function  for the entanglement entropies in the harmonic lattices based on \cite{br-04}, the property (\ref{E_A sum_i}) holds, but the positivity condition (\ref{positiv}) has not been proved \cite{frerot-roschilde}.

In this manuscript we construct new contour functions for the entanglement entropies in generic harmonic lattices which fulfil (\ref{E_A sum_i}) and (\ref{positiv}).
For a specific proposal, we prove that it satisfies also a weaker version of the three further requirements introduced in \cite{chen-vidal}.
A detailed numerical analysis of this contour function is performed for harmonic chains in one spatial dimension.

The continuum limit of the harmonic lattice in the massless regime is the CFT given by the free massless boson
and in one spatial dimension it has central charge $c=1$.
In two-dimensional CFTs, some time-independent examples have been found where the entanglement hamiltonian $K_A$ of a finite interval $A$ can be written as an integral over $A$ of the $T_{00}$ component of the energy-momentum tensor multiplied by a suitable local weight function \cite{chm, klich-13, ct-16}.
Focusing on these configurations, for some examples we find that, in the massless regime and in the scaling limit, 
the contour function highlighted above is in good agreement with the inverse of the local weight function multiplying $T_{00}$ in the corresponding entanglement hamiltonian, specified to the $c=1$ case.

The manuscript is organised as follows. In \S\ref{sec:entHC} the usual method to compute the entanglement entropies in harmonic lattices is reviewed, highlighting the role of the Williamson's theorem.
In \S\ref{sec:contour} a set of contour functions $s_A^{(n)}(i)$ satisfying (\ref{E_A sum_i}) and (\ref{positiv}) is constructed. 
Then, we focus on a specific proposal belonging to this set, showing that it also fulfils a weaker version of the three requirements introduced in \cite{chen-vidal}; therefore it is our best candidate for the contour function for the entanglement entropies in harmonic lattices. 
In \S\ref{sec:cft} we consider this contour function in the massless regime and in the scaling limit for various configurations, comparing the lattice data with the inverse of the local weight function which multiplies the local operator $T_{00}$ in the corresponding entanglement hamiltonian (in the $c=1$ case), already derived through a CFT analysis. 
In \S\ref{sec:other-proposals} alternative constructions are discussed, including the one based on \cite{br-04}.
In \S\ref{sec:conclusions} we draw our conclusions and mention some open problems. 
In appendix \ref{app:williamson} some technical issues concerning the Williamson's theorem are discussed.
In appendix \ref{app:properties} we show that the contour constructed in \S\ref{sec:contour euler} satisfies the properties (a), (b) and (c), in the weaker version discussed in \S\ref{sec:contour properties}.
In appendix \ref{app:correlators} we collect all the correlators employed in our numerical analysis.


\section{Entanglement entropies in the harmonic lattices}
\label{sec:entHC}

In this section we briefly recall the standard method to compute the entanglement entropies in harmonic lattices, which involves the symplectic spectrum of the covariance matrix associated to the reduced density matrix $\rho_A$ of the subsystem \cite{ee-lattice-boson, br-04, ee-lattice-rev}.
Although throughout this manuscript we mainly consider the one-dimensional case, the method described in this section holds for a  generic number of spatial dimensions.

The hamiltonian of the one-dimensional harmonic chain with $L$ lattice sites and nearest neighbour spring-like interaction reads
\be
\label{HC ham}
H = \sum_{i=0}^{L-1} \left(
\frac{1}{2m}\,\hat{p}_i^2+\frac{m\omega^2}{2}\,\hat{q}_i^2 +\frac{\kappa}{2}(\hat{q}_{i+1} -\hat{q}_i)^2
\right),
\ee
where we impose either periodic ($\hat{q}_L =\hat{q}_0$ and $\hat{p}_L =\hat{p}_0$) or Dirichlet ($\hat{q}_L =\hat{q}_0 = \hat{p}_L =\hat{p}_0 =0$) boundary conditions. 
The canonical commutation relations $[\hat{q}_i, \hat{q}_j] = [\hat{p}_i, \hat{p}_j]  =0$ and $[\hat{q}_i, \hat{p}_j] = \textrm{i} \delta_{ij}$ are satisfied.

By arranging the operators $\hat{q}_i$ and $\hat{p}_i$ into the vector $\hat{\boldsymbol{r}} \equiv (\hat{q}_1 , \dots , \hat{q}_L, \hat{p}_1, \dots, \hat{p}_L)^{\textrm{t}}$, the canonical commutation relations can be written as $[\hat{r}_i, \hat{r}_j] = \textrm{i} J_{ij}$ (here we set $\hbar =1$), being $J$ the standard symplectic matrix 
\be
\label{Jmat}
J \equiv
\bigg( \hspace{-.1cm} \begin{array}{cc}
 \boldsymbol{0} &  \boldsymbol{1} \\
 - \,\boldsymbol{1} &  \boldsymbol{0} \\
\end{array}  \hspace{-.05cm}  \bigg)\,,
\ee
where the square matrices $ \boldsymbol{1}$ and $ \boldsymbol{0}$ are the identity matrix and the matrix made by zeros respectively.
Notice that $J^{\textrm{t}} = -J$ and $J^2 = - \boldsymbol{1}$.

The real symplectic group $\textrm{Sp}(L)$ is made by the $2L \times 2L$ real matrices $M$ such that $M J M^{\textrm{t}} = J$.
Given a symplectic matrix $M$, we have that $\textrm{det}(M) =  1$ and $M^{\textrm{t}} \in \textrm{Sp}(L)$ \cite{deGosson}.
The real symplectic group is very important for the analysis of the harmonic lattices because the linear transformations $\hat{\boldsymbol{r}} \to \hat{\boldsymbol{r}}' = M \hat{\boldsymbol{r}} $ constructed through symplectic matrices $M$ preserve the canonical commutation relations.

A canonical rescaling of the variables allows to write the hamiltonian (\ref{HC ham}) as the hamiltonian of a free boson with mass $\omega$ discretised on a lattice with spacing $a=\sqrt{m/\kappa}$.
In the numerical computations on the lattice presented in this manuscript we have set $m=\kappa=1$ without loss of generality.
The continuum limit of this model provides the free scalar boson with mass $\omega$ in two spacetime dimensions. 
It is well known that in the massless case $\omega=0$ this quantum field theory is a CFT with central charge $c=1$.

In this manuscript we deal only with some Gaussian states associated to the harmonic chain (\ref{HC ham}), which can be either pure (e.g. the ground state) or mixed (e.g. the thermal state) (see \cite{weedbrook} for a review).
A Gaussian state for the harmonic chain (\ref{HC ham}) is  completely characterised by the correlators $\langle \hat{r}_i \rangle$ (first moments) and $\langle \hat{r}_i \, \hat{r}_j\rangle$ (second moments).
A shift in the first moments corresponds to a unitary transformation which preserves the Gaussian nature of the state. 
We will consider Gaussian states with vanishing first moments, which are fully described by the 
 $2L \times 2L$ covariance matrix $\gamma \equiv \textrm{Re} \langle \boldsymbol{r} \, \boldsymbol{r}^{\textrm{t}}\rangle$  collecting all the second moments 
 \cite{ee-lattice-boson, br-04, ee-lattice-rev, deGosson}.

The covariance matrix $\gamma$ is real and symmetric.
It can be shown that the uncertainty principle coming from the canonical commutation relations implies that $\gamma + \iu  J/2 $ is positive semidefinite 
and that this result is equivalent to impose that $\gamma$ is positive definite and $\sigma_k \geqslant 1/2$ for its symplectic eigenvalues \cite{Simon-94, deGosson}.
Moreover, the covariance matrix corresponding to a pure Gaussian state satisfies the relation $(\textrm{i} J \, \gamma)^2 = \tfrac{1}{4}\,\boldsymbol{1}$ \cite{lindblad-00}.
It is straightforward to observe that the linear map  $\hat{\boldsymbol{r}} \to \hat{\boldsymbol{r}}' = M \hat{\boldsymbol{r}} $ characterised by the symplectic matrix $M$ induces on the covariance matrix the transformation $\gamma \to \gamma' = M \gamma M^{\textrm t}$.

In this manuscript we are interested in the entanglement associated to spatial bipartitions where the whole harmonic lattice $A\cup B$ is in the Gaussian state characterised by the covariance matrix $\gamma$.
Since we are considering the states with $\langle \hat{r}_i \rangle=0$, the group of linear transformations which preserves the Gaussian character of a state is $\textrm{Sp}(L)$.

The reduced density matrix $\rho_A$ associated to a spatial subsystem $A$ made by $\ell$ sites corresponds to a mixed state also when the state of the whole system $A \cup B$ is pure. 
For the harmonic chain (\ref{HC ham}) and its higher dimensional generalisations, the reduced density matrix $\rho_A$ remains Gaussian for any choice of the subsystem $A$.
Because of this property, $\rho_A$ is characterised by the  $2\ell \times 2\ell$ reduced covariance matrix $\gamma_A$ obtained by extracting from the covariance matrix $\gamma$ of the entire system the rows and the columns corresponding to the lattice sites which belong to $A$.
Let us consider the reduced covariance matrices $\gamma_A$ with the following block structure
\be
\label{cov mat}
\gamma_A =
\bigg( \hspace{-.1cm} \begin{array}{cc}
 Q & R   \\
 R^{\textrm t}    & P \\
\end{array}  \hspace{-.1cm}  \bigg)\,,
\ee
where $Q$, $P$ and $R$ are the $\ell \times \ell$ correlation matrices  corresponding to the subsystem $A$
(the subindex $A$ has been dropped for these matrices in order to lighten the forthcoming expressions).
The reduced covariance matrix $\gamma_A$ is real, symmetric and positive definite.
We remark that $Q$ and $P$ are symmetric and strictly positive, while $R$ is not constrained in general.

A crucial result for the quantitative analysis of the entanglement in the harmonic lattice is the Williamson's theorem \cite{Williamson}.
This theorem holds for a generic real, symmetric and positive matrices, but in this context we are interested in its consequences for the covariance matrices $\gamma_A$.

According to the Williamson's theorem, a symplectic matrix $W\in \textrm{Sp}(\ell)$ exists such that 
\be
\label{williamson th gammaA}
\gamma_A = W^{\textrm t} \big( D \oplus D \big) W\,,
\ee
where $D=\textrm{diag} (\sigma_1 , \dots , \sigma_\ell)$ and $\sigma_k > 0$.
The set $\{ \sigma_1 , \dots , \sigma_\ell \}$ is given by the positive square roots of the spectrum of the matrix $(\iu J \gamma_A)^2$ and it is known as the symplectic spectrum of $\gamma_A$.
The symplectic spectrum is invariant under canonical transformations. 

The Williamson's theorem has been proved in various ways \cite{williamson-other proofs}, but in our analysis we will employ some steps of the proof found in \cite{simon-99}.
In the appendix \ref{app:williamson} we briefly discuss some issues related to the Williamson's theorem, including also the part of the proof given by \cite{simon-99}
that will be useful for our discussions.

The entanglement entropies in the harmonic lattices are obtained from the symplectic spectrum of $\gamma_A$ as 
\cite{ee-lattice-boson, br-04, ee-lattice-rev}
\be
\label{SA and SAn from nuk}
S_A = \sum_{k=1}^\ell s(\sigma_k)\,,
\qquad
S_A^{(n)} = \sum_{k=1}^\ell s_n(\sigma_k)\,,
\ee
where $s(y)$ and $s_n(y)$ are the following analytic functions:
\be
\label{sx def}
s(y) \equiv (y+1/2)\log(y+1/2) - (y-1/2)\log(y-1/2)\,,
\ee
and 
\be
\label{snx def}
s_n(y) \equiv \frac{1}{n-1} \, \log\big[(y+1/2)^n - (y-1/2)^n\big]\,.
\ee
Notice that $\lim_{n \to 1} s_n(y)  = s(y) $ for $y \geqslant 1/2$, as expected from the replica limit. 
This naturally leads us to adopt the notation $s_1(y)  \equiv s(y)$.
Notice that $s_n(y) \geqslant 0$ when $y \geqslant 1/2$, being $n \geqslant 1$.
Thus, the symplectic eigenvalues $\sigma_k =1/2$ do not contribute to the entanglement entropies, i.e. the non vanishing terms in the sums (\ref{SA and SAn from nuk}) correspond only to the symplectic eigenvalues $\sigma_k >1/2$.


\section{A contour for the entanglement entropies}
\label{sec:contour}

In this section we describe the construction of our proposal for the contour function for the entanglement entropies in generic harmonic lattices. 
In \S\ref{sec:orthog-mat}, by adapting the observations of \cite{chen-vidal}, we give a procedure to write contour functions satisfying (\ref{E_A sum_i}) and (\ref{positiv}) from real orthogonal matrices.
In \S\ref{sec:contour euler} we focus on the contour function obtained from an orthogonal matrix related to the Euler decomposition of the symplectic matrix $W$ occurring in the Williamson's theorem (\ref{williamson th gammaA}).
Then, in \S\ref{sec:contour properties} we show that the contour function for the entanglement entropies constructed in \S\ref{sec:contour euler} fulfils also a weaker formulation of the three constraints introduced in \cite{chen-vidal}.

\subsection{Contour functions from orthogonal matrices}
\label{sec:orthog-mat}

In \S\ref{sec:entHC} we reviewed that the entanglement entropies in the harmonic lattice are given by the sums (\ref{SA and SAn from nuk}) over the symplectic spectrum of the covariance matrix $\gamma_A$ characterising the domain $A$ containing $\ell$ lattice sites.

Following \cite{br-04, chen-vidal}, we construct the contour function $s_A^{(n)}(i)$ satisfying (\ref{E_A sum_i}) by assuming that we can associate $\ell$ real numbers $p_k(i)$ to every symplectic eigenvalue $\sigma_k$  (here $1\leqslant i \leqslant \ell$)  such that
\be
\label{p sum rule}
\sum_{i\,=\,1}^\ell  p_k(i) = 1\,,
\qquad
1\leqslant k \leqslant \ell\,.
\ee 
Indeed, by inserting (\ref{p sum rule}) into (\ref{SA and SAn from nuk}) and inverting the sums, 
it is immediate to recognize that the entanglement entropies can be written in the form (\ref{E_A sum_i})
with the contour function given by
\be
\label{contour pki}
s_A^{(n)}(i) = \sum_{k=1}^\ell p_k(i) \,s_n(\sigma_k) \,,
\ee
where the function $s_n(x)$ is given by (\ref{sx def}) for the entanglement entropy and by (\ref{snx def}) for the R\'enyi entropies.
We call {\it mode participation function} the function $p_k(i)$ in  (\ref{contour pki}), as done in \cite{br-04}.
It encodes information about the contribution of the $i$-th site in $A$ to the term associated to the $k$-th eigenvalue $\sigma_k$ of the symplectic spectrum in the sums (\ref{SA and SAn from nuk}) providing the entanglement entropies.
If we also require that $p_k(i) \geqslant 0$ for all  $i$ and $k$, then Eq.\,(\ref{p sum rule}) naturally leads to interpret the mode participation function  as a set of probabilities. 
In this case the positivity condition (\ref{positiv}) is guaranteed because $s_n(\sigma_k)  \geqslant 0 $ for $\sigma_k \geqslant 1/2$.
Any set of probabilities $p_k(i)$ provides a contour function (\ref{contour pki}) fulfilling the constraints (\ref{E_A sum_i}) and (\ref{positiv}).
However, we are interested in finding mode participation functions which are based on the method underlying the computation of the entanglement entropies. 

In order to identify some mode participation functions from (\ref{SA and SAn from nuk}), let us write the entanglement entropies as traces of suitable matrices.
In particular, given a $2\ell \times 2\ell$ real orthogonal matrix $O \in  O(2\ell)$, let us introduce
\be
\label{Phi_A def}
\Phi_A
\equiv
O^{\textrm{t}} \big(D \oplus D\big) \,O\,,
\ee
being $D$ the diagonal matrix containing the symplectic spectrum of the reduced covariance matrix $\gamma_A$ (see Eq.\,(\ref{williamson th gammaA})).
Then, the entanglement entropies (\ref{SA and SAn from nuk}) can be written as
\be
\label{SAn Phi_A}
S_A^{(n)} 
=   \,\Tr \big[  s_n(  D )\big]
=   \frac{1}{2} \, \Tr \big[  s_n(  D \oplus D )\big]
=   \frac{1}{2} \, \Tr \big[  s_n(  \Phi_A )\big]\,,
\ee
where the analytic functions $s_n(x)$ are given by (\ref{sx def}) and (\ref{snx def}).
In the last step we have employed (\ref{Phi_A def}), the cyclic property of the trace and the fact that the functions $s_n(x)$ are analytic.

The matrix $O$ defines a linear mapping sending the Williamson's mode, labelled by $k$, into another set of modes, that we will label by the index $\alpha$.
In order to isolate the contribution  to the entanglement entropies (\ref{SAn Phi_A}) due to a specific mode characterised by a fixed value of $\alpha$, 
let us introduce a family $\{X^{(\alpha)},1\leqslant \alpha \leqslant \ell \} $ of orthogonal projectors.
These operators are represented by $2\ell \times 2\ell$ matrices which are symmetric, semi-positive definite and they satisfy the property $X^{(\alpha)} X^{(\beta)} = \delta_{\alpha\beta } X^{(\alpha)}$ and  $ \sum_{\alpha=1}^\ell X^{(\alpha)} = \boldsymbol{1} $. 
In the base defined by the map $O$, the projector $X^{(\alpha)}$ can be written as  $X^{(\alpha)} = \delta^{(\alpha)} \oplus \,\delta^{(\alpha)}$,
being $\delta^{(\alpha)}$ the $\ell \times \ell$ matrix whose elements are $\delta^{(\alpha)}_{ab} = \delta_{a\alpha} \delta_{b\alpha}$.

Plugging the identity matrix written in the form $ \sum_{\alpha=1}^\ell X^{(\alpha)} = \boldsymbol{1} $ into the argument of the trace occurring in the last step of (\ref{SAn Phi_A}) and employing the linearity of the trace, it is straightforward to realise that
\be
\label{entropies alpha}
S_A^{(n)} = \sum_{\alpha\,=\,1}^\ell s_A^{(n)}(\alpha)\,,
\ee
where 
\be
\label{contour sn_A alpha}
s^{(n)}_A(\alpha) = \frac{1}{2} \, \Tr \big[  X^{(\alpha)} s_n(  \Phi_A)\big]\,.
\ee

By employing the expression $X^{(\alpha)}=\delta^{(\alpha)} \oplus \,\delta^{(\alpha)}$ into (\ref{contour sn_A alpha}), together with the fact that $s_n(x)$ are analytic functions and the cyclic property of the trace, it becomes
\bea
\label{sn(i) phiA step1}
& &
\hspace{-.4cm}
s^{(n)}_A(\alpha) 
\;=\;
 \frac{1}{2} \, \Tr \big[  X^{(\alpha)} s_n(  O^{\textrm{t}} (D \oplus D) O )\big]
 =
 \frac{1}{2} \, \Tr \big[  OX^{(\alpha)} O^{\textrm{t}} \, s_n(  D \oplus D )\big]
 \\
 \label{sn(i) phiA step1a}
 \rule{0pt}{.7cm}
 & &
 \hspace{.7cm}
 \;=\;
  \frac{1}{2} \, \Tr \big[  
  O \big( \delta^{(\alpha)} \oplus \delta^{(\alpha)} \big) O^{\textrm{t}} 
  \big( s_n(D) \oplus s_n(D)\big)\big]\,.
\eea
In order to write (\ref{sn(i) phiA step1a}) in the form (\ref{contour pki}) and read the corresponding mode participation function, let us partition the orthogonal matrix $O$ introduced in (\ref{Phi_A def}) in four $\ell \times \ell$ blocks
\be
\label{O-mat block}
O= \bigg( 
\begin{array}{cc}
U_O & Y_O \\ Z_O & V_O
\end{array}
\bigg)\,.
\ee
Plugging this block partitioned matrix into (\ref{sn(i) phiA step1a}), for the contour function we find
\be
\label{sn(alpha) phiA step2}
s^{(n)}_A(\alpha) 
=
  \frac{1}{2} \, \Big(
   \Tr \big[ \big( U_O \, \delta^{(\alpha)}U_O^{\textrm{t}}  + Y_O \, \delta^{(\alpha)} Y_O^{\textrm{t}} \big) s_n(D)  \big]
   + 
   \Tr \big[ \big( Z_O \, \delta^{(\alpha)} Z_O^{\textrm{t}} + V_O \, \delta^{(\alpha)}V_O^{\textrm{t}}\big) s_n(D)  \big]
   \Big)\,.
\ee
Writing explicitly the four terms occurring in this expression, we find that (\ref{sn(alpha) phiA step2}) becomes
\be
s_A^{(n)}(\alpha) = \sum_{k=1}^\ell p_k(\alpha) \,s_n(\sigma_k) \,,
\ee
with the mode participation function given by 
\be
\label{pkalpha general}
p_k(\alpha) \,=\, \frac{1}{2} \Big( 
\big[(U_O)_{k\alpha}\big]^2 + \big[(Y_O)_{k\alpha}\big]^2 + \big[(Z_O)_{k\alpha}\big]^2 + \big[(V_O)_{k\alpha}\big]^2 
\Big)\,,
\ee
which is positive by construction. 

In order to check  that the mode participation function (\ref{pkalpha general}) satisfies (\ref{p sum rule}),
let us start from the orthogonality condition $O O^{\textrm{t}}= \boldsymbol{1} $ for the block partitioned matrix (\ref{O-mat block}), i.e.
\be
\label{orto-cond}
O O^{\textrm{t}}= \bigg( 
\begin{array}{cc}
U_O U_O^{\textrm{t}} + Z_O Z_O^{\textrm{t}} & U_O Y_O^{\textrm{t}} + Z_O V_O^{\textrm{t}} \\ 
Y_O U_O^{\textrm{t}} + V_O Z_O^{\textrm{t}} & Y_O Y_O^{\textrm{t}} + V_O V_O^{\textrm{t}}
\end{array}
\bigg)
= 
\bigg( 
\begin{array}{cc}
\boldsymbol{1} & \boldsymbol{0} \\ \boldsymbol{0} & \boldsymbol{1}
\end{array}
\bigg)\,.
\ee
By considering the $k$-th element (with $1\leqslant k \leqslant \ell$) and the $(\ell+k)$-th element along the diagonal of (\ref{orto-cond}), we obtain respectively
\be
\label{ort-cond off-diag}
\sum_{\alpha=1}^\ell  \Big( \big[(U_O)_{k\alpha}\big]^2 + \big[(Z_O)_{k\alpha}\big]^2 \Big) = 1\,,
\qquad
\sum_{\alpha=1}^\ell  \Big(  \big[(Y_O)_{k\alpha}\big]^2 + \big[(V_O)_{k\alpha}\big]^2 \Big) = 1\,,
\ee
which tell us that  (\ref{pkalpha general}) fulfils the constraint 
$\sum_{\alpha=1}^\ell p_k(\alpha) = 1$ for any integer $k \in [1,\ell]$.

By employing the orthogonality condition $O^{\textrm{t}} O  = \boldsymbol{1}$ instead of (\ref{orto-cond}) and following
similar steps, one finds  that (\ref{pkalpha general}) satisfies also the further property
$\sum_{k=1}^\ell p_k(\alpha) = 1$ for any integer $\alpha \in [1,\ell]$.

Since we are interested in constructing contour functions, let us consider those cases where the index $\alpha$ labels the sites of $A$, namely $\alpha = i \in A$, according to the notation adopted throughout this manuscript. 
Henceforth we will employ the projector $X^{(i)}$ corresponding to the $i$-th site of the region $A$.

Summarising, the entanglement entropies (\ref{SA and SAn from nuk}) can be written in the form (\ref{E_A sum_i}) with the following contour function
\be
\label{contour sn_A Xi}
s^{(n)}_A(i) = \frac{1}{2} \, \Tr \big[  X^{(i)} s_n(  \Phi_A)\big]\,,
\ee
which can be expressed as in (\ref{contour pki}) with the mode participation function given by
\be
\label{pki general}
p_k(i) \,=\, \frac{1}{2} \Big( 
\big[(U_O)_{ki}\big]^2 + \big[(Y_O)_{ki}\big]^2 + \big[(Z_O)_{ki}\big]^2 + \big[(V_O)_{ki}\big]^2 
\Big)\,,
\ee
in terms of the elements of the orthogonal matrix (\ref{O-mat block}).
Since the mode partition function (\ref{pki general}) is positive and fulfils the constraint (\ref{p sum rule}), the corresponding contour function (\ref{contour sn_A Xi}) satisfies the properties (\ref{E_A sum_i}) and (\ref{positiv}).

We find it worth remarking that, while the above discussion is based on the fact that the matrix $O$ in (\ref{Phi_A def}) is orthogonal, 
a canonical transformation is implemented by a symplectic matrix. 
By requiring that $O\in O(2\ell) \cap \textrm{Sp}(\ell)$, further constraints for the blocks in (\ref{O-mat block}) coming from the condition $O J O^{\textrm{t}} = J$ can be employed.
An explicit example belonging to this class is considered in the next subsection.

\subsection{A proposal based on the Williamson's theorem and the Euler decomposition}
\label{sec:contour euler}

The discussion in \S\ref{sec:orthog-mat} allows to conclude that, given an orthogonal matrix (\ref{O-mat block}), we can construct the mode participation function (\ref{pki general}) and, consequently, the corresponding contour function (\ref{contour pki}) satisfies (\ref{E_A sum_i}) and (\ref{positiv}). 

Some particular orthogonal matrices are more relevant for the physics of our problem, which is encoded in the Gaussian reduced density matrix $\rho_A$.
Being $\rho_A$ fully described by the reduced covariance matrix $\gamma_A$ in our cases, let us focus on the orthogonal matrices related to $\gamma_A$.
In particular, since $p_k(i)$ provides the contribution of the $i$-th site in $A$ to the term associated to the $k$-th symplectic eigenvalue in (\ref{SA and SAn from nuk}), 
we find it worth looking for a meaningful orthogonal matrix within the linear transformation which relates the canonical variables $(\hat{q}_i,\hat{p}_i)$, labelled by the index $i$ of the lattice sites, to the canonical variables labelled by the index $k$ associated to the symplectic spectrum. 
This particular canonical transformation is implemented by the real symplectic matrix $W$
associated to $\gamma_A$ through the Williamson's theorem (\ref{williamson th gammaA}).

The Euler decomposition (also known as Bloch-Messiah decomposition) \cite{euler-dec} of the real symplectic matrix $W$ introduced in (\ref{williamson th gammaA}) reads
\be
\label{euler-dec-W}
W = K_{\textrm{\tiny L}} \,E \, K_{\textrm{\tiny R}}\,,
\qquad
E \,=\,  e^{\chi} \oplus e^{-\chi}\,,
\qquad
\chi \equiv \textrm{diag}(\chi_1 , \dots , \chi_\ell) \,,
\ee
where $\chi_j \geqslant 0$ and the real matrices $K_{\textrm{\tiny L}}$ and $K_{\textrm{\tiny R}}$ are symplectic and orthogonal.
The non-uniqueness of the decomposition (\ref{euler-dec-W}) is due only to the freedom to order the elements along the diagonal of $\chi$.
The set containing the matrices of the form given by $E$ is a subgroup of $\textrm{Sp}(\ell)$ corresponding to the single-mode squeezing operations. 
Combining the polar decomposition of the real symplectic matrix $W$ and its Euler decomposition (\ref{euler-dec-W}), 
we find that it can be written as follows
\be
\label{polar-dec-W}
W
\,=\,  E_{\textrm{\tiny L}}  \, K
\,=\, K  \, E_{\textrm{\tiny R}}\,,
\ee
where
\be
\label{K-mat def}
K \,\equiv\, K_{\textrm{\tiny L}} \, K_{\textrm{\tiny R}} \,,
\qquad
\hspace{.4cm} \textrm{and} \hspace{.4cm} 
\qquad
E_{\textrm{\tiny L}} \,\equiv\, K_{\textrm{\tiny L}}\, E \, K_{\textrm{\tiny L}}^{\textrm t} \,,
\qquad
E_{\textrm{\tiny R}} \,\equiv\, K_{\textrm{\tiny R}}^{\textrm t} \, E \, K_{\textrm{\tiny R}}\,.
\ee
The real matrix $K$ is symplectic and orthogonal, while the real matrices 
$E_{\textrm{\tiny R}}$ and  $E_{\textrm{\tiny L}}$ are symplectic, symmetric and positive definite. 
The orthogonal matrix $K$ is obtained by removing the squeezing matrix $E$ in the Euler decomposition of $W$, which is the factor making $W$ non-orthogonal. 
In the following we will employ the polar decomposition $W= K  E_{\textrm{\tiny R}}$ and this factorisation is unique \cite{gilmore}.

Symplectic matrices which are also orthogonal form a subgroup of $\textrm{Sp}(\ell)$ which is isomorphic to the group of the $\ell \times \ell$ unitary matrices (see e.g. Proposition\,2.12 of \cite{deGosson}).
These matrices preserve the trace of the covariance matrix and  they correspond to passive unitary transformations for its density matrix.
Instead, when a symplectic transformation is not orthogonal, the trace of the covariance matrix changes and the corresponding unitary transformations for the density matrix are called active \cite{weedbrook}.

By employing the properties of the matrices highlighted above, we can easily write $E_{\textrm{\tiny L}}$ and $E_{\textrm{\tiny L}}$ in terms of $W$ as follows
\be
\label{ELandER}
 E_{\textrm{\tiny L}}^2 \,=\, W  \,W^{\textrm t} \,,
 \qquad
 E_{\textrm{\tiny R}}^2 \,=\, W^{\textrm t}  \, W\,.
\ee
Also the orthogonal matrix $K$ can be written in terms of $W$ by using $K=E_{\textrm{\tiny L}}^{-1} \, W = W E_{\textrm{\tiny R}}^{-1}$ and the relations in (\ref{ELandER}). The result reads
\be
K 
\,= \,  \big( W  \,W^{\textrm t}  \big)^{-1/2} \,W
\,= \,  W\,\big( W^{\textrm t}  \, W  \big)^{-1/2}\,.
\ee

The factorisations (\ref{euler-dec-W}) and (\ref{polar-dec-W})  hold for any real symplectic matrix.
In our analysis we are interested in the real symplectic matrix entering in the Williamson's theorem.
Among the orthogonal matrices occurring in the decompositions (\ref{euler-dec-W}) and (\ref{polar-dec-W}), 
we think that the matrix $K$ is the most natural one to consider in order to construct a mode participation function. 
Plugging the polar decomposition $W= K \, E_{\textrm{\tiny R}}$ into (\ref{williamson th gammaA}), we find that
\be
\label{williamson th gammaA-K}
E_{\textrm{\tiny R}}^{-1}\,\gamma_A\, E_{\textrm{\tiny R}}^{-1}
\, =\, 
K^{\textrm t} \big( D \oplus D \big) K\,.
\ee
This is an explicit realisation of (\ref{Phi_A def}) with 
$\Phi_A = E_{\textrm{\tiny R}}^{-1}\,\gamma_A\, E_{\textrm{\tiny R}}^{-1}$ and $O=K$.
Notice that in this case $K$ is also symplectic.
The relation (\ref{williamson th gammaA-K}) tells us that $K$ is the orthogonal matrix which diagonalises the symmetric matrix $E_{\textrm{\tiny R}}^{-1}\,\gamma_A\, E_{\textrm{\tiny R}}^{-1}$.

The next step consists in writing $E_{\textrm{\tiny R}}^{-1}$ in (\ref{williamson th gammaA-K}) in terms of the covariance matrix $\gamma_A$.
This can be done by employing some steps of the constructive proof of the Williamson's theorem found in \cite{simon-99}, which have been briefly recalled in the appendix \ref{app:williamson}.
In particular, given the real, symmetric and positive definite covariance matrix $\gamma_A$, 
one introduces the following antisymmetric matrix 
\be
\label{hatgamma def}
\hat{\gamma}_A \equiv \gamma_A^{1/2} J  \, \gamma_A^{1/2}\,,
\ee 
where $J$ is the standard $2\ell \times 2\ell$ symplectic matrix (\ref{Jmat}).
Being the matrix $\hat{\gamma}_A$ antisymmetric, an orthogonal matrix $\widetilde{O}$ exists such that
\be
\label{hatOmega diag}
\widetilde{O}  \,\hat{\gamma}_A  \,\widetilde{O}^{\textrm t} = 
\bigg( \hspace{-.1cm} \begin{array}{cc}
 \boldsymbol{0}  &  D \\
 - D  &  \boldsymbol{0} \\
\end{array}  \hspace{-.05cm}  \bigg)\,,
\ee
where $D=\textrm{diag} (\sigma_1 , \dots , \sigma_\ell)$ is the diagonal matrix containing the symplectic spectrum of $\gamma_A$ introduced in (\ref{williamson th gammaA}).
By extracting $\hat{\gamma}_A$ from (\ref{hatOmega diag}), we find that
\be
\label{hatgamma2}
\hat{\gamma}_A  \hat{\gamma}_A^{\textrm{t}}
\,=\, 
-\, \hat{\gamma}_A^2
\,=\, 
\widetilde{O} ^{\textrm{t}} \big(D^2 \oplus D^2\big) \,\widetilde{O} \,,
\ee
which is positive definite, being $M M^{\textrm{t}}$ positive definite for any real invertible matrix $M$. 
From (\ref{hatgamma2}) one obtains $ | \hat{\gamma}_A | $, which reads
\be
\label{abs-hatgamma}
| \hat{\gamma}_A |
\equiv
\big( \hat{\gamma}_A \hat{\gamma}_A^{\textrm{t}} \big)^{1/2}
=\,
\widetilde{O} ^{\textrm{t}} (D \oplus D) \widetilde{O} \,.
\ee
This relation tells us that the symmetric and positive definite matrix $ | \hat{\gamma}_A | $ is diagonalised by the orthogonal matrix $\widetilde{O} $ and its spectrum coincides with the symplectic spectrum.

It is  worth noticing that (\ref{abs-hatgamma}) provides a realisation of (\ref{Phi_A def}) different from (\ref{williamson th gammaA-K}).
Indeed, $\Phi_A = | \hat{\gamma}_A |$ and $O=\widetilde{O}$ in this case. 
Let us remark that $\widetilde{O}$ is not necessarily symplectic. 
In \S\ref{sec:other-proposals} the contour function associated to the orthogonal matrix $\widetilde{O}$ will be discussed.

In order to express $E_{\textrm{\tiny R}}^{-1}$ in (\ref{williamson th gammaA-K}) in terms of $\gamma_A$, we employ
 a crucial step occurring in the proof of the Williamson's theorem found in \cite{simon-99} (see also the appendix\,\ref{app:williamson}), where the symplectic matrix $W$ satisfying (\ref{williamson th gammaA}) is constructed as follows
\be
\label{W-gammaA}
W = \big(D^{-1/2} \oplus D^{-1/2}\big) \,\widetilde{O} \, \gamma_A^{1/2}\,.
\ee
By using this expression, the second relation in (\ref{ELandER}) becomes
\be
E_{\textrm{\tiny R}}^2 
\,=\,
\gamma_A^{1/2}  \,\widetilde{O} ^{\textrm t} \big( D^{-1} \oplus D^{-1} \big) \widetilde{O} \,\gamma_A^{1/2}
\,=\,
\gamma_A^{1/2}  \,| \hat{\gamma}_A |^{-1}\,\gamma_A^{1/2}\,,
\ee
where in the last step (\ref{hatgamma2}) has been employed. 
Thus, $E_{\textrm{\tiny R}}^{-1}$ reads
\be
\label{ER gammaA}
E_{\textrm{\tiny R}}^{-1}
\,=\,
\Big( \gamma_A^{-1/2}  \,| \hat{\gamma}_A |\,\gamma_A^{-1/2} \Big)^{1/2}.
\ee
This result allows us to write the l.h.s. of (\ref{williamson th gammaA-K}) from the covariance matrix $\gamma_A$.

By specialising (\ref{SAn Phi_A}) to the explicit case given by (\ref{williamson th gammaA-K}),
we can write the entanglement entropies (\ref{SA and SAn from nuk}) as follows
\be
\label{ee s function K}
S^{(n)}_A = \frac{1}{2} \, \Tr \big[  s_n( E_{\textrm{\tiny R}}^{-1} \gamma_A E_{\textrm{\tiny R}}^{-1}  )\big]\,,
\ee
where the matrix $E_{\textrm{\tiny R}}^{-1}$ is the function of the covariance matrix $\gamma_A$ in (\ref{ER gammaA}).
The corresponding contour function is obtained by specialising the expression (\ref{contour sn_A Xi}) to the case given by (\ref{williamson th gammaA-K}) 
and the result reads
\be
\label{contour s_A project - K}
s^{(n)}_A(i) = \frac{1}{2} \, \Tr \big[  X^{(i)} s_n( E_{\textrm{\tiny R}}^{-1} \gamma_A E_{\textrm{\tiny R}}^{-1}  )\big]\,.
\ee
Once the orthogonal matrix $K$ introduced in (\ref{K-mat def}) is written in its block form
\be
\label{K-mat block}
K = \bigg( 
\begin{array}{cc}
U_K & Y_K \\ Z_K & V_K
\end{array}
\bigg)\,,
\ee
the  mode participation function corresponding to the contour function (\ref{contour s_A project - K})
is obtained by specialising (\ref{pki general}) to the case $O=K$, namely
\be
\label{pki general - K}
p_k(i) \,=\, \frac{1}{2} \Big( 
\big[(U_K)_{ki}\big]^2 + \big[(Y_K)_{ki}\big]^2 + \big[(Z_K)_{ki}\big]^2 + \big[(V_K)_{ki}\big]^2 
\Big)\,,
\ee
which fulfils the constraint (\ref{p sum rule}) because of the orthogonality of $K$, as shown in general in (\ref{orto-cond}) and (\ref{ort-cond off-diag}) for the orthogonal matrix $O$.

The symplectic condition $K J K^{\textrm{t}} = J$ for the block matrix (\ref{K-mat block}) tells us that $U_K Y_K^{\textrm{t}}$ and $V_K Z_K^{\textrm{t}}$ are symmetric matrices and that $U_K V_K^{\textrm{t}} - Y_K Z_K^{\textrm{t}} = \boldsymbol{1}$.
These relations do not simplify (\ref{pki general - K}) in the general case.

The expressions in (\ref{contour s_A project - K}) and (\ref{pki general - K}) are the main result of this section and
they provide our proposal for the contour function in generic harmonic lattices. 
This proposal fulfils the constraints (\ref{E_A sum_i}) and (\ref{positiv}).
In the next subsection we prove that it also satisfies three further requirements which correspond to a weaker version of the properties introduced in \cite{chen-vidal} for the contour functions for the entanglement entropies.

In the remaining part of this section, we discuss the cases where the reduced covariance matrix (\ref{cov mat}) is block diagonal, 
namely $R =\boldsymbol{0}$ and therefore $\gamma_A = Q \oplus P$,
being $Q_{ij} = \langle \hat{q}_i \hat{q}_j\rangle$ and $P_{ij} = \langle \hat{p}_i \hat{p}_j\rangle$ the $\ell \times \ell$ correlation matrices restricted to the subsystem $A$.
All the examples considered in our numerical analysis belong to this class.

When $\gamma_A$ is block diagonal, we have that $(\textrm{i} J \gamma_A)^2 = (PQ)\oplus (QP)$.
This implies that the symplectic spectrum can be found by first computing the spectrum of $QP$ (which is equal to the spectrum of its transpose $PQ$) and then taking its positive square root.
The antisymmetric matrix $\hat{\gamma}_A$ defined in (\ref{hatgamma def}) simplifies to
\be
\label{hat gamma_A QP}
\hat{\gamma}_A =
\bigg( \hspace{-.1cm} \begin{array}{cc}
 \boldsymbol{0}  &  Q^{1/2} \,P^{1/2} \\
 -\,P^{1/2} \, Q^{1/2}  &  \boldsymbol{0} \\
\end{array}  \hspace{-.05cm}  \bigg)\,.
\ee
By specifying (\ref{hatgamma2}) and (\ref{abs-hatgamma}) to this case, we find that also $| \hat{\gamma}_A |$ is block diagonal 
\be
\label{hat-gamma qp}
| \hat{\gamma}_A |
=
\big( \hat{\gamma}_A \hat{\gamma}_A^{\textrm{t}} \big)^{1/2}
\,=\,
 \big[Q^{1/2} \,P \,Q^{1/2}\big]^{1/2} 
\oplus
 \big[P^{1/2} \,Q \,P^{1/2}\big]^{1/2} .
\ee

Given a block diagonal covariance matrix $\gamma_A$, the symplectic matrix $W$ occurring in the Williamson's theorem (\ref{williamson th gammaA}) is block diagonal as well. 
Thus, also the factorisations (\ref{euler-dec-W}) and (\ref{polar-dec-W}) for $W$ are made by block diagonal matrices. 
In particular, considering the symplectic, symmetric and positive definite matrix  $E_{\textrm{\tiny R}}^{-1} = \Xi \oplus \Pi$, we have that $\Xi$ are $\Pi$ are symmetric and positive definite matrices.
The fact that $E_{\textrm{\tiny R}}^{-1} $ is symplectic becomes the condition  $\Xi \, \Pi^{\textrm{t}}  = \boldsymbol{1}$, which tells us that $\Xi $ and $\Pi$ are not orthogonal.
The matrices $\Xi$ and $\Pi$ can be written in terms of $Q$ and $P$ by plugging (\ref{hat-gamma qp}) into (\ref{ER gammaA})  and exploiting the block diagonal structure of $\gamma_A $. The result reads
\be
\label{XiPi from QP}
\Xi^2 \,= \, Q^{-1/2} \, \big(Q^{1/2} \,P \,Q^{1/2}\big)^{1/2} \, Q^{-1/2} \,,
\hspace{.1cm}  \qquad  \hspace{.1cm} 
\Pi^2 \,= \, P^{-1/2} \, \big(P^{1/2} \,Q \,P^{1/2}\big)^{1/2} \, P^{-1/2} \,.
\ee
Also the matrix $E_{\textrm{\tiny R}}^{-1}  \gamma_A  E_{\textrm{\tiny R}}^{-1} $ occurring in the contour function (\ref{contour s_A project - K}) becomes block diagonal in this case.
In particular, it reads
\be
\label{ERgammaER block-diag}
E_{\textrm{\tiny R}}^{-1} \, \gamma_A \, E_{\textrm{\tiny R}}^{-1} 
\,=\,
\big( \Xi \, Q\, \Xi \big)
\oplus
\big( \Pi \, P\, \Pi \big)\,,
\ee
where the blocks on the diagonal are complicated functions of $Q$ and $P$ provided by (\ref{XiPi from QP}).

Since the matrices involved in the factorisations (\ref{euler-dec-W}) and (\ref{polar-dec-W}) are block diagonal,
the orthogonal and symplectic $K$ in (\ref{K-mat block}) becomes $K= U_K \oplus V_K$, namely $Y_K = Z_K = \boldsymbol{0}$.
The orthogonality condition for this $K$ is equivalent to require that both $U_K$ and $V_K$ are orthogonal,
while the symplectic condition leads to the relation $U_K  V_K^{\textrm{t}}  = \boldsymbol{1}$. 
Combining these observations, we can conclude that $U_K=V_K$.
By employing this result and (\ref{ERgammaER block-diag}), the matrix relation (\ref{williamson th gammaA-K}) simplifies to
\be
\label{xi-pi diagonalization}
\Xi \, Q\, \Xi
\,=\,
U_K^{\textrm t}  \,D\,  U_K
\,=\,
\Pi \, P\, \Pi\,,
\ee
which tell us that the orthogonal matrix $U_K$ diagonalises the symmetric matrix $\Xi \, Q\, \Xi=\Pi \, P\, \Pi$.
This observation and (\ref{XiPi from QP}) allow us to compute $U_K$ from the correlation matrices $Q$ and $P$.

The mode participation function for $\gamma_A = Q \oplus P$ 
is obtained by first specialising (\ref{pki general - K}) to this simpler case and then employing $U_K=V_K$. The result reads
\be
\label{pki general - K - pq}
p_k(i) 
\,=\,
\big[(U_K)_{ki}\big]^2
\,.
\ee

 \begin{figure}[t!]
\vspace{.2cm}
\begin{center}
\includegraphics[width=.8\textwidth]{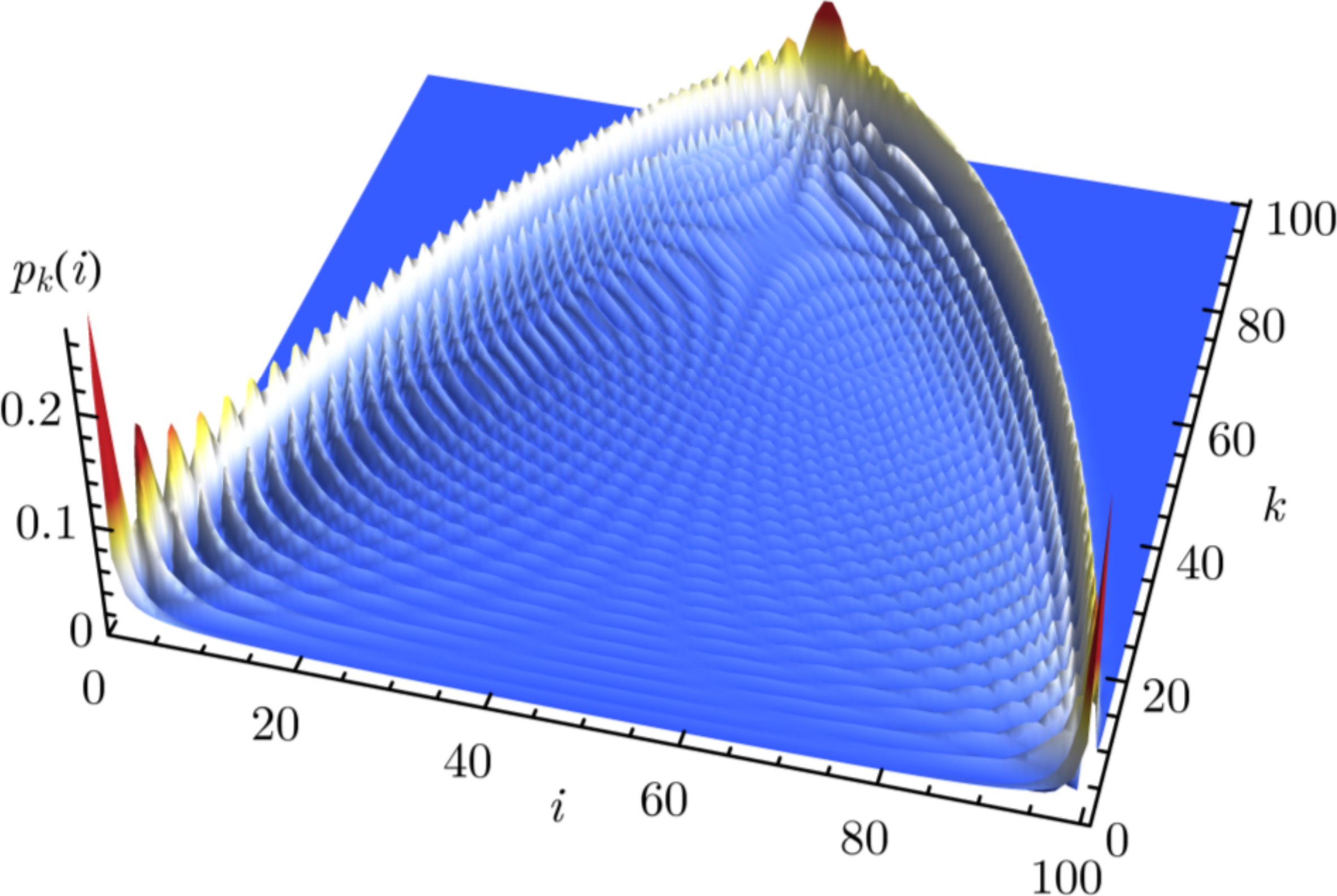}
\end{center}
\vspace{.1cm}
\caption{
The mode participation function (\ref{pki general - K - pq}) for the single interval $A$  made by 100 sites in the harmonic chain (\ref{HC ham}) with $\omega = 10^{-8}$  in the thermodynamic limit.
}
\label{fig:pki3D}
\end{figure}

 \begin{figure}[t!]
\vspace{.2cm}
\hspace{-.4cm}
\includegraphics[width=1.07\textwidth]{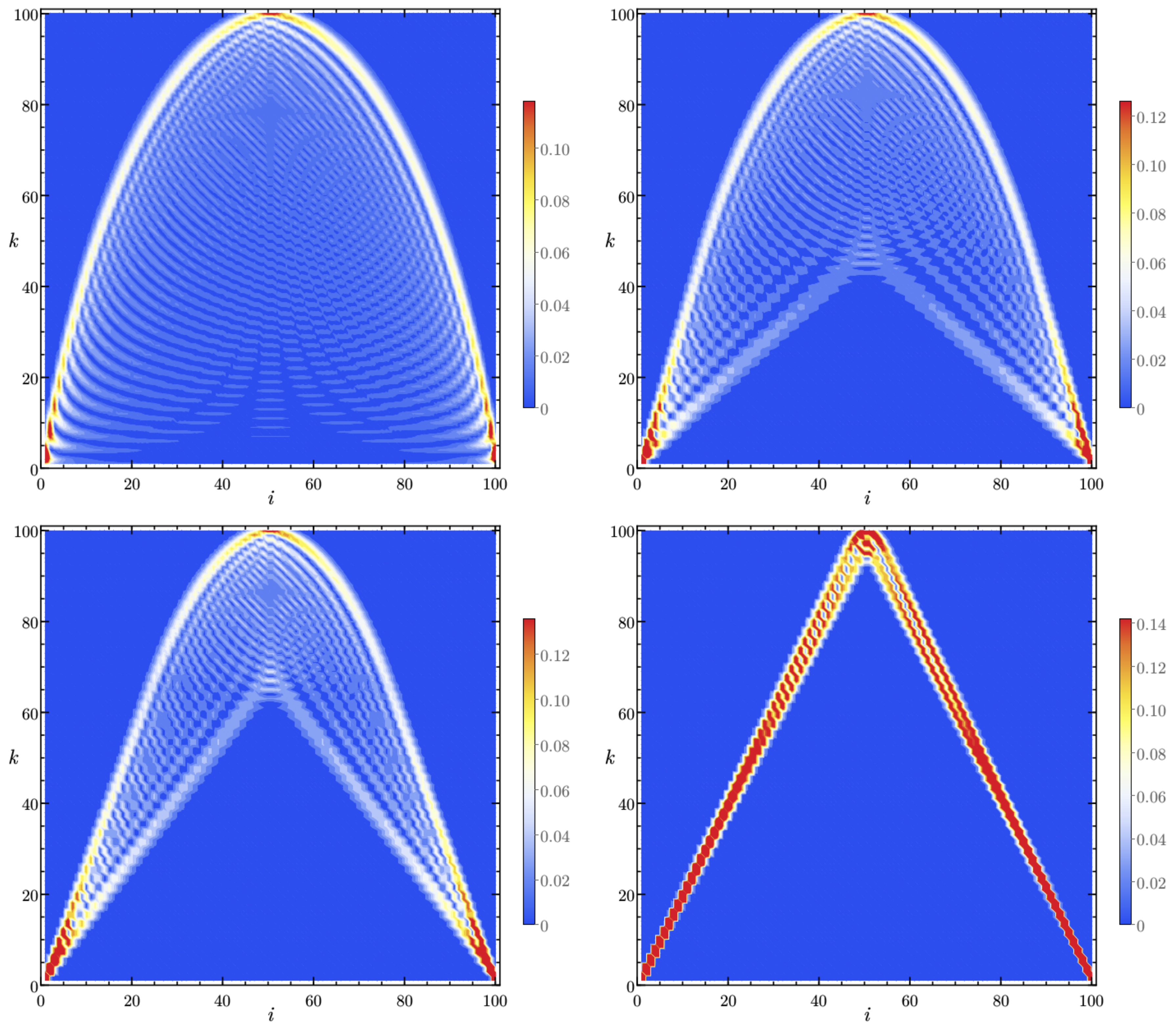}
\vspace{-.3cm}
\caption{
The mode participation function (\ref{pki general - K - pq}) for the single interval $A$  made by 100 sites
in the harmonic chain in the thermodynamic limit.
The value of the mass $\omega$ changes in the different panels:
$\omega = 10^{-8}$ (top left), 
$\omega = 0.5$ (top right), 
$\omega = 1$ (bottom left) 
and $\omega = 4$ (bottom right). 
}
\label{fig:pkiMass}
\end{figure}

 \begin{figure}[t!]
\vspace{.2cm}
\hspace{-1.75cm}
\includegraphics[width=1.15\textwidth]{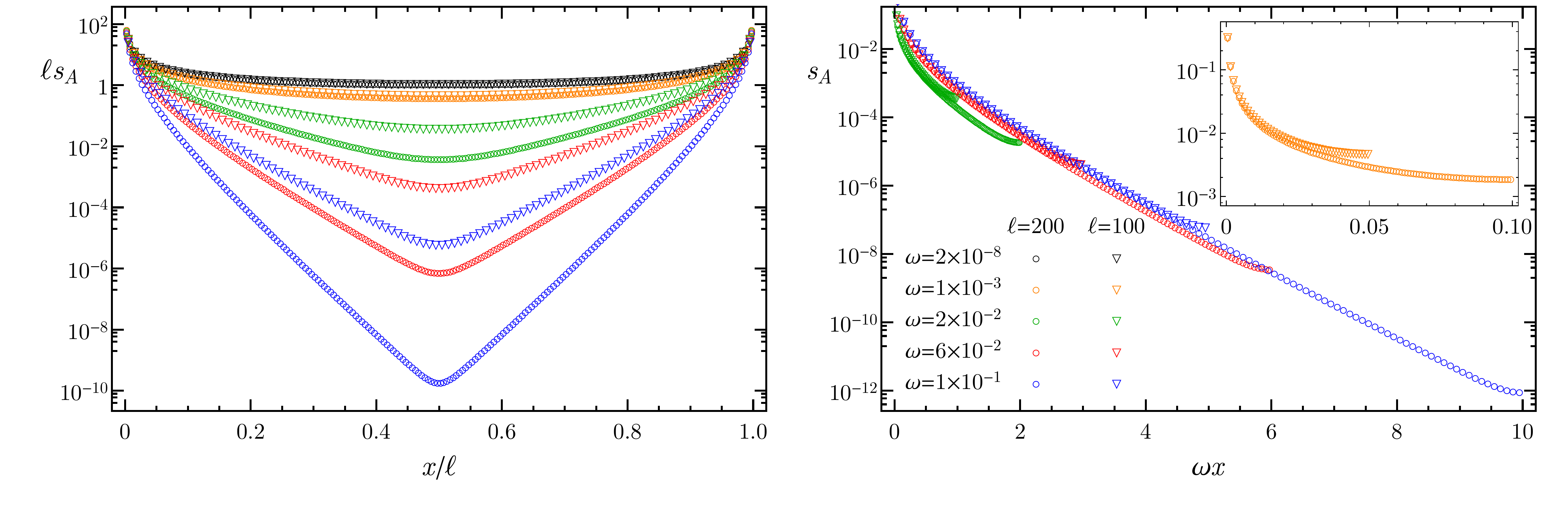}
\vspace{-.8cm}
\caption{
The contour function computed from (\ref{contour pki}) and (\ref{pki general - K - pq}) for a single interval of length $\ell$ in the harmonic chain in the thermodynamic limit
and for different values of the mass $\omega$.
}
\label{fig:ContMass}
\end{figure}

Henceforth we focus on the contour function for the entanglement entropies in the harmonic chain (\ref{HC ham}),
setting $m=\kappa=1$ without loss of generality.
Since all the examples explored in this manuscript have a block diagonal reduced covariance matrix $\gamma_A= Q \oplus P$, the mode participation function is given by (\ref{pki general - K - pq}).
The boundary conditions imposed at the endpoints of the harmonic chain are crucial in the computation of $p_k(i)$ because they determine the correlators entering in the matrices $Q$ and $P$.
All the correlators employed in our explicit examples have been collected in the appendix\,\ref{app:correlators}.
In the numerical analysis the symplectic eigenvalues $\{\sigma_k, \, 1\leqslant k \leqslant \ell \}$ have been arranged in decreasing order in terms of the label $k$, namely $\sigma_1 \geqslant \sigma_2 \geqslant \dots \geqslant \sigma_\ell$.

The correlators providing the matrix elements of $Q$ and $P$ for the harmonic chain with periodic boundary conditions in the vacuum
are given by (\ref{corrs per T=0}) and in the thermodynamic limit $L \to \infty$ they become (\ref{corrs per qq thermo}) and (\ref{corrs per pp thermo}).
Considering this regime, in Fig.\,\ref{fig:pki3D} we show the mode participation function (\ref{pki general - K - pq}) for a single interval $A$ made by $100$ sites when the mass parameter is very small ($\omega = 10^{-8}$).
We remark that we cannot set $\omega = 0$ for periodic boundary conditions because the correlator $\langle \hat{q}_i \hat{q}_j  \rangle$ diverges when $\omega \to 0$. This divergence is due to the occurrence of the zero mode, which is a consequence of the invariance under translations.

In Fig.\,\ref{fig:pki3D} the peaks of the mode participation function close to the endpoints of the interval and for small values of $k$, which correspond to the large symplectic eigenvalues, tell us that a large contribution to the entanglement entropies comes from the sites close to the endpoints of the interval.
We find it worth noticing that the mode participation function is symmetric with respect to the center of the interval and
also the occurrence of a profile  $\tilde{k}(i)$ which delimits the modes providing a non vanishing contribution to the mode participation function, namely $p_k(i) \simeq 0$ for $k > \tilde{k}(i)$ for the $i$-th site.

In Fig.\,\ref{fig:pkiMass} we have considered the dependence on the mass $\omega$ in $p_k(i)$ for the same setup described for Fig.\,\ref{fig:pki3D}.
The density plots for the mode participation function correspond to different values of $\omega$: 
the top left panel contains the same data of Fig.\,\ref{fig:pki3D}, while in the other panels $\omega$ is larger.
As the mass increases, the set of modes  providing a non vanishing contribution to $p_k(i)$ at fixed $i$ is sharper, becoming very localised for large values of $\omega$ (see the bottom right panel).
In the massive regime, where an area law occurs, the contribution of the large symplectic eigenvalues, 
which correspond to small values of $k$ and provide the largest part of the entanglement entropies, 
is localised close to the endpoints of the interval, as one observes by comparing the top panels of Fig.\,\ref{fig:pkiMass}. 
The symmetry of the profile with respect to the center of the interval is not influenced by the value of the mass, as expected.
As for the profile of the function $\tilde{k}(i)$, it clearly depends on $\omega$.

In Fig.\,\ref{fig:ContMass} we show the contour function for the entanglement entropy for  increasing values of the mass $\omega$.
As the mass increases, the power law behaviour of the contour function in the massless case becomes exponential in the regions between an endpoint and the center of the interval. 
In the right panel only the first half of the interval is considered and the collapse of the data tells us that, in the massive regime, there is a domain in $A$ where $s_A(i) \propto e^{- b \omega i}$ with the parameter $b>0$ independent of $\ell$.

\subsection{Three further requirements}
\label{sec:contour properties}

The constraints (\ref{E_A sum_i}) and (\ref{positiv}) are the minimal requirements for the contour function for the entanglement entropies.
In \cite{chen-vidal} other three reasonable properties have been introduced.
In the following, we show that (\ref{contour s_A project - K}) satisfies a weaker version of these further constraints.

Before discussing the statements of these properties, let us motivate the fact that we restrict our analysis to the special class of canonical transformations corresponding to symplectic matrices which are also orthogonal.

Given a canonical transformation implemented by a symplectic matrix $M$, the covariance matrix transforms as
$ \gamma_A \to \gamma'_A = M \gamma_A M^{\textrm t}$.
From this transformation rule and the Williamson's theorem (\ref{williamson th gammaA}), it is straightforward to realise that 
$\gamma_A' = (W')^{\textrm t} \big( D \oplus D \big) W'$, where for the symplectic matrix $W'$ we have $W' = W M^{\textrm t}$.
This shows that the symplectic spectrum is invariant under canonical transformations.
Plugging the polar decompositions $W=K E_{\textrm{\tiny R}} $ and $W' = K' E'_{\textrm{\tiny R}} $ into the relation $W' = W M^{\textrm t} $, one finds
\be
\label{polar chains}
K' E'_{\textrm{\tiny R}} 
= W M^{\textrm t} 
= K E_{\textrm{\tiny R}} M^{\textrm t} 
= \big(K M^{-1} \big) \big( M E_{\textrm{\tiny R}} M^{\textrm t}  \big)\,.
\ee
The last step is obtained by employing the identity matrix in the form $\boldsymbol{1} = M^{-1} M$, in order to recognise the symmetric matrix $M E_{\textrm{\tiny R}} M^{\textrm t}$, which is also positive definite, being $E_{\textrm{\tiny R}}$ positive definite. 
The last expression in (\ref{polar chains}) does not provide a polar decomposition for $W'$ because the symplectic matrix $K M^{-1}$ is not orthogonal for a generic symplectic matrix $M$, unless we restrict to the class of symplectic matrices $M$ which are also orthogonal.
In this case, from (\ref{polar chains}) we have
\be
\label{polar dec prime}
K' = K M^{-1}\,,
\qquad
E'_{\textrm{\tiny R}} = M E_{\textrm{\tiny R}} M^{\textrm t} \,,
\hspace{.5cm} \qquad \hspace{.5cm} 
M \in \textrm{Sp}(\ell) \cap O(2\ell)\,,
\ee
consistently with the properties of the matrices entering  in the polar decompositions of $W'$ and $W$.
By employing (\ref{polar dec prime}), for the matrix $E_{\textrm{\tiny R}}^{-1} \gamma_A E_{\textrm{\tiny R}}^{-1} $ occurring in the contour function (\ref{contour s_A project - K})
we find 
\be
\label{trans gamma_tilde}
(E_{\textrm{\tiny R}}')^{-1}\, \gamma'_A \,(E_{\textrm{\tiny R}}')^{-1}
=\,
M \big( E_{\textrm{\tiny R}}^{-1} \gamma_A E_{\textrm{\tiny R}}^{-1} \big) M^{\textrm t} \,,
\ee
which tells us that, if we restrict to the class of canonical transformations implemented by $M \in \textrm{Sp}(\ell) \cap O(2\ell)$, then the matrix $E_{\textrm{\tiny R}}^{-1} \gamma_A E_{\textrm{\tiny R}}^{-1} $ transforms like $\gamma_A $.

In the following (see (a), (b) and (c) below) we introduce a weaker version of the three requirements proposed in \cite{chen-vidal} beside (\ref{E_A sum_i}) and (\ref{positiv}) 
by considering only the transformations characterised by the subgroup $\textrm{Sp}(\ell) \cap O(2\ell)$ instead of the entire symplectic group $\textrm{Sp}(\ell)$.

{\bf (a)} {\it Spatial symmetry.}
If $\rho_A$ is invariant under a transformation relating the sites $i$ and $j$ in the subsystem $A$, 
then $s_A^{(n)}(i)=s_A^{(n)}(j)$.

This requirement is due to the possible occurrence of a spatial symmetry, which depends both on the underlying lattice model and on the choice of the spatial region $A$. Typical examples of spatial symmetries could be related to the invariance under translations, rotations or space reflections.


In order to formulate the remaining properties, we need to introduce  also the contour $s_A^{(n)}(G) $ of a subregion $G \subseteq A$ as follows
\be
\label{def contour region}
s_A^{(n)}(G)   \equiv \sum_{i  \,\in\, G} s_A^{(n)}(i)\,.
\ee
In the special case of $G = A$, from (\ref{E_A sum_i}) and (\ref{def contour region}) we find $s_A^{(n)}(A) = S_A^{(n)}$.
The contour $s_A^{(n)}(G)$ is clearly additive: for any two non intersecting spatial subsets $G \subsetneq A $ and $\tilde{G} \subsetneq A$ we have $s_A^{(n)}(G \cup \tilde{G}) = s_A^{(n)}(G) + s_A^{(n)}(\tilde{G})$.
Moreover, the contour $s_A^{(n)}(G)$ is monotonous, i.e. for $G \subseteq \tilde{G} \subseteq A $ the inequality $ s_A^{(n)}(G) \leqslant s_A^{(n)}(\tilde{G})$ holds.

{\bf (b)} {\it Invariance under local unitary transformations.}
Given a system in the state characterised by the density matrix $\rho$ and a unitary transformation $U_{G}$ acting non trivially only on $G \subseteq A$, denoting by $\rho'$ the state of the system after such transformation, the same contour $s_A^{(n)}(G)$ should be found for $\rho$ and $\rho'$.

The property (b) is motivated by the expectation that the contribution of a subregion $G$ to the entanglement should not be modified by a change of basis 
restricted to $G$.

{\bf (c)} {\it A bound.}
Given a system in the pure state $| \Psi \rangle $ and the bipartition $\mathcal{H}= \mathcal{H}_A \otimes \mathcal{H}_B$, let us assume that the further decompositions $\mathcal{H}_A = \mathcal{H}_{\Omega_A} \otimes \mathcal{H}_{\bar{\Omega}_A} $ and
$\mathcal{H}_B = \mathcal{H}_{\Omega_B} \otimes \mathcal{H}_{\bar{\Omega}_B} $ lead to the following factorisation of the state
\be
\label{factor state hyp}
| \Psi \rangle = | \Psi_{\Omega_A \Omega_B} \rangle \otimes | \Psi_{\bar{\Omega}_A \bar{\Omega}_B} \rangle\,.
 \ee
Considering a subregion $G \subseteq A$ such that $\bigotimes_{i \in G} \mathcal{H}_i \subseteq  \mathcal{H}_{\Omega_A}$, we must have that 
\be
s^{(n)}_A(G) \leqslant S^{(n)}(\Omega_A)\,,
\ee
where $S^{(n)}(\Omega_A)$ are the entanglement entropies corresponding to the reduced density matrix $\rho_{\Omega_A}$, obtained by tracing over the degrees of freedom of $\mathcal{H}_{\bar{\Omega}_A}  \otimes \mathcal{H}_B $.

The upper bound discussed in the requirement (c) can be also presented as a lower bound. 
This observation can be examined by adapting  the corresponding discussion made in \cite{chen-vidal} to our case in a straightforward way. 
We refer the interested reader to \cite{chen-vidal} for a more detailed discussion on the motivations leading to this property.

In appendix \ref{app:properties} we have shown that the contour function (\ref{contour s_A project - K}) fulfils also the constraints (a), (b) and (c) for the restricted class of transformations
characterised by symplectic and orthogonal matrices. 
It would be interesting either to extend this analysis to the whole group of the symplectic matrices or to find physical motivations leading to this restriction.

As emphasized in \cite{chen-vidal}, let us remark  that the constraints (\ref{E_A sum_i}) and (\ref{positiv}) together with the three further requirements presented in this subsection (even in the version formulated in \cite{chen-vidal}) do not characterise the contour function for the entanglement entropies in a unique way. 
Inequivalent expressions for $s_A^{(n)}(i)$ satisfying the above five properties could be found.
It would be very interesting to list a set of constraints which identify a unique construction of the contour function for the entanglement entropies.


\section{Massless regime and entanglement hamiltonians in 2d CFTs}
\label{sec:cft}

In this section we study the contour function for the entanglement entropies proposed in  \S\ref{sec:contour euler} in the massless regime of the harmonic chains where either periodic or Dirichlet boundary conditions are imposed.
For some particular configurations of the subsystem $A$ made by a single interval, the corresponding entanglement hamiltonians in two dimensional CFTs suggest a candidate for the continuum limit of the contour function for the entanglement entropies.

\begin{figure}[t!]
\vspace{.2cm}
\hspace{-.3cm}
\includegraphics[width=1.\textwidth]{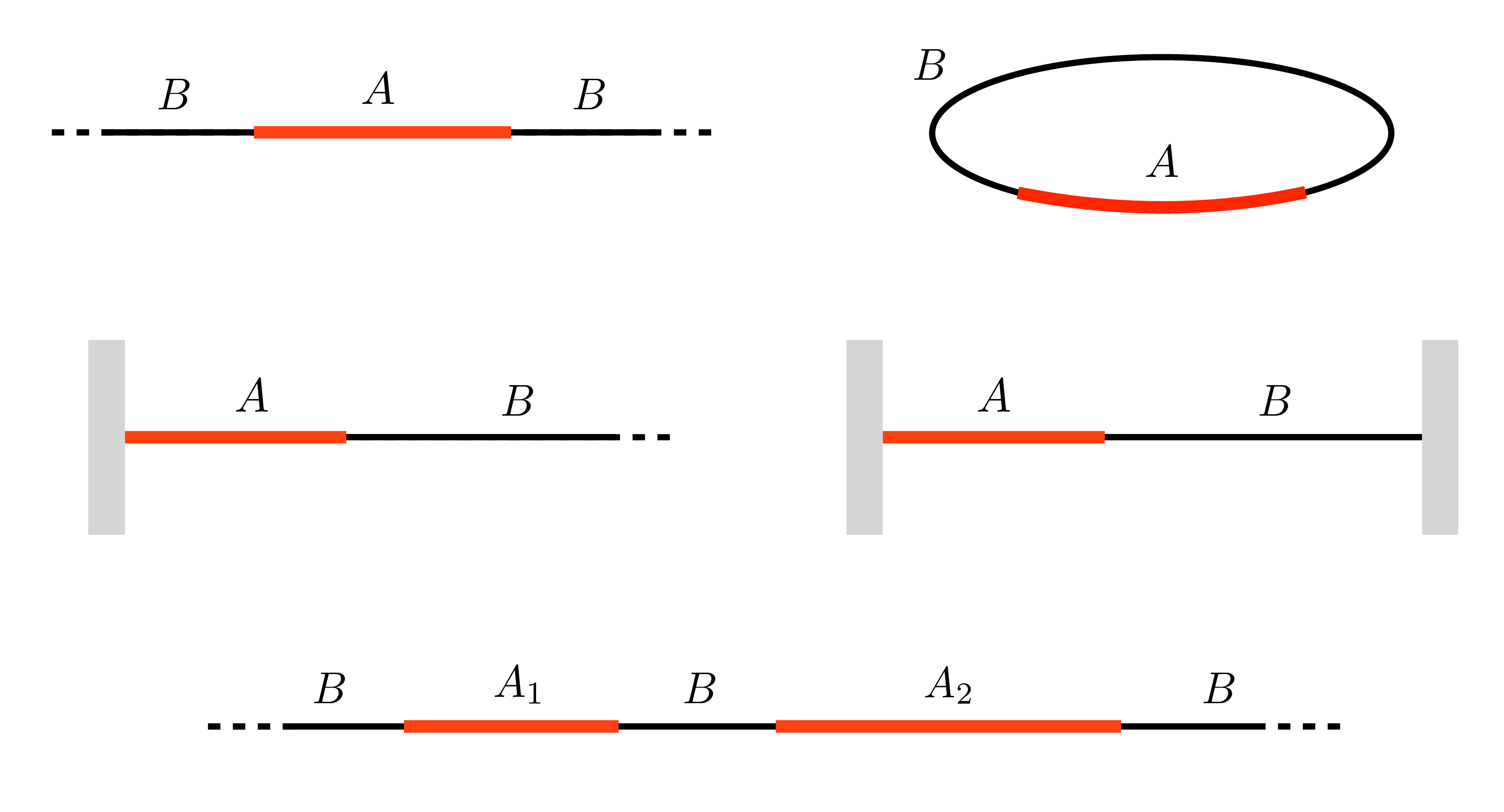}
\vspace{.1cm}
\caption{
The spatial configurations of the subsystem $A$ considered in \S\ref{sec:cft}.
Top: Single interval $A$ either in the infinite line (left) or in a finite segment with periodic boundary conditions (right).
Middle: Single interval $A$ at the beginning of either a semi-infinite line (left) or a finite segment (right), with Dirichlet boundary conditions.
Bottom: Subsystem $A = A_1 \cup A_2$ made by two disjoint intervals in the infinite line. 
}
\label{fig:ContConfigurations}
\end{figure}

In \S\ref{sec:local-weights-cft} we  briefly review the static cases where the entanglement hamiltonians can be written as an integral over the domain $A$ of the $T_{00}$ component of the energy-momentum tensor multiplied by a suitable local weight function.
We focus our attention on the results obtained in two-dimensional CFTs, where the weight function entering in the entanglement hamiltonian provides a candidate for the continuum limit of the contour function for the entanglement entropies. 
In \S\ref{sec:single-int} we explicitly study the static cases shown in the top and middle panels of Fig.\,\ref{fig:ContConfigurations}, where this analysis can be applied by performing a comparison with the corresponding numerical results from the lattice. 
In \S\ref{sec:2intervals} we consider the contour function for the entanglement entropy in the massless regime when $A= A_1 \cup A_2$ is made by two disjoint intervals in the infinite line (see the bottom panel in Fig.\,\ref{fig:ContConfigurations}). 
We emphasize that in this case a reliable candidate for the contour function in the continuum limit coming from a CFT analysis is not known.

\subsection{Weight function in some entanglement hamiltonians in 2d CFTs}
\label{sec:local-weights-cft}

The entanglement hamiltonian $K_A$ defined in (\ref{ent ham def}) provides the reduced density matrix of the subsystem; therefore in principle it contains more information than the entanglement entropies.
For free models on the lattice the entanglement hamiltonians have been studied in \cite{peschel-03-modham}.
In relativistic quantum field theories the entanglement hamiltonian is usually non-local but in certain cases it can be written in terms of integrals of local operators multiplied by suitable local weight functions. 
In a spacetime with a generic number of dimensions, this happens for very few known examples. 
Instead, for two-dimensional CFTs the large symmetry allows to enlarge the class of the configurations whose entanglement hamiltonian can be written in a local way as explained above, by including both static and time dependent examples. 
In the following we will focus on the static cases.

The most important example where the entanglement hamiltonian can be written as an integral of a local operator multiplied by a suitable weight function is the vacuum state of a QFT in flat $d$-dimensional Minkowski space ($\mathbb{R}^d$ in euclidean space) with the spatial subsystem $A$ given by the half space $x_1 >0$ and $x_i \in \mathbb{R}$ for $2\leqslant i \leqslant d-1$.
For this configuration the following crucial result due to Bisognano and Wichmann holds \cite{bw}
\be
\label{K_A bw}
K_A = \int_A x_1 \, T_{00}\, d^{d-1}x\,,
\ee 
which tells us that  $K_A$ is the generator of the Lorentz boosts along the $x_1$-direction, or of the euclidean rotations around $x_1 =0$ in the euclidean space. 

When the QFT is a $d$-dimensional CFT, a conformal transformation can be employed to map (\ref{K_A bw}) into the entanglement hamiltonian $K_A$ for a ball $A$ with radius $R$. The result of this mapping reads \cite{chm}
\be
\label{K_A ball}
K_A = \int_A \frac{R^2 - x^2}{2R}\; T_{00}\, d^{d-1}x\,.
\ee
In the special case of $d=2$, the subsystem $A$ is a single interval $A=(-R, R)$ in the infinite line (see Fig.\,\ref{fig:ContConfigurations}, top left panel) at zero temperature and (\ref{K_A ball}) 
gives the corresponding entanglement hamiltonian.
The main difference between (\ref{K_A bw}) and (\ref{K_A ball})  is the local weight function multiplying $T_{00}$ within the integrand.

For two-dimensional CFTs, the entanglement hamiltonian corresponding to other interesting configurations where the subsystem $A$ is a single interval can be written by conformal mapping the result (\ref{K_A bw}) in a suitable way. 
In \cite{ct-16} these cases have been studied within a unifying framework that we briefly review in the following.

Consider the two-dimensional euclidean spacetime (which can also have boundaries) describing the physical system.
The subsystem $A$ is an interval that can be infinite or reach the boundary of this euclidean space. 
A suitable regularization procedure consists in removing small discs of radius $\epsilon \ll 1$ around the endpoints of the interval $A$ \cite{wilczek}.
When the subsystem $A$ touches the boundary of the spacetime, then only one endpoint has to be regularised in this way. 
In \cite{ct-16} it has been argued that, after this regularisation, whenever the regularised domain can be conformally mapped into an annulus the entanglement hamiltonian corresponding to the initial configuration can be identified with the generator of the translations around the annulus along the direction orthogonal to the 
direction which connects the two boundaries of the annulus. Denoting by $w$ the complex variable parameterising the annulus and by $z$ the complex coordinate of the original domain where the regularisation discs have been removed, we have that $w=f(z)$.

All the examples that we consider in this manuscript with $A$ made by a single interval (see \S\ref{sec:single-int}) are static and fall into the class of configurations just described.
They are shown in the top and middle panels of Fig.\,\ref{fig:ContConfigurations}.
For these configurations, we can map back the generator of the translations around the annulus into the $z$-domain, finding that the entanglement hamiltonian can be written as follows \cite{ct-16}
\be
\label{K_A from f}
K_A = \int_A \frac{T_{00}(x)}{f'(x)}\, dx\,.
\ee

The analysis of  \cite{ct-16} allows to relate the local weight function $1/f'(x)$ multiplying $T_{00}(x)$ to the corresponding entanglement entropies.
The result reads
\be
\label{renyi_W}
S_A^{(n)} = \frac{c}{12}\left(1 + \frac{1}{n} \right) \mathcal{W} + C\,,
\ee
where $c$ is the central charge of the underlying CFT and $\mathcal{W}$ is the width of the annulus in the $w$-domain.
The width $\mathcal{W}$  can be computed from the above mentioned conformal transformation $f(z)$ mapping the $z$-domain into the annulus as follows
\be
\label{W from f}
\mathcal{W}  = \int_{A_\epsilon} f'(x) \, dx\,,
\ee
where we have introduced the notation $A_\epsilon$ to denote the subsystem $A$ after the removal of the small discs around the endpoints. 
The width $\mathcal{W} $ in (\ref{W from f}) is divergent as $\epsilon \to 0$.
The simplest example is a single interval $A =(0,\ell)$ in the infinite line at zero temperature (see Fig.\,\ref{fig:ContConfigurations}, top left panel).
In this case the integration domain in (\ref{W from f}) is $A_{\epsilon}=(\epsilon,\ell - \epsilon)$ and one finds that  $\mathcal{W}=2\log(\ell/\epsilon)$.

The constant $C$ in (\ref{renyi_W}) is subleading as $\epsilon \to 0$ and it is related to the boundary entropy \cite{affleck-ludwig} associated to the conformally invariant boundary conditions imposed at the boundaries of the $z$-domain, which include also the boundaries due to the small discs removed during the regularization procedure \cite{tachikawa-14, ct-16}.

The expressions in (\ref{renyi_W}) and (\ref{W from f}), which provide the entanglement entropies of those configurations where (\ref{K_A from f}) holds,
suggest a natural candidate for the contour function for the entanglement entropies in the scaling limit. 
Indeed, plugging (\ref{W from f}) into (\ref{renyi_W}) and neglecting terms which are infinitesimal as $\epsilon \to 0$, it is straightforward 
to consider the following function 
\be
\label{sAn(x) generic}
s_{A}^{(n)}(x) 
=
\frac{c}{12}\left(1 + \frac{1}{n} \right) f'(x) + \frac{C}{\ell}\,,
\ee
where $\ell$ is the length of the interval $A$.
We remark that $f(x)$ depends both on the euclidean spacetime defining the model and on the configuration of the interval.

In the following we restrict our analysis to the massless scalar field in two spacetime dimensions, which has $c=1$.
Our main goal is to compare the expression (\ref{sAn(x) generic}) with $c=1$ to the scaling limit of the contour function for the entanglement entropies proposed in \S\ref{sec:contour euler} in the special cases of one-dimensional harmonic chains in the massless regime.
In particular, we are interested in the space dependent term containing $f'(x)$ in (\ref{sAn(x) generic}) 
and not in the vertical shift characterised by the constant $C$, which is influenced by non universal features (see \cite{jin-korepin} for an exact computation of this term in a specific spin chain).

\subsection{Single interval}
\label{sec:single-int}

The first case we consider is a single interval $A$ of length $\ell$ in the infinite line (top left panel of Fig.\,\ref{fig:ContConfigurations}), when the whole system is in the ground state. 
Because of the invariance under translations, we can fix the origin in the first endpoint of the interval and therefore $A = (0,\ell)$.
The map transforming the configuration obtained by removing small discs of radius $\epsilon$ around the endpoints of $A$
 into the annulus is given by 
\be
f(x) = \log \bigg( \frac{x}{\ell - x} \bigg)\,,
\qquad
x\in (0,\ell)\,.
\ee 
Taking the derivative of this expression and plugging the result into (\ref{sAn(x) generic}), for $c=1$ we get
\be
\label{contour 1int cft}
\ell \, s_A^{(n)}(x) = 
\frac{1}{12}\left(1 + \frac{1}{n} \right)
\frac{1}{\big(1-x/\ell \big)\,  x/\ell } + C\,.
\ee

 \begin{figure}[t!]
\vspace{.2cm}
\hspace{-1.6cm}
\includegraphics[width=1.15\textwidth]{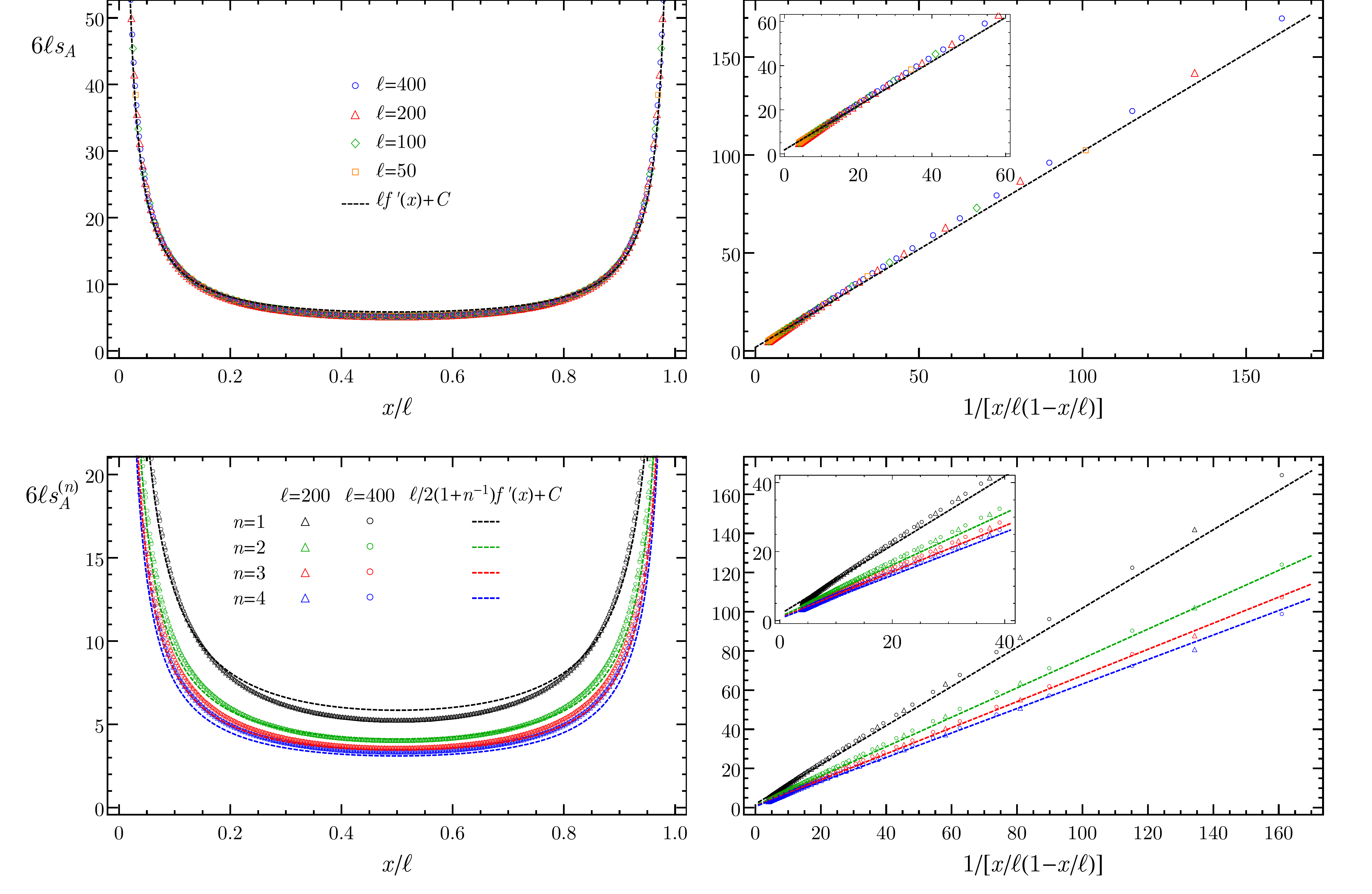}
\vspace{-.8cm}
\caption{
The contour function for the entanglement entropies described in \S\ref{sec:contour euler} (see (\ref{contour pki}) and (\ref{pki general - K - pq})) for a single interval of length $\ell$ in the periodic chain in the thermodynamic limit with $\omega \ell = 4\times 10^{-4}$.
The dashed curves correspond to the CFT formula (\ref{contour 1int cft}).
}
\label{fig:ContLinf}
\end{figure}

In Fig.\,\ref{fig:ContLinf} we show the contour function for the entanglement entropies constructed in \S\ref{sec:contour euler} when $A$ is a single interval of various lengths in the infinite line.
The data have been found by considering the harmonic chain (\ref{HC ham}) in the thermodynamic regime, whose correlators are (\ref{corrs per qq thermo}) and (\ref{corrs per pp thermo}). 
The data are obtained for very small but non vanishing mass $\omega$. 
Indeed, this parameter cannot be set to zero because the correlator $\braket{\hat q_i \hat q_j} $ diverges in this limit. This is due to the occurrence of the zero mode, which is a consequence of the invariance under translations of the model. 

In Fig.\,\ref{fig:ContLinf} the panels on the right have the same vertical axis of the corresponding ones on the left, but
the parameterisations of the horizontal axis are different.
In particular, we have employed the function suggested by the CFT result (\ref{contour 1int cft}), 
in order to show more clearly the behaviour of the lattice data nearby the endpoints of the interval. 
This way to display the data will be adopted also in other figures corresponding to the subsequent examples. 

The dashed curves in Fig.\,\ref{fig:ContLinf} are obtained from the CFT formula (\ref{contour 1int cft}), 
where the constant $C$ is not universal.
In order to fix $C$ through a method applicable to all the examples considered throughout  this subsection, we can employ (\ref{sAn(x) generic}).
In particular, given the lattice points for the contour function, we impose that 
$\sum_{i=1}^{\ell_{\textrm{\tiny max}}} s_A^{(n)}(i) 
= C + \tfrac{1}{12}(1+1/n) \sum_{i=1}^{\ell_{\textrm{\tiny max}}} f'(i-1/2)\big|_{\ell_{\textrm{\tiny max}}}$, where 
 $\ell_{\textrm{\tiny max}}$ is the size of the largest interval considered in the corresponding numerical analysis and
 the shifted argument for $f'$ is introduced because this function diverges at the endpoints of the interval.
A deeper analysis that we leave for future work could lead to an insightful method to fix the constant $C$ in (\ref{sAn(x) generic}) from the numerical data.

 A very good collapse of the numerical data corresponding to different values of $\ell$ is observed in the top panels of Fig.\,\ref{fig:ContLinf}, already for small intervals.  
 Few points close to the endpoints of the interval, where the contour function diverges, have not been shown because non universal features due to the lattice are expected very close to the endpoints of the interval. 
 The agreement between the CFT expression (\ref{contour 1int cft}) and the lattice data is very good nearby the endpoints of the interval and gets worse around the center. 
 This agrees with the expectation that the universal part of the entanglement entropies is determined by the regions close to the endpoints of the interval. 
 The disagreement between the CFT formula and the lattice data around the center is due to non universal contributions.

    \begin{figure}[t!]
\vspace{.2cm}
\hspace{-1.6cm}
\includegraphics[width=1.15\textwidth]{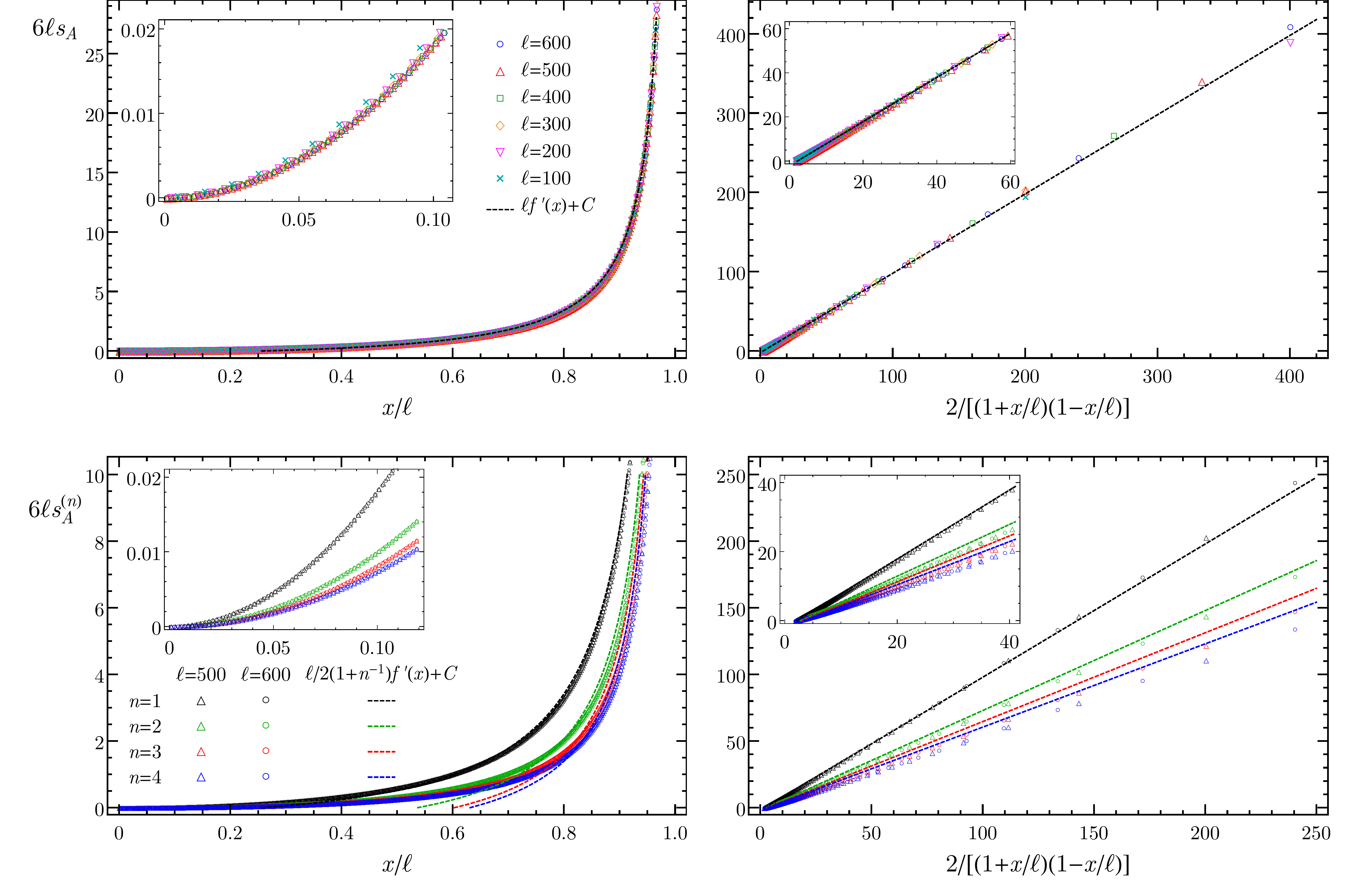}
\vspace{-.7cm}
\caption{
The contour function for the entanglement entropies described in \S\ref{sec:contour euler} for a single interval of length $\ell$ at the beginning of an open chain with Dirichlet boundary conditions in the thermodynamic limit with $\omega=0$.
The dashed curves are obtained from the CFT formula (\ref{contour 1int cft semiinf}).
}
\label{fig:Contsemiinf}
\end{figure}

  In the next example the whole system is on a semi-infinite line, therefore boundary conditions must be imposed on the physical boundary.
  The subsystem $A= (0,\ell)$ is a single interval of length $\ell$ placed at the beginning of the semi-infinite line (middle left panel of Fig.\,\ref{fig:ContConfigurations}).
  The regularizing procedure requires to remove a small disc of radius $\epsilon$ only around the second endpoint of $A$. 
  Mapping the resulting configuration into the annular geometry, it is worth remarking that in this case the boundary conditions on the two boundaries of the annulus may be different.
The conformal map reads
\be
\label{map f bcft}
f(x) = \log \bigg( \frac{x + \ell}{\ell - x} \bigg)\,,
\qquad
x\in (0,\ell)\,.
\ee 
Given this function, for this example the expression (\ref{sAn(x) generic}) specialised to $c=1$ becomes
\be
\label{contour 1int cft semiinf}
\ell \, s_A^{(n)}(x) = 
\frac{1}{6}\left(1 + \frac{1}{n} \right)
\frac{1}{\big(1-x/\ell \big) \big(1+x/\ell \big)}   + C\,.
\ee

As for the numerical analysis on the lattice, for this configuration we consider the harmonic chain defined on a segment with Dirichlet boundary conditions imposed to both its endpoints and then we take the thermodynamic limit.
In this case the invariance under translations is broken because of the presence of the physical boundary; therefore the zero mode does not occur in the correlators (\ref{qq open}) and (\ref{pp open}), which become (\ref{corr qq dirichlet thermo}) and (\ref{corr pp dirichlet thermo}) respectively in the thermodynamic limit. This fact allows us to set $\omega =0$ in the numerical analysis. 

In Fig.\,\ref{fig:Contsemiinf} we show the contour function for the entanglement entropies described in \S\ref{sec:contour euler} with 
$\omega =0$.
Since we have imposed Dirichlet boundary conditions, the curves for the contour function obtained from the lattice data pass through the origin, as highlighted in the insets of the left panels. 

The dashed curves in Fig.\,\ref{fig:Contsemiinf} (only their positive part are shown) correspond to the CFT formula (\ref{contour 1int cft semiinf}) where the constant $C$ has been fixed as explained above by employing the map (\ref{map f bcft}).
The CFT curves nicely reproduce the divergent behaviour of the lattice data nearby the endpoint of the interval, while they cannot capture the lattice data for the contour function around the boundary, which introduces non universal features.

  \begin{figure}[t!]
\vspace{.2cm}
\hspace{-1.7cm}
\includegraphics[width=1.15\textwidth]{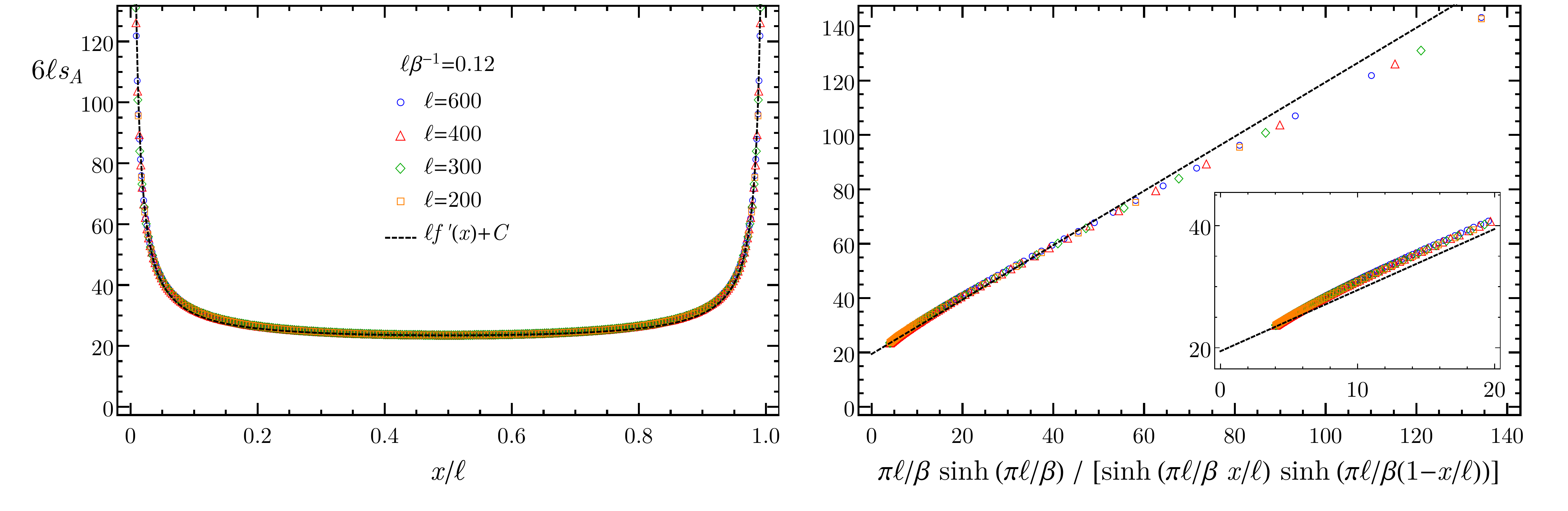}
\vspace{-.8cm}
\caption{
The contour function for the entanglement entropy described in \S\ref{sec:contour euler} for a single interval of length $\ell$ in a periodic chain at finite temperature $T=1/\beta$ in the thermodynamic limit with $\omega\beta=10^{-3}$. 
The dashed curve corresponds to the CFT formula (\ref{contour cft finite T}) specialised to $n=1$.
}
\label{fig:ContTemp}
\end{figure}

The above analysis can be applied also for a single interval $A$ of length $\ell$ in the infinite line when the whole system is in a thermal state at temperature $T=1/\beta$ (top left panel of Fig.\,\ref{fig:ContConfigurations}).
Setting the interval in $A=(0,\ell)$  and removing a small disc around both the endpoints of $A$,
the conformal map relating this configuration to the annular domain reads
  \be
  \label{mapping temp}
f(x) = \log \bigg( \frac{e^{2\pi  \, x/\beta} -1}{e^{2\pi  \, \ell/\beta}  - e^{2\pi  \, x/\beta} } \bigg)\,,
\qquad
x\in (0,\ell) \,.
\ee 
Given this expression and fixing $c=1$, one finds that (\ref{sAn(x) generic}) becomes
\be
\label{contour cft finite T}
\ell \, s_A^{(n)}(x) = 
\frac{1}{12}\left(1 + \frac{1}{n} \right)
\frac{( \pi \ell/\beta)  \sinh(\pi \ell/\beta )}{\sinh((x/\ell)\,  \pi \ell/\beta) \, \sinh((1- x/\ell)\,  \pi \ell/\beta)} + C\,,
\ee
which has been written in a form highlighting the dependence on the ratios $x/\ell \in (0,1)$ and $\ell/\beta$ in the
universal term.

In Fig.\,\ref{fig:ContTemp} we show the contour function for the entanglement entropy of a single interval of length $\ell$ in the harmonic lattice at finite temperature $T\equiv 1/\beta$ and in the thermodynamic regime.
The contour function has been evaluated by employing (\ref{contour pki}) and (\ref{pki general - K - pq}), with the correlators (\ref{qq per T therm}) and (\ref{pp per T therm}). 
Since this model is invariant under translations, the zero mode occurs in the correlators and this prevents us from setting $\omega = 0$.
The data shown in the figure are obtained for very small non vanishing mass $\omega$.

The dashed curve in Fig.\,\ref{fig:ContTemp} corresponds to the CFT formula (\ref{contour cft finite T}) specialised to $n=1$, 
with the vertical shift given by the constant $C$ fixed through the function (\ref{mapping temp}), as explained in the first example of this subsection.

   In the remaining two examples we consider finite length systems in their ground state. 
   Different boundary conditions (either periodic or Dirichlet) are imposed.

   \begin{figure}[t!]
\vspace{.2cm}
\hspace{-1.7cm}
\includegraphics[width=1.15\textwidth]{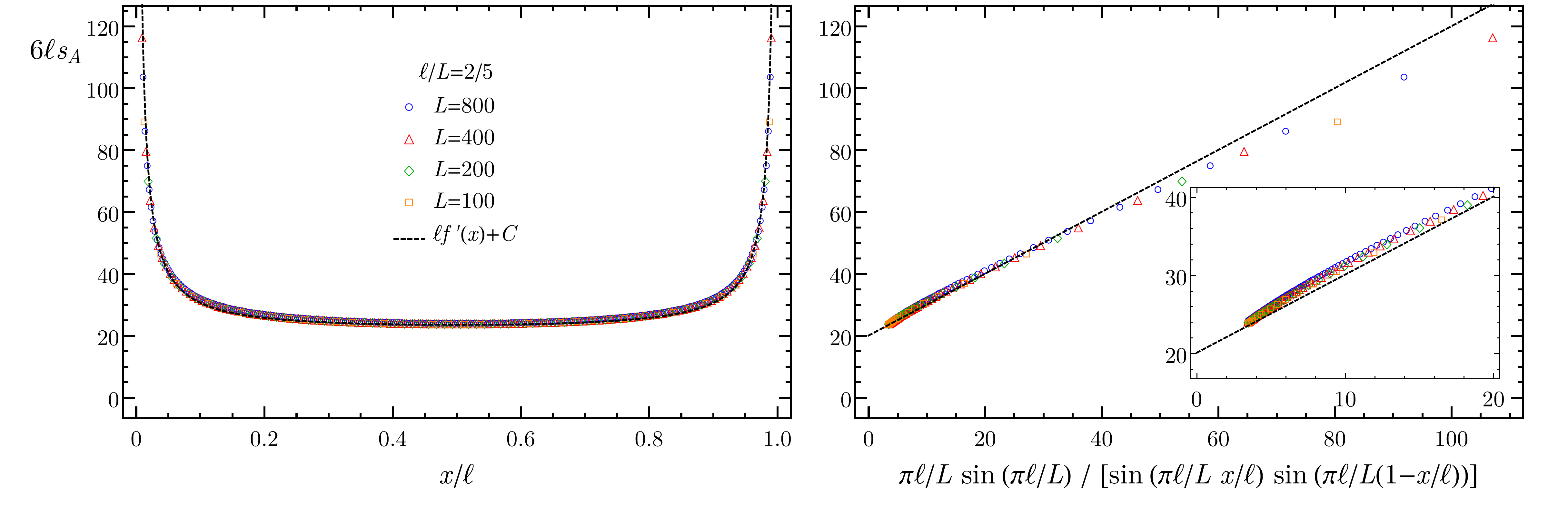}
\vspace{-.8cm}
\caption{
The contour function for the entanglement entropy described in \S\ref{sec:contour euler} for a single interval of length $\ell$ in a periodic chain of finite length $L$ with $\omega L = 8 \times 10^{-4}$. 
The dashed curve corresponds to the CFT formula (\ref{contour cft finite L}) specialised to $n=1$.
}
\label{fig:ContPBC}
\end{figure}
   
   An interesting configuration to study is given by a single interval $A$ of length $\ell$ in a spatial circle whose length is $L>\ell$ (top right panel of Fig.\,\ref{fig:ContConfigurations}).
   Setting the system in the finite segment $(0,L)$ and fixing the position of the interval to $A=(0,\ell)$, periodic boundary conditions are imposed at $x=0$ and $x=L$. The system is invariant under translations along the circle. 
   After the removal of the small discs at the endpoints of $A$ in the euclidean spacetime describing this system (an infinite cylinder), the conformal map which sends the resulting domain into the annulus reads
   \be
f(x) = \log \bigg( \frac{e^{2\pi \textrm{i} \, x/L} -1}{e^{2\pi \textrm{i} \, \ell/L}  - e^{2\pi \textrm{i} \, x/L} } \bigg)\,,
\qquad
x\in (0,\ell) \subseteq (0,L)\,.
\ee 
Applying the formula (\ref{sAn(x) generic}) for this map and for $c=1$, one obtains
\be
\label{contour cft finite L}
\ell \, s_A^{(n)}(x) = 
\frac{1}{12}\left(1 + \frac{1}{n} \right)
\frac{(\pi \ell / L)  \sin(\pi \ell / L)}{\sin((x/\ell)\, \pi \ell / L) \, \sin((1- x/\ell)\, \pi \ell / L )}   + C\,.
\ee
This expression  has been written in a form highlighting the fact that the universal term in the r.h.s. is a function of the two ratios $x/\ell \in (0,1)$ and $\ell/L \in (0,1)$.

In Fig.\,\ref{fig:ContPBC} we plot the contour function for the entanglement entropy discussed in \S\ref{sec:contour euler} for a single interval of length $\ell$ in a periodic harmonic chain of finite length $L$ in its ground state. 
The correlators employed for this numerical analysis have been written in (\ref{corrs per T=0}).
Notice that $\langle \hat{q}_i \hat{q}_j  \rangle$ diverges as $\omega \to 0$  because of the occurrence of the zero mode $k=0$; therefore the mass $\omega$ must be very small but non vanishing. 
The dashed curve in Fig.\,\ref{fig:ContPBC} has been obtained from the CFT formula (\ref{contour cft finite L}) in the special case of $n=1$, with the constant $C$ fixed by employing the function in (\ref{contour cft finite L}) as explained for the previous cases.

      \begin{figure}[t!]
\vspace{.2cm}
\hspace{-1.6cm}
\includegraphics[width=1.15\textwidth]{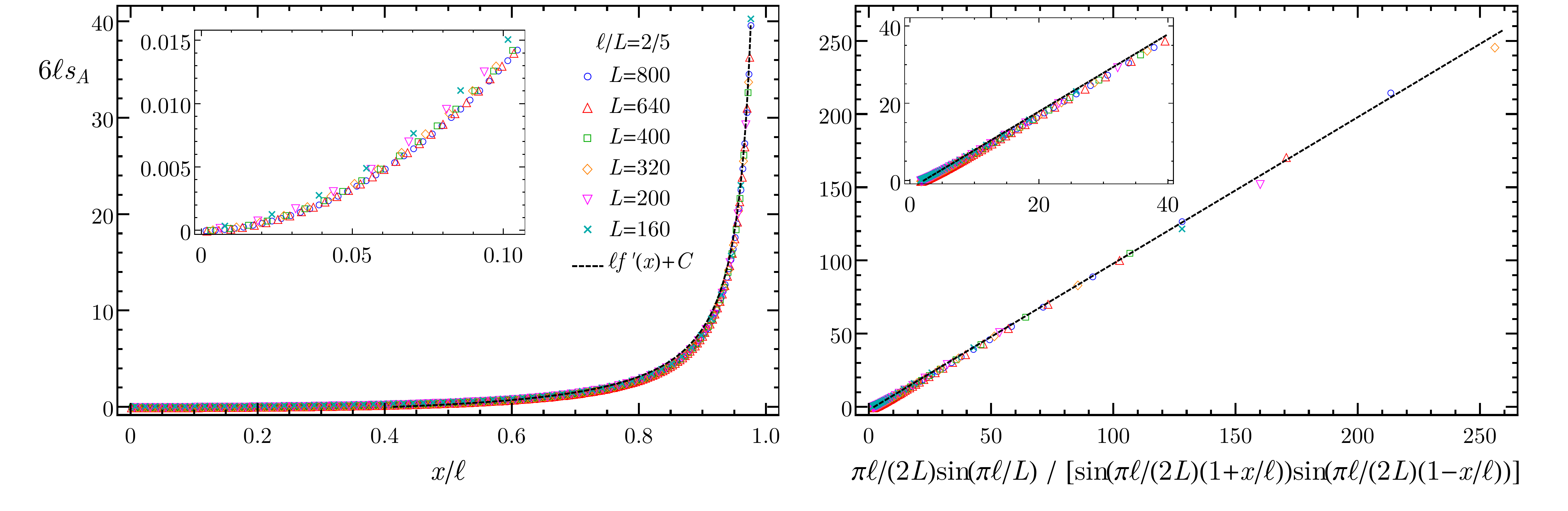}
\vspace{-.6cm}
\caption{
The contour function for the entanglement entropy described in \S\ref{sec:contour euler} for a single interval of length $\ell$ at the beginning of an open chain of finite length $L$ with Dirichlet boundary conditions and $\omega=0$.
The dashed curve corresponds to the CFT formula (\ref{contour obc cft}) with $n=1$.
}
\label{fig:ContOBC}
\end{figure}
  
Our last example in the class of configurations whose entanglement hamiltonian in the continuum can be written in the form (\ref{K_A from f}), 
is given by a system in a finite segment $(0,L)$ where the same boundary conditions are imposed at both its endpoints. 
The subsystem $A = (0,\ell)$ is a single interval of length $\ell < L$ sharing an endpoint with the entire system (middle right panel of Fig.\,\ref{fig:ContConfigurations}).
By adapting the procedure described above to this case through the map given in \cite{ct-16}, one finds that
\be
\label{contour obc cft}
\ell \, s_A^{(n)}(x) = 
\frac{1}{12}\left(1 + \frac{1}{n} \right)
\frac{( \pi \ell/2L)  \sin(\pi \ell/L )}{\sin((1+x/\ell)\,  \pi \ell/2L) \, \sin((1- x/\ell)\,  \pi \ell/2L)}  + C \,.
\ee    

In Fig.\,\ref{fig:ContOBC} we compare this CFT formula, with $n=1$ (dashed curve) and the constant $C$ fixed by adapting to this case the method explained above, to  the contour function for the entanglement entropy constructed in \S\ref{sec:contour euler} for a single interval made by $\ell$ sites at the beginning of the massless harmonic chain in a segment of finite length $L$ with Dirichlet boundary conditions imposed at its endpoints. 
The correlators employed for this numerical analysis are given by (\ref{qq open}) and (\ref{pp open}) with $\omega = 0$.
Because of the Dirichlet boundary conditions, the contour obtained from the lattice data pass through the origin, as highlighted in the inset of the panel on the left. 

In this case, where the massless regime can be considered without approximation, the agreement between the lattice data and the CFT curve is quite remarkable nearby the second endpoint of the interval and it gets worse close to the boundary. 
This is expected from the fact that the universal part in the CFT expression (\ref{contour obc cft}) does not contain information about the specific boundary conditions imposed at the endpoints of the segment $(0,L)$.
This non universal information is encoded in the constant $C$.

\subsection{Two disjoint intervals}
\label{sec:2intervals}

      A very interesting configuration to study involves a subsystem $A=A_1 \cup A_2$ made by the union of two disjoint intervals $A_1$ and $A_2$.
      In an infinite system, we denote by  $\ell_1$ and $\ell_2$  the lengths of $A_1$ and $A_2$ respectively, while $d$ is the distance separating the intervals (see the bottom panel of Fig.\,\ref{fig:ContConfigurations}).

 \begin{figure}[t!]
\vspace{.2cm}
\hspace{-.8cm}
\includegraphics[width=1.14\textwidth]{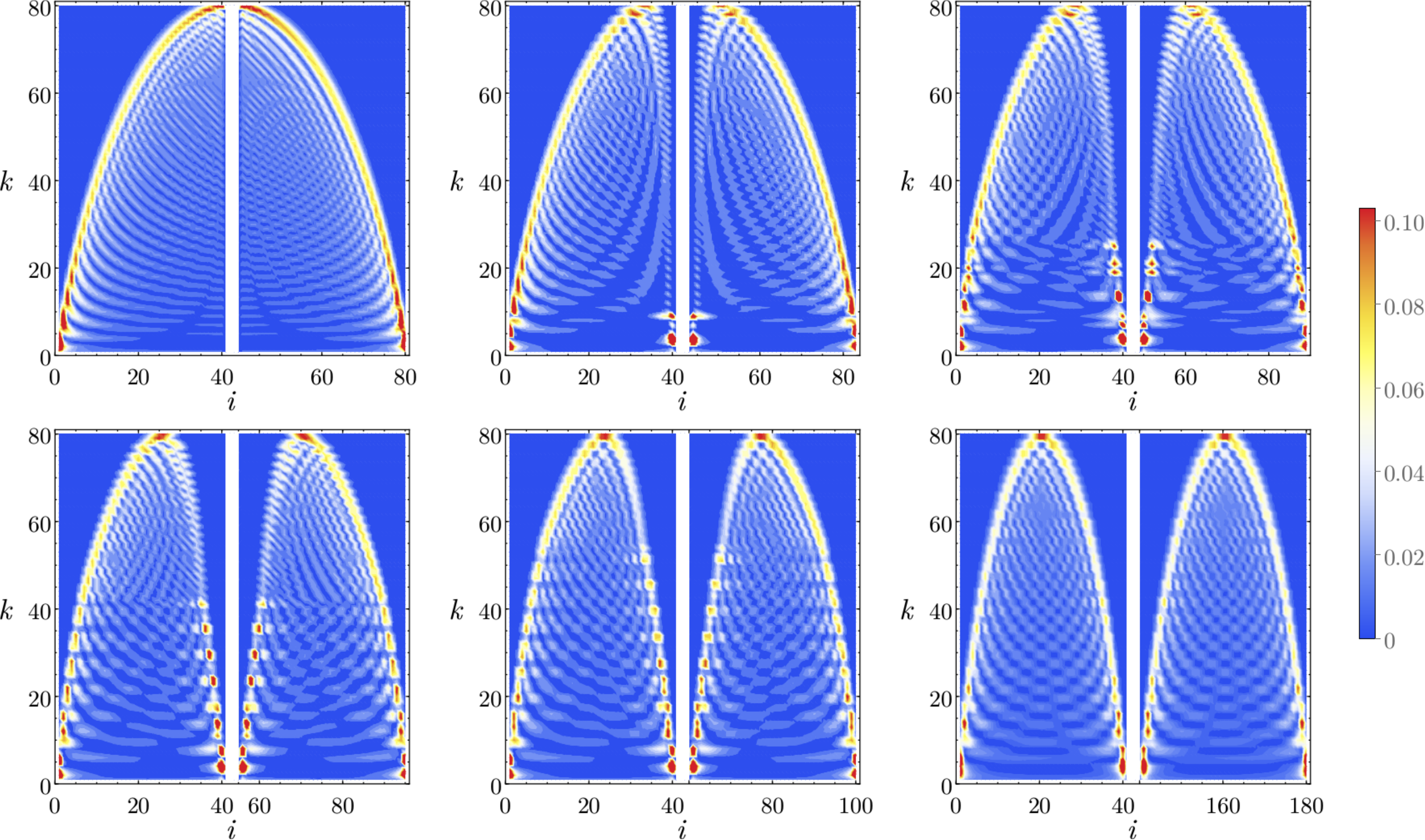}
\vspace{-.3cm}
\caption{
The mode participation function described in \S\ref{sec:contour euler} for the harmonic chain in the thermodynamic regime with mass  
$\omega=10^{-8}$ when the subsystem $A$ is made by two equal intervals $\ell_1=\ell_2=40$ separated by a distance $d$.
The panels correspond to different values of $d$, which can be inferred  from the values of the index $i$ labelling the sites in $A$ on the horizontal axis. 
Top panels: $d= 0$ (left), $d= 3$ (middle) and $d= 9$ (right). 
Bottom panels: $d=15$ (left), $d= 20$ (middle) and $d= 100$ (right). 
}
\label{fig:Part2int}
\end{figure}

Considering the harmonic chain in the thermodynamic regime and for a very small mass ($\omega = 10^{-8}$),
in Fig.\,\ref{fig:Part2int} we show the mode participation function (\ref{pki general - K - pq}) for two disjoint and equal intervals with $\ell_1=\ell_2=40$ separated by a distance $d$. The panels correspond to increasing values of $d$, starting from $d=0$, when $A_1 \cup A_2$ is a single interval of length $\ell_1 +\ell_2=80$, 
until a large value of $d$ (in the bottom right panel $d=100$), when the intervals are very far apart. 
For small $d$ the profile of the mode participation function restricted to one of the two intervals is clearly influenced by the presence of the other interval, 
while for large distances it becomes qualitatively like the mode participation function of a single interval (see the top left panel of Fig.\,\ref{fig:pkiMass} 
or the top left panel of Fig.\,\ref{fig:Part2int}).

\begin{figure}[t!]
\vspace{.2cm}
\hspace{-1.7cm}
\includegraphics[width=1.15\textwidth]{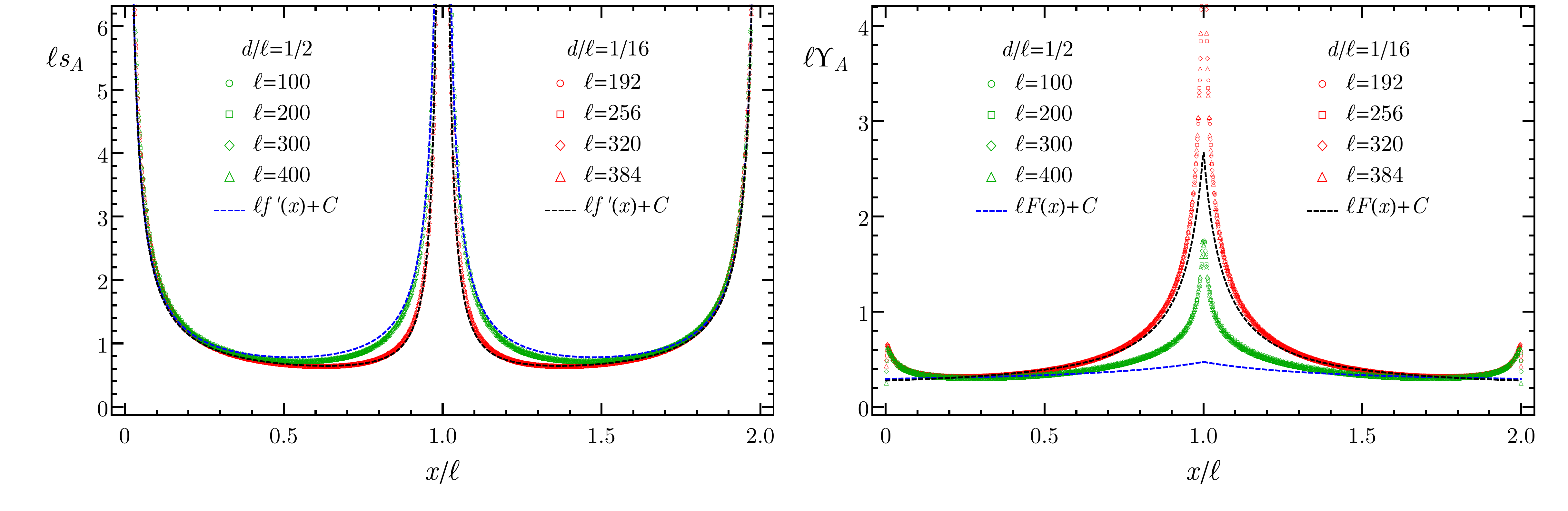}
\vspace{-.6cm}
\caption{
Left: The contour function for the entanglement entropy described in \S\ref{sec:contour euler} for the periodic chain in the thermodynamic regime when $A$ is made by two equal intervals of length $\ell$ separated by a distance $d$ and $\omega=10^{-8}$.
The dashed curves are obtained from (\ref{sAn(x) generic}) and (\ref{f(x) 2int fake}), where
the constant $C$ has been fixed by maximising the overlap with the lattice data for a given ratio $d/\ell$.
Right: The contour function for the mutual information (see Eq.\,(\ref{contour MI lattice}) for $n=1$) in the setup of the left panel. 
The dashed curves correspond to (\ref{contour MI cft}) with $n=1$.
}
\label{fig:dis}
\end{figure}

Given the lattice setup described for Fig.\,\ref{fig:Part2int}, in Fig.\,\ref{fig:dis} we show the contour function for the entanglement entropy of two equal and disjoint intervals ($\ell_1 = \ell_2 \equiv \ell$) for various lengths $\ell$ and for two fixed values of the dimensionless ratio $d/\ell$.
All the data corresponding to the same value of $d/\ell$ nicely collapse on the same curve. 

It would be very interesting to find an analytic function through a CFT analysis such that its integral over $A=A_1 \cup A_2$ provides the entanglement entropy of two disjoint intervals for the massless scalar in two spacetime dimensions.
This function, which  is not known in the literature, would be a natural candidate to compare against the numerical results for the contour function for the entanglement entropy shown in Fig.\,\ref{fig:dis} (left panel).

Despite the lack of a candidate function derived from CFT methods, we find it worth employing a function which captures the expected divergencies close to the endpoints of the two intervals. This function provides only part of the expected CFT result for the entanglement entropies of two disjoint intervals. 

Inspired by the results of \cite{cc-04,casini-2int}, let us consider 
\be
\label{f(x) 2int fake}
f(x) \,=\, \log \bigg( \frac{(x-u_1)(x-u_2)}{(v_1 - x)(v_2 - x)} \bigg)
= f_{A_1}(x) + f_{A_2}(x)\,,
\ee
where $x\in (u_1, v_1) \cup (u_2, v_2)$ and we have introduced the following notation
\be
\label{f(x) 1int A_j}
f_{A_j}(x) = \log \bigg( \frac{x-u_j}{v_j - x} \bigg)\,.
\ee

We stress that the expression obtained by plugging $f'(x)$ from (\ref{f(x) 2int fake}) into (\ref{sAn(x) generic}) does not
provide the expected result for the massless free boson.
Indeed, by removing the small discs of radius $\epsilon$ around the four endpoints of the intervals and integrating $f'(x)$ from (\ref{f(x) 2int fake}) over the remaining domain $x\in A_\epsilon=(u_1+\epsilon, v_1-\epsilon) \cup (u_2+\epsilon, v_2-\epsilon)$, one obtains
\be
\label{SA 2int wrong}
\frac{c}{12}\left(1 + \frac{1}{n} \right) 
\int_{A_\epsilon} f'(x)  \, dx 
\,=\,
 \frac{2\Delta_n}{1-n}  \,\log \mathcal{P}_{A}\,,
\ee
where 
\be
 \Delta_n \equiv \frac{c}{12}\left(n-\frac{1}{n}\right) ,
 \qquad
\mathcal{P}_{A} \equiv
 \frac{\epsilon^2\,(u_2-u_1)(v_2-v_1)} {(v_1-u_1)(v_2-u_2)(v_2-u_1)(v_1-u_2)}\,.
\ee
This is not the results found in \cite{furukawa-09, cct-09} for the entanglement entropies of two disjoint interval for this model because an important additive term is missing in (\ref{SA 2int wrong}).
In particular, by adopting the notation of \cite{cct-09}, the missing term is given by $\tfrac{1}{1-n} \log \mathcal{F}_n$, where $\mathcal{F}_n$ is a function of the harmonic ratio of the four endpoints which has been computed analytically for $n\geqslant 2$ in \cite{furukawa-09, cct-09}.
In \cite{cct-09}, an explicit expression written as an integral of an analytic function has been found for the $n=1$ case.

In the left panel of Fig.\,\ref{fig:dis}, the dashed curves correspond to the r.h.s. of (\ref{sAn(x) generic}) with the function $f(x)$ given by (\ref{f(x) 2int fake}).
The vertical shift due to the constant $C$ in (\ref{sAn(x) generic}) has been fixed by trying to maximise the overlap with the lattice data. 
We cannot employ the method adopted in the previous single interval cases because for two disjoint intervals a non trivial function obtained from CFT providing the entanglement entropies once integrated over $A_1 \cup A_2$ is not available for the free massless scalar.  
Despite the fact that (\ref{f(x) 2int fake}) is not expected to provide the correct result, the agreement between the dashed curves and the corresponding lattice data is quite satisfactory as shown in the left panel of Fig.\,\ref{fig:dis}.

In order to highlight the missing contribution due to the term $\tfrac{1}{1-n} \log \mathcal{F}_n$ in the R\'enyi entropies,
let us introduce the contour function corresponding to the combination of entanglement entropies in (\ref{MI renyi def}), which gives the contour function for the mutual information when $n \to 1$.
It reads
\be
\label{contour MI lattice}
\Upsilon^{(n)}_{A}(i)
\,\equiv\,
s^{(n)}_{A_1}(i) \, \theta_{A_1}(i) 
+ s^{(n)}_{A_2}(i) \, \theta_{A_2}(i) 
- s^{(n)}_{A}(i) \,,
\qquad
i \in A= A_1 \cup A_2\,,
\ee
where the contour functions are constructed as explained in \S\ref{sec:contour euler}.
The functions $s^{(n)}_{A_j}(i)$ and $\theta_{A_j}(i) $ in (\ref{contour MI lattice}) are respectively the contour function and the Heaviside step function corresponding to the interval $A_j$.

In the continuum, let us consider the following CFT expression 
\be
\label{contour MI cft}
\Upsilon^{(n)}_{A}(x)
=
\frac{c}{12}\left(1+\frac{1}{n}\right) F(x)  + C\,,
\qquad
x\in (u_1, v_1) \cup (u_2, v_2)\,,
\ee
being $c$ is the central charge ($c=1$ for the massless free boson) and
\bea
& & \hspace{-.7cm}
F(x) 
\,\equiv\, 
 \theta_{A_1}(x) \,f'_{A_1}(x)
 +  \theta_{ A_2}(x) \,f'_{ A_2}(x)
-  f'(x)
\\
\label{FF step2}
\rule{0pt}{.8cm}
& & \hspace{-.7cm}
\phantom{F_A(x)}
\,=\,
 -\, \theta_{A_1}(x) \left( \frac{1}{x-u_2} + \frac{1}{v_2 - x}  \right)
 - \theta_{ A_2}(x)  \left( \frac{1}{x-u_1} + \frac{1}{v_1 - x}  \right),
\eea
where the function $f(x)$ is given by (\ref{f(x) 2int fake}), the function $f_{ A_j}(x)$ by (\ref{f(x) 1int A_j}) and $ \theta_{A_j}(x)$ is the Heaviside step function with support in $A_j$.
Because of the invariance under translations on the infinite line, we can set $u_1=0$, $v_1=\ell_1$, $u_2=\ell_1+d$ and $v_2=\ell_1+d+\ell_2$ in (\ref{FF step2}).
This leads to the following expression
\be
\label{F-elle_tot}
\ell \,F(x)
 \,=\,
 -   \, \frac{(\ell_2/\ell) \, \theta_{A_1}(x) }{
  \big[x/\ell-(\ell_1+d)/\ell\big]\big[(\ell_1+d+\ell_2)/\ell-x/\ell\big]}
 -
 \,\frac{(\ell_1/ \ell)\,  \theta_{ A_2}(x)  }{
 x/\ell\, (\ell_1/\ell - x/\ell)}\,,
\ee
where $\ell$ is a generic length.
For instance, we can set either
$\ell \equiv \ell_1 + d + \ell_2$ 
or $\ell \equiv \ell_1 + \ell_2$.

In the right panel of Fig.\,\ref{fig:dis}, by considering the harmonic chain with mass $\omega=10^{-8}$ in the thermodynamic regime,
we show the contour function for the mutual information, which is given by (\ref{contour MI lattice}) with $n=1$.
The dashed curves are obtained from (\ref{contour MI cft}) and (\ref{F-elle_tot}), where the constant $C$ has been fixed by trying to maximise the overlap with the corresponding curves found from the lattice data, as done in the left panel for the reason discussed above. 
The agreement is reasonable for $d/\ell = 1/16$ while it gets worse for $d/\ell = 1/2$. 
This is expected from the CFT expression for $\mathcal{F}_n$ \cite{cct-09}, a function of the harmonic ratio of the four endpoints, which is $(1+d/\ell)^{-2}$ for two equal intervals of length $\ell$.
Indeed, $\mathcal{F}_n \to 1$ when $d/\ell \to 0$ and $d/\ell \to \infty$, while it reaches its maximum for $(1+d/\ell)^{-2} =1/2$, i.e. when $d/\ell = \sqrt{2}-1$, which is close to $d/\ell = 1/2$.

We find it worth remarking that a CFT candidate for the contour function for the entanglement entropies when $A$ is made by disjoint intervals is available in the case of free fermions on the infinite line \cite{casini-2int}. 
In particular, for two disjoint intervals it is given by the function employed to plot the dashed curves in the left panel of Fig.\,\ref{fig:dis}.
It would be interesting to compare this function with the lattice data coming from the contour function constructed in \cite{chen-vidal} for free fermions.


\section{Alternative proposals}
\label{sec:other-proposals}

In this section we discuss two alternative constructions for the contour function which are different from the one presented in \S\ref{sec:contour euler}.
The first one is based again on the symplectic matrix $W$ entering in (\ref{williamson th gammaA}) and it has been inspired by the proof of the Williamson's theorem 
found in \cite{simon-99}. 
The second one is based on the mode participation function proposed by Botero and Reznik \cite{br-04} and it has been studied more recently in \cite{frerot-roschilde}.
 In \S\ref{sec:deformation} we describe a deformation of the procedure explained in \S\ref{sec:orthog-mat} which also provides a positive mode participation function satisfying the sum rule (\ref{p sum rule}).

\subsection{A proposal based on a proof of the Williamson's theorem}
\label{sec:otilde contour}

In \S\ref{sec:contour} we have shown that, given a real orthogonal matrix (\ref{O-mat block}), 
the expression (\ref{contour sn_A Xi}) provides a natural candidate for the contour function of the entanglement entropies and the corresponding mode participation function is (\ref{pki general}).
We find it worth focusing our attention on the orthogonal matrices which naturally occur in the analysis of the symplectic spectrum of $\gamma_A$.
In \S\ref{sec:contour} we discussed the orthogonal and symplectic matrix $K$ in (\ref{K-mat def}) coming from the Euler decomposition of the symplectic matrix $W$ of the Williamson's theorem (\ref{williamson th gammaA}).
As already remarked in \S\ref{sec:contour}, another orthogonal matrix naturally related to the symplectic matrix $W$ involved in the Williamson's theorem (\ref{williamson th gammaA}) is the orthogonal matrix $\widetilde{O}$ defined by (\ref{hatOmega diag}), which enters in the factorisation (\ref{W-gammaA}) of $W$ provided in the constructive proof of the Williamson's theorem found in \cite{simon-99}.

\begin{figure}[t!]
\vspace{.2cm}
\hspace{-1.7cm}
\includegraphics[width=1.15\textwidth]{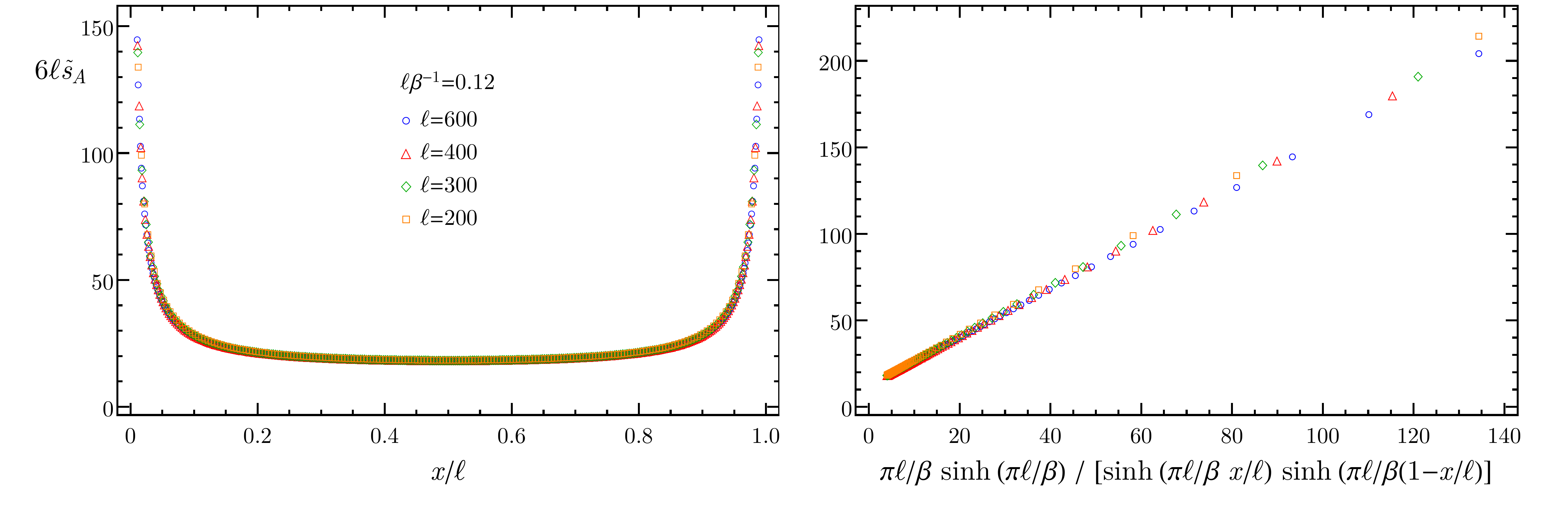}
\vspace{-.8cm}
\caption{
The contour function for the entanglement entropy described in \S\ref{sec:otilde contour} for a single interval of length $\ell$ in a periodic chain at finite temperature $T=1/\beta$ in the thermodynamic limit with $\omega\beta=10^{-3}$.
The slope of the straight line in the right panel disagrees with the one coming from the CFT formula (\ref{contour cft finite T}) 
for $n=1$, which is equal to one. 
}
\label{fig:ContTemp_old}
\end{figure}

The expression (\ref{abs-hatgamma}) tells us that  $\widetilde{O}$ is the orthogonal matrix diagonalising $ | \hat{\gamma}_A |$.
Moreover, by specialising (\ref{SAn Phi_A}) and (\ref{contour sn_A Xi}) to this case where $\Phi_A = | \hat{\gamma}_A |$, 
one finds respectively that the entanglement entropies can be written as
\be
S^{(n)}_A = \frac{1}{2} \, \Tr \big[  s_n(  | \hat{\gamma}_A |)\big]\,,
\ee
and that the corresponding contour function reads
\be
s^{(n)}_A(i) = \frac{1}{2} \, \Tr \big[  X^{(i)} s_n(  | \hat{\gamma}_A |)\big]\,.
\ee
Decomposing $\widetilde{O}$ in blocks like in (\ref{O-mat block}), i.e.
\be
\label{Otilde-mat block}
\widetilde{O} = \bigg( 
\begin{array}{cc}
U_{\widetilde{O}} & Y_{\widetilde{O}} \\ Z_{\widetilde{O}} & V_{\widetilde{O}}
\end{array}
\bigg)\,,
\ee
we have that the mode participation function in this case is (\ref{pki general}) with $O = \widetilde{O}$, namely
 \be
\label{pki Otilde}
\tilde{p}_k(i) \,=\, \frac{1}{2} \Big( 
\big[(U_{\widetilde{O}})_{ki}\big]^2 + \big[(Y_{\widetilde{O}})_{ki}\big]^2 
+ \big[(Z_{\widetilde{O}})_{ki}\big]^2 + \big[(V_{\widetilde{O}})_{ki}\big]^2 
\Big)\,.
\ee

We find it instructive to discuss in some detail  the simpler examples of reduced covariance matrices which are block diagonal, namely  $\gamma_A = Q \oplus P$.
In these cases, (\ref{W-gammaA}) tells us that $ \widetilde{O} = U_{\widetilde{O}} \oplus V_{\widetilde{O}}$ and 
$W = U \oplus V$ are also block diagonal.

Being $\widetilde{O}$ orthogonal, both the matrices $U_{\widetilde{O}}$ and $V_{\widetilde{O}}$ on its diagonal are orthogonal.
By plugging (\ref{hat gamma_A QP}) and the block diagonal form of $\widetilde{O}$ into the square of (\ref{abs-hatgamma}), one finds that
\be
\label{qpq diag}
 Q^{1/2} \,P \,Q^{1/2}  = U_{\widetilde{O}}^{\textrm{t}} \,D^2  \,U_{\widetilde{O}}\,,
\qquad
 P^{1/2} \,Q \,P^{1/2}  = V_{\widetilde{O}}^{\textrm{t}} \,D^2  \,V_{\widetilde{O}}\,,
\ee
which tell us that $U_{\widetilde{O}}$ and $V_{\widetilde{O}}$ are the orthogonal matrices diagonalising the symmetric matrices 
$ Q^{1/2} \,P \,Q^{1/2} $ and $P^{1/2} \,Q \,P^{1/2} $ respectively.
Once $U_{\widetilde{O}}$ and $V_{\widetilde{O}}$ have been computed, the mode participation function in these cases is given by (\ref{pki Otilde}) with $Y_{\widetilde{O}} =Z_{\widetilde{O}} =\boldsymbol{0}$, namely
\be
\label{pki}
\tilde{p}_k(i) \,=\, \frac{1}{2} \Big( \big[(U_{\widetilde{O}})_{ki}\big]^2 + \big[(V_{\widetilde{O}})_{ki}\big]^2 \Big)\,.
\ee
The contour function $\tilde{s}^{(n)}_{A}(i) $ for the entanglement entropies can be constructed like in (\ref{contour pki}), with the mode participation function $ \tilde{p}_k(i)$ instead of $p_k(i)$.

\begin{figure}[t!]
\vspace{.4cm}
\hspace{-1.7cm}
\includegraphics[width=1.15\textwidth]{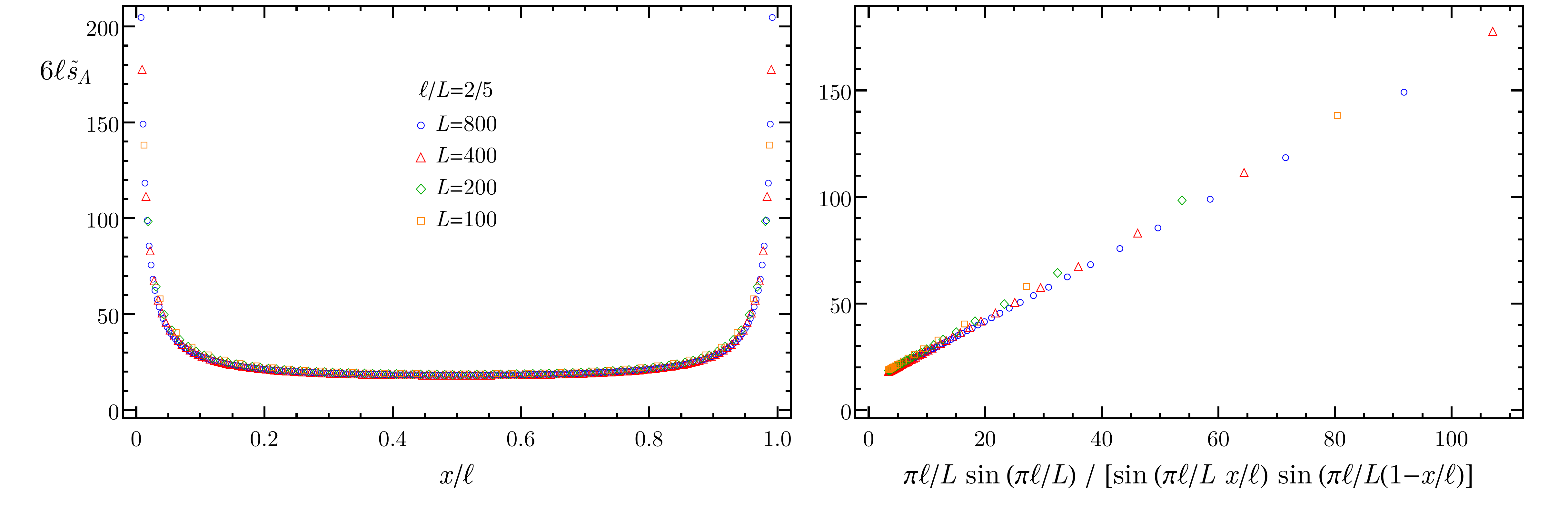}
\vspace{-.8cm}
\caption{
The contour function for the entanglement entropy described in \S\ref{sec:otilde contour} for a single interval of length $\ell$ in a periodic chain of finite length $L$ with $\omega L = 8 \times 10^{-4}$.
The slope of the straight line in the right panel disagrees with the one coming from the CFT formula (\ref{contour cft finite L}) 
for $n=1$, which is equal to one. 
}
\label{fig:ContPBC_old}
\end{figure}

We have repeated the numerical analysis performed in \S\ref{sec:contour} and \S\ref{sec:cft} by employing the mode participation function (\ref{pki}).
All the examples presented in \S\ref{sec:single-int} and \S\ref{sec:2intervals} have been considered and basically the same curves have been found, 
except for the harmonic chain in thermodynamic limit at finite temperature $T=1/\beta$ and for the periodic chain with finite length $L$. 
The contour function for the entanglement entropy evaluated through the mode participation function (\ref{pki}) in these two cases is shown in
Fig.\,\ref{fig:ContTemp_old} e Fig.\,\ref{fig:ContPBC_old} respectively.
From the right panels of these figures one can clearly observe the disagreement with the corresponding CFT formulas, which provide straight lines 
whose slope is equal to one.

\subsection{A proposal based on the eigenvectors of $(\textrm{i}J \gamma_A)^2$}
\label{sec:BR}

The first proposal we found in the literature of a mode participation function in harmonic lattices was suggested by Botero and Reznik \cite{br-04}.
The contour function associated to this mode participation function was studied in \cite{frerot-roschilde} and in this subsection we further discuss this construction.
For simplicity, we focus only on block diagonal covariance matrices $\gamma_A$.

When $\gamma_A = Q\oplus P$, we have that $(\textrm{i} J \gamma_A)^2 = (PQ)\oplus (QP)$ and 
 the symplectic spectrum is given by the positive square root of the spectrum of $QP$, as already mentioned in \S\ref{sec:contour}.
In \cite{br-04} the right eigenvectors of $Q P$ and $P Q$ have been employed to construct a mode participation function. 
In the following we show that these eigenvectors can be read from the rows of the symplectic matrix $W$ entering in the 
Williamson's theorem (\ref{williamson th gammaA}).

Given a block diagonal covariance matrix $\gamma_A = Q\oplus P$, the symplectic matrix $W= U \oplus V$ occurring in (\ref{williamson th gammaA}) is block diagonal as well and the symplectic condition for $W$ is equivalent to the relations $U V^{\textrm{t}} = \boldsymbol{1} = V U^{\textrm{t}}$, which tell us that $U^{-1}= V^{\textrm{t}} $ and $V^{-1}= U^{\textrm{t}} $.
In particular, $U$ and $V$ are not orthogonal matrices. 
Then, specialising (\ref{W-gammaA}) to this case, one gets
\be
\label{UandV}
U = D^{-1/2} \,U_{\widetilde{O}}\, Q^{1/2}\,,
\qquad
V = D^{-1/2} \,V_{\widetilde{O}} \, P^{1/2}\,,
\ee
where the block diagonal structure of $\widetilde{O} = U_{\widetilde{O}} \oplus V_{\widetilde{O}}$ has been employed. 
From (\ref{hat gamma_A QP}) and the block diagonal structure of $\widetilde{O}$, one finds that the relation (\ref{hatOmega diag}) specified to this simpler case becomes $ U_{\widetilde{O}} \,Q^{1/2} \,P^{1/2} \,V_{\widetilde{O}}^{\textrm{t}}  =   D  $.
By using this expression, one can check that $U V^{\textrm{t}} = \boldsymbol{1}$ holds for (\ref{UandV}), as expected.

By employing (\ref{UandV}) and the fact that $U_{\widetilde{O}}$ and $V_{\widetilde{O}}$ are orthogonal, we can write that
\be
\label{UdU and VdV}
 Q = U^{\textrm{t}} D \,U\,,
\qquad
 P = V^{\textrm{t}} D \,V\,,
\ee
which do not provide the diagonalization of the real and symmetric matrices $Q$ and $P$ because $U$ and $V$ are invertible but not orthogonal. 
Expressions relating $Q$ and $P$ to the orthogonal matrices $U_{\widetilde{O}}$ and $V_{\widetilde{O}}$ respectively are obtained by inverting the relations (\ref{UandV}). They read  $ Q^{1/2} = U_{\widetilde{O}}^{\textrm{t}}  \, D^{1/2}\,U = U^{\textrm{t}}  D^{1/2}\,U_{\widetilde{O}}$
and $P^{1/2} = V_{\widetilde{O}}^{\textrm{t}}  \, D^{1/2} \,V = V^{\textrm{t}}  D^{1/2}\,V_{\widetilde{O}}$.
From these results, (\ref{UdU and VdV}) and the condition $U V^{\textrm{t}} = \boldsymbol{1} = V U^{\textrm{t}}$, one can check that the expressions in (\ref{qpq diag}) 
are recovered, as expected.

Finally, by using (\ref{UdU and VdV}) and $U V^{\textrm{t}} = \boldsymbol{1} = V U^{\textrm{t}}$, one finds
\be
\label{QP diag UV}
Q P = U^{\textrm{t}} D^2 \,V = V^{-1} D^2 \,V\,,
\qquad
P Q = V^{\textrm{t}} D^2 \,U = U^{-1} D^2 \,U\,,
\ee
which tell us that the invertible matrices $U$ and $V$ diagonalise $P Q$ and $Q P$ respectively.

The first relation in (\ref{QP diag UV}) can be written as $QP \, V^{-1} = V^{-1} D^2$, which means  that the $k$-th column of $V^{-1}= U^{\textrm{t}} $ is the right eigenvector $\boldsymbol{v}_k$ of $QP$ corresponding to the eigenvalue $\sigma_k^2$, namely $QP\, \boldsymbol{v}_k= \sigma_k^2 \,\boldsymbol{v}_k$.
In the same way, from the second relation in (\ref{QP diag UV}) one concludes that the $k$-th column of $U^{-1}= V^{\textrm{t}} $ provides the right eigenvector $\boldsymbol{u}_k$ of $PQ$ corresponding to the eigenvalue $\sigma_k^2$, 
i.e. $PQ\, \boldsymbol{u}_k= \sigma_k^2 \,\boldsymbol{u}_k$.
Denoting by $v_k(i)$ and $u_k(i)$ the $i$-th element of $\boldsymbol{v}_k$ and $\boldsymbol{u}_k$ respectively, 
in \cite{br-04} the following expression has been proposed for the mode participation function
\be
\label{BR mode part}
\check{p}_k(i) \,= \, v_k(i) \,  u_k(i)\,.
\ee
By inverting $U V^{\textrm{t}} = \boldsymbol{1}$, one finds $(V^{-1})^{\textrm{t}} \, U^{-1}= \boldsymbol{1}$, and the diagonal elements of this relation provide the normalization condition $\sum_{k=1}^\ell \check{p}_k(i)  = 1$ with $1\leqslant k \leqslant \ell$.
Given (\ref{BR mode part}), the corresponding contour function $\check{s}^{(n)}_{A}(i) $ for the entanglement entropies can be constructed like in (\ref{contour pki}), 
with the mode participation function $ \check{p}_k(i)$ instead of $p_k(i)$.

For subsystems made by two disjoint intervals, one can also define $\check{\Upsilon}_A^{(n)}$ like in (\ref{contour MI lattice}), by replacing 
$s^{(n)}_{A}(i) $ with $\check{s}^{(n)}_{A}(i) $. 

 \begin{figure}[t!]
\vspace{.2cm}
\hspace{-1.6cm}
\includegraphics[width=1.15\textwidth]{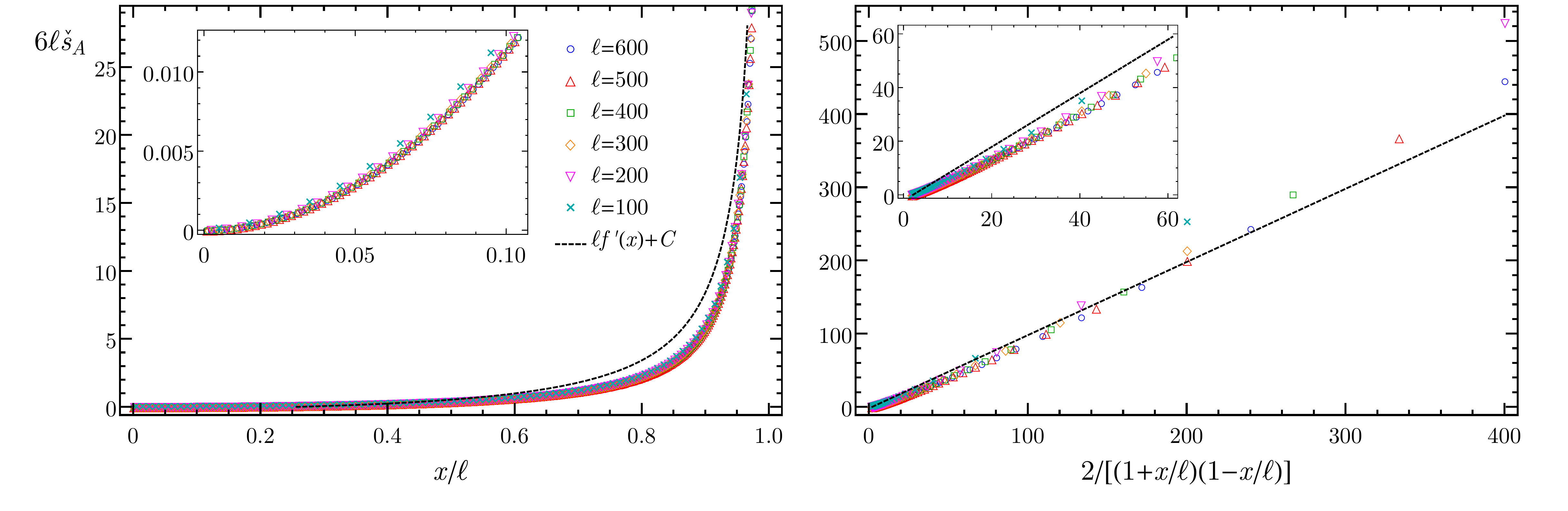}
\vspace{-.7cm}
\caption{
The contour function for the entanglement entropy described in \S\ref{sec:BR} for a single interval of length $\ell$ at the beginning of an open chain with Dirichlet boundary conditions in the thermodynamic limit with $\omega=0$.
The dashed curve corresponds to the CFT formula (\ref{contour 1int cft semiinf}) for $n=1$ and it is the same dashed curve showed in the top panels of Fig.\,\ref{fig:Contsemiinf}.
}
\label{fig:contourBR}
\end{figure}

We remark that the property $\check{p}_k(i)  \geqslant 0$ is not obvious for the proposal  (\ref{BR mode part}), as also emphasised in \cite{frerot-roschilde}.
Actually, for the configuration where $A$ is made by two disjoint intervals, we have found several cases where $\check{p}_k(i)$ becomes negative.
For instance, given two disjoint intervals with the same length $\ell_1 = \ell_2 = 30$ separated by $d=10$ in the harmonic chain in the thermodynamic limit 
with $\omega = 10^{-8}$, we found that the mode participation function (\ref{BR mode part}) for $k=14$ reaches small negative values of order $10^{-4}$ for some sites, while the largest positive ones are of order $10^{-1}$. 
Nonetheless, the contour function obtained from this $\check{p}_k(i)$ is positive. 
Moreover, we have always found $\check{s}^{(n)}_{A}(i) \geqslant 0$ for all the sites in all the examples that we have considered.

In Fig.\,\ref{fig:contourBR} we show the contour function constructed through (\ref{BR mode part}) when $A$ is a single interval of length $\ell$ at the beginning of a massless open chain with Dirichlet boundary conditions in the thermodynamic limit.
The dashed curve comes from the CFT expression (\ref{contour 1int cft semiinf}) for $n=1$ and it is the same dashed curve showed in the top panels of Fig.\,\ref{fig:Contsemiinf}.
Thus, the contour function corresponding to the mode participation function (\ref{BR mode part}) does not seem to provide the CFT expressions discussed in \S\ref{sec:single-int} in the scaling limit.

\begin{figure}[t!]
\vspace{.2cm}
\hspace{-1.6cm}
\includegraphics[width=1.15\textwidth]{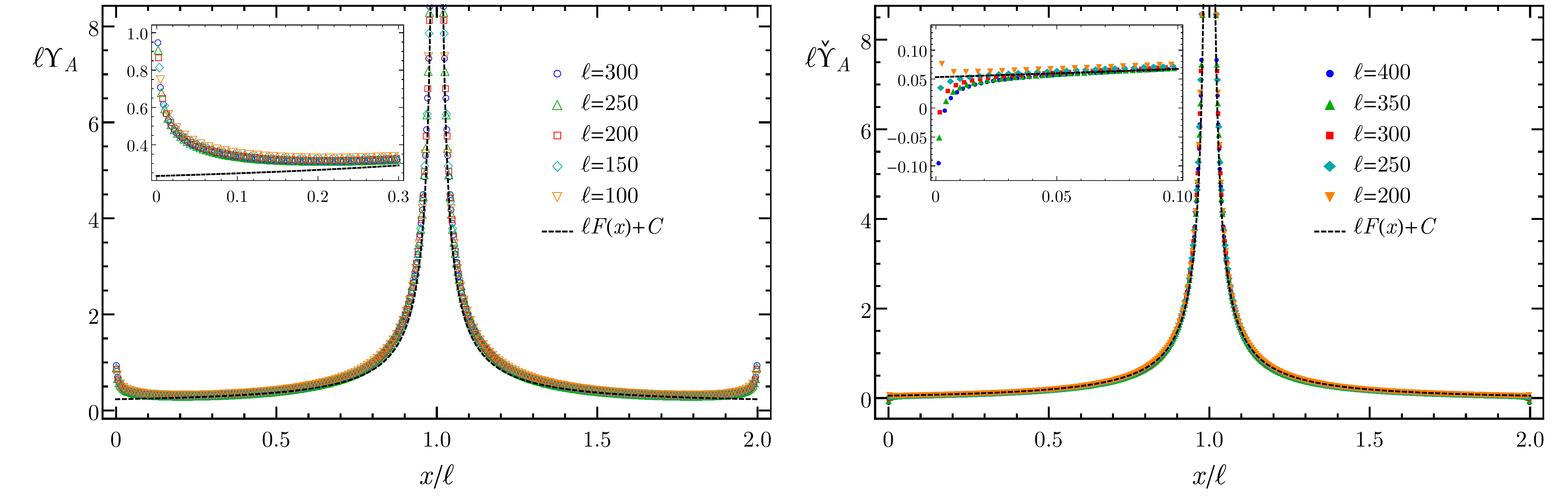}
\vspace{-.7cm}
\caption{
The contour function for the mutual information when $A$ is made by two equal and adjacent intervals of length $\ell$ in the periodic chain in the thermodynamic regime with $\omega=10^{-8}$. 
The dashed curves correspond to (\ref{contour MI cft}) with $n=1$ and the constant $C$ fixed by maximising the overlap with the lattice data.
The data are obtained from the mode participation functions described in \S\ref{sec:contour euler} (left panel) and \S\ref{sec:BR} (right panel).
In the insets a zoom of the region close to the endpoints of $A_1 \cup A_2$ is shown and a better collapse of the lattice data is observed in the left panel. 
}
\label{fig:MIcomparisonBR}
\end{figure}

In Fig.\,\ref{fig:MIcomparisonBR} we have considered the contour function for the mutual information (namely Eq.\,(\ref{contour MI lattice}) for $n=1$) for the periodic chain in the thermodynamic regime when $\omega=10^{-8}$, with the subsystem $A$ made by two equal and adjacent intervals of length $\ell$.
The results obtained from the mode participation functions (\ref{pki general - K - pq}) (left panel) and (\ref{BR mode part}) (right panel) are shown,
together with the CFT expression (\ref{contour MI cft}) for $n=1$ (dashed curves).
Comparing the insets of the two panels, we notice that the data collapse is better for the contour function constructed in \S\ref{sec:contour euler} and we also observe that $\check{\Upsilon}_A^{(n=1)}$ becomes negative close to the two endpoints of the single interval $A_1 \cup A_2$ for large values of $\ell$.

\subsection{A deformation of the contour}
\label{sec:deformation}

We find it instructive to discuss briefly a deformation of construction described in \S\ref{sec:contour} through a parameter $\xi \in (0,1)$.
The resulting mode participation function provides a contour function which satisfies (\ref{E_A sum_i}) and (\ref{positiv}).

Given two real and positive parameters $\xi$ and $\eta$, let us revisit the construction described in \S\ref{sec:orthog-mat} by considering
the operator $ \breve{X}^{(i)} \,\equiv\, 2\big[ \xi \,\delta^{(i)}  \oplus \eta\,\delta^{(i)}\big]$ associated to the $i$-th site instead of the projector $X^{(i)}$, 
which is recovered in the special case of  $\xi=\eta=1/2$.
The contour function in (\ref{contour sn_A Xi}) with $X^{(i)}$ replaced by  $\breve{X}^{(i)}$ gives
\be
\label{contour sn_A Xi def}
\breve{s}^{(n)}_{A}(i)  = \frac{1}{2} \, \Tr \big[  \breve{X}^{(i)} s_n(  \Phi_A)\big]\,,
\ee
where $\Phi_A$ is  (\ref{Phi_A def}) with the orthogonal matrix $O$ written in terms of its blocks like in (\ref{O-mat block}). 
By adapting the steps in (\ref{sn(i) phiA step1}) to this case, for (\ref{contour sn_A Xi def}) we find 
\bea
& &
\hspace{-2.38cm}
\breve{s}^{(n)}_{A}(i) 
 \;=\;
  \Tr \big[  
  O \big( \xi\,\delta^{(i)} \oplus \eta\, \delta^{(i)} \big) O^{\textrm{t}} 
  \big( s_n(D) \oplus s_n(D)\big)\big]
 \\
 \rule{0pt}{.6cm}
 & &
 \hspace{-1.25cm}
 \;=\;
   \Tr \big[ \big( \xi\,U_O \, \delta^{(i)}U_O^{\textrm{t}} 
   + \eta \,Y_O \, \delta^{(i)}Y_O^{\textrm{t}}\big) s_n(D) \big]
   + \Tr \big[ \big( \xi\,Z_O \, \delta^{(i)}Z_O^{\textrm{t}}
       + \eta\,V_O \, \delta^{(i)}V_O^{\textrm{t}}\big) s_n(D)  \big]
   \nonumber \\
 \rule{0pt}{.6cm}
 & &
  \hspace{-1.2cm}
 \,=\,
  \sum_{k=1}^\ell \,\breve{p}_k(i) \, s_n(\sigma_k)\,,
  \nonumber
\eea
where in the last expression we have introduced 
\be
\label{pki general xi}
\breve{p}_k(i) \,=\, 
\xi \big[(U_O)_{ki}\big]^2 + \eta \big[(Y_O)_{ki}\big]^2 + \xi \big[(Z_O)_{ki}\big]^2 + \eta  \big[(V_O)_{ki}\big]^2 \,,
\ee
which reduces to the mode participation function (\ref{pki general}) for $\xi=\eta=1/2$, as expected. 

Being $\xi$ and $\eta$ positive, we have that $\breve{p}_k(i)  \geqslant 0$  by construction and this implies that $\breve{s}^{(n)}_{A}(i)  \geqslant 0$ for the function in (\ref{contour sn_A Xi def}).
Moreover, because of the relations (\ref{ort-cond off-diag}) coming from the orthogonality condition $O^{\textrm{t}} O = \boldsymbol{1}$,
we find that the property $\sum_{k=1}^\ell \breve{p}_k(i) =1$ is satisfied only when $\xi+\eta=1$.
Thus, the contour function (\ref{contour sn_A Xi def}) fulfils the constraints (\ref{E_A sum_i}) and (\ref{positiv}) for $ \breve{X}^{(i)} \,\equiv\, 2\big[ \xi \delta^{(i)}  \oplus (1-\xi) \delta^{(i)}\big]$ with $\xi \in (0,1)$.

Whenever $\xi\neq 1/2$, we have that $ \sum_{i\in A} \widetilde{X}^{(i)} = 2[\xi \,\boldsymbol{1}  \oplus (1-\xi)\,\boldsymbol{1}] \neq \boldsymbol{1}$ 
and that $ \breve{X}^{(i)}$ is not a projector because $\big(\breve{X}^{(i)} \big)^2 \neq \breve{X}^{(i)}$.
Since in \S\ref{sec:contour properties} the properties of $X^{(i)}$ have been employed in a crucial way, this proposal for the contour function could be ruled out by the requirements introduced in \cite{chen-vidal}. 
Notice that, by employing the orthogonality condition $O O^{\textrm{t}} = \boldsymbol{1}$, one finds that
 the constraint $\sum_{i=1}^{\ell} \breve{p}_k (i) = 1$ for any integer $k \in [1,\ell]$, 
 which is not required for the mode participation function,
 can be fulfilled only for $\xi=1/2$.


\section{Conclusions}
\label{sec:conclusions}

In this paper we studied the contour function for the entanglement entropies in generic harmonic lattices.
Our proposal is given by the expression in (\ref{contour s_A project - K}), which can be written also in the form (\ref{contour pki}) in terms of the particular mode participation function in (\ref{pki general - K}).
This proposal fulfils the basic constraints (\ref{E_A sum_i}) and (\ref{positiv}) and also three further requirements 
corresponding to a weakened version of  the properties introduced in \cite{chen-vidal}.

Focusing on one-dimensional harmonic chains with either periodic or Dirichlet boundary conditions in the massless regime, we have studied 
the configurations whose entanglement hamiltonian found though CFT methods can be written in the form (\ref{K_A from f}), namely as an integral over the interval $A$ of the component $T_{00}$ of the energy-momentum tensor multiplied by a suitable local weight function \cite{chm, klich-13, ct-16}.
Comparing the scaling limit of our contour function for the entanglement entropies with the inverse of the local weight function multiplying $T_{00}$ in the entanglement hamiltonian for various configurations, a good agreement has been observed in the regions of the interval nearby the endpoints, where a universal behaviour is expected. 
It remains to be understood more quantitatively how to extract the non-universal information contained in the contour function.

We have also considered the contour function for the entanglement entropy for a massless harmonic chain in the thermodynamic regime 
in some cases where $A$ is made by two disjoint intervals, 
finding well defined curves from the collapse of the numerical data in the scaling limit.
It would be very interesting to find an analytic expression for these curves through a CFT analysis.

Other contour functions for the entanglement entropies satisfying the properties (\ref{E_A sum_i}) and (\ref{positiv}) have been considered (see \S\ref{sec:other-proposals})
and we observed that these contour functions do not always agree with the functions coming from the corresponding entanglement hamiltonians obtained through CFT methods. 
We leave for future work to check whether they fulfil also the three requirements introduced in \cite{chen-vidal} or at least their weaker version discussed in \S\ref{sec:contour properties}.

The contour functions for the entanglement entropies provide a way to characterise the spatial structure of entanglement within the spatial region $A$.
Few analysis of  the contour functions for bipartite entanglement  have been performed in the literature \cite{br-04, chen-vidal, frerot-roschilde} and many issues could be explored in the future.

Finding a complete list of properties which allows to define the contour function for the entanglement entropies in a unique way is a very interesting open problem.

Contour functions are naturally defined also in higher dimensions.
Since they give a more refined information on the entanglement structure than entanglement entropy, it could be interesting to study them in higher dimensional systems.
As already noticed in \cite{chen-vidal}, the contour function for fermions in two spatial dimensions can discriminate between the ground state of a massive theory and the one of a critical system with a finite number of zero modes.
On the contrary, the entanglement entropy scales in both cases with the length of the boundary of the region $A$ and therefore it cannot distinguish the two cases.
Moreover, in higher dimensions the shape of the spatial region $A$ plays a crucial role.
For example, for critical models in two spatial dimensions, when $A$ has corners logarithmic terms are present which contain universal information \cite{corners}.
It would be interesting to explore how the contour function encodes this and maybe further information.

We also find it worth studying further the contour function for the free fermions proposed by Chen and Vidal \cite{chen-vidal}: in the massless case one could compare the lattice results with the CFT expressions found in \cite{ct-16} in the case of $c=1/2$,  as done in \S\ref{sec:single-int} for massless harmonic chain  in the case of $c=1$. 
For the single interval on the infinite line this has been already done in \cite{chen-vidal}
and it would be interesting to explore other configurations. 
We remark that, for free massless fermions and for subsystems $A$ made by a generic number of disjoint intervals, a natural CFT candidate for scaling limit of the contour function for the entanglement entropies is available in the literature \cite{casini-2int}.

Finding contour functions which fulfil all the proper requirements also for more interesting lattice models (e.g. the Ising spin chain) is an important open problem for future studies.
The construction of contours for the entanglement entropies could be helpful to improve the current understanding about the role played by the entanglement spectrum \cite{ent-spectrum} and by the eigenvectors of the reduced density \cite{eigenvector, ep-rev} in the characterisation of the bipartite entanglement.
The interesting question to understand is whether and how is possible to find a general way to obtain a contour function directly from the modular hamiltonian.
Some results for specific configurations and particular free models have been found in \cite{chen-vidal} and in this paper, but more complicated configurations and more interesting models (the interacting ones in particular) should be addressed.  
Recently, the modular hamiltonian for free lattice models has been employed to suggest another way to recover the universal function in (\ref{contour 1int cft}) for $n=1$ through a lattice calculation \cite{casini-16}.
Moreover, also analytical results for the modular hamiltonian of an interval on the infinite line for the free fermion have been found \cite{eisler-peschel-17}.

Tensor networks are very successful in describing the low energy physics of lattice models \cite{tensor-network1}.
The scaling of entanglement in the ground state of lattice systems has been used as input in constructing the tensor networks algorithms which allow to simulate those systems \cite{tensor-network2}.
Studying the spatial distribution of entanglement could potentially improve such tensor networks algorithms.

The problem of providing a list of properties which characterises the contour function in a unique way can be extended also to other measures of the bipartite entanglement.
For entanglement measures different from the entanglement entropies explicit constructions of the corresponding contour functions are missing.
A measure of the bipartite entanglement for mixed states which has attracted some attention is the logarithmic negativity \cite{neg papers}.
A QFT approach to this measure of entanglement has been proposed in \cite{cct-neg}, with particular focus on CFTs
(see also \cite{doyon-neg-mass} for the massive case).
Numerical calculations of the  logarithmic negativity have been performed in lattice models, both in one spatial dimension \cite{neg-after-1d} and in two-dimensional lattices \cite{neg-2dim}.
In a forthcoming publication \cite{cont-neg} 
we apply the approach described in this manuscript to study a contour function for the logarithmic negativity 
in generic harmonic lattices.

\section*{Acknowledgements}
We are grateful to Marcus Cramer, Mihail Mintchev, Guifre Vidal 
and in particular to Pasquale Calabrese, Robert Myers and Luca Tagliacozzo for helpful discussions or correspondence. 
AC thanks SISSA, Trieste, and the University of Strathclyde, Glasgow, for warm hospitality and support during part of this work. 
ET acknowledges YITP, Kyoto, and NORDITA, Stockholm, for warm hospitality and support during part of this work. 
AC acknowledges financial support from MINECO (grant MTM2014-54240-P) and Comunidad de Madrid (grant QUITEMAD+-CM, ref. S2013/ICE-2801).
AC has been also supported by European Research Council (ERC) under the European Union's Horizon 2020 research and innovation programme (grant agreement No 648913).
ET has been supported by the ERC under Starting Grant  279391 EDEQS.
\\


\begin{appendices}

\section*{Appendices}

\section{On the Williamson's theorem}
\label{app:williamson}

In this appendix we first review part of the proof of the Williamson's theorem found in \cite{simon-99}, 
which has been employed to construct the mode participation functions in \S\ref{sec:contour} and in \S\ref{sec:otilde contour}.
Then, some issues related to the uniqueness of the symplectic matrix occurring in the Williamson's theorem are briefly discussed.

{\it Williamson's theorem} \cite{Williamson}. 
Given a $2 m \times 2 m$ real matrix $M$ which is also symmetric and positive definite, 
a real symplectic matrix $W$ exists such that
\be
\label{williamson th}
M = W^{\textrm t} \big( D \oplus D \big) W\,,
\ee
where $D=\textrm{diag} (\sigma_1 , \dots , \sigma_m)$ with $\sigma_j > 0$. 
The sequence $\{ \sigma_1 , \dots , \sigma_m\}$ is called {\it symplectic spectrum} of $M$. 
The symplectic spectrum can be obtained by taking the modulus of the spectrum of the matrix $J M$.

Various proofs of the Williamson's theorem have been found \cite{williamson-other proofs} and we focus on the one given in \cite{simon-99}.
In this proof the matrix $\hat{M} \equiv M^{1/2} J M^{1/2}$ is introduced.
It is straightforward to realise that $\hat{M}$ and $JM$ have the same spectrum. 
Since $\hat{M} $ is real and antisymmetric, an orthogonal matrix $O$ exists such that
\be
\label{hatH diag}
O \hat{M}  O^{\textrm t} = 
\bigg( \hspace{-.1cm} \begin{array}{cc}
 \boldsymbol{0}  &  D \\
 - D &  \boldsymbol{0} \\
\end{array}  \hspace{-.05cm}  \bigg)\,,
\ee
being $D=\textrm{diag} (\sigma_1 , \dots , \sigma_m)$ with $\sigma_j > 0$.

By employing the orthogonal matrix $O$ defined in (\ref{hatH diag}), the symplectic matrix $W$ entering in (\ref{williamson th}) is constructed as follows
\be
\label{Wdef}
W = (D^{-1/2} \oplus D^{-1/2}) \, O M^{1/2}\,.
\ee
It is straightforward to check that the orthogonality condition for $O$ implies that the matrix $W$ in (\ref{Wdef}) satisfies (\ref{williamson th}).
Moreover, by employing (\ref{hatH diag}) one easily finds that $W$ in (\ref{Wdef}) is symplectic. 
Notice that the non uniqueness of the matrix $O$ is determined by the diagonalization problem defined by (\ref{hatH diag}) and therefore it is related to the degeneracy of the symplectic eigenvalues.

In this manuscript the Williamson's theorem is applied to the reduced covariance matrix $\gamma_A$ characterising the subsystem $A$.
The symplectic matrix $W$ corresponding to $\gamma_A$ is crucial for the constructions of the contour functions for the entanglement entropies discussed in the main text.
Indeed, the Euler decomposition of the symplectic matrix $W$ corresponding  to $\gamma_A$ is employed to define the mode participation function (\ref{pki general - K}) described  in \S\ref{sec:contour euler}.
Moreover, the orthogonal matrix defined through the relation (\ref{hatH diag}) in the special case of $M = \gamma_A$ (see Eq.\,(\ref{hatOmega diag})) is the key object occurring in the construction of the mode participation function (\ref{pki}) discussed in \S\ref{sec:otilde contour}.

Given a matrix $M$ satisfying the hypothesis of the Williamson's theorem and assuming that $W$ and $W'$ are two symplectic matrices such that (\ref{williamson th}) holds for both of them with the same $D$, it can proved that,  $W' \, W^{-1} \equiv \mathcal{U}$ is symplectic and orthogonal  (Proposition 8.12 of \cite{deGosson}).

On one side, by considering the decomposition (\ref{Wdef}) for $W$ and the analogue one for $W'$, 
which involves the same $D$ and $M$ but a different orthogonal matrix $O'$,
and plugging these decompositions into the relation $W' \, W^{-1} \equiv \mathcal{U}$, it is straightforward to observe that 
$ O' = \big[ (D^{1/2} \oplus D^{1/2})  \,\mathcal{U}\,  (D^{-1/2} \oplus D^{-1/2})  \big] O$.
On the other side, both $O$ and $O'$ fulfil (\ref{hatH diag}), and this leads to observe that $O |\hat M| O^\textrm{t} = D\oplus D = O' |\hat M| (O')^\textrm{t}$. 
Thus, the orthogonal matrices $O$ and $O'$ are related through an orthogonal matrix $\mathcal{V}$ which does not mix eigenspaces corresponding to different eigenvalues in $D\oplus D$.
Plugging $O' = \mathcal{V} O$ into the relation between $O$ and $O'$ written right above, one finds $\mathcal{V} = (D^{1/2}\oplus D^{1/2})\, \mathcal{U}\, (D^{-1/2}\oplus D^{-1/2})$. By first isolating $\mathcal{U} $ in the r.h.s. of the latter equation and then employing that $\mathcal{V}$ commutes with $D^{1/2}\oplus D^{1/2}$, we obtain that $\mathcal{V} = \mathcal{U}$.

The above observations support the claim that, ultimately, the non uniqueness of the orthogonal matrix $O$ depends on the degeneracies within the symplectic spectrum. 
Nonetheless, this freedom does not influence the contour function.
Indeed, in the diagonalization problem defined by $O |\hat M| O^\textrm{t} = D\oplus D$ there is an obvious degeneracy due to the structure of the diagonal matrix in the right hand side. This allows to mix the operators $\hat{q}_i$ and $\hat{p}_i$ corresponding to the same site, but such freedom does not change the contour function (see the property (b) in \S\ref{sec:contour properties} in the special case where $G$ is just the $i$-th site).
When degeneracies occur in the symplectic spectrum, the contour (\ref{contour pki}) can be written as $\sum_{\kappa=1}^{\ell' < \ell} \mathcal{P}_\kappa(i) s_n(\sigma_\kappa)$, where the sum is performed over eigenspaces corresponding to different symplectic eigenvalues and $\mathcal{P}_\kappa(i) = \sum_{\kappa'} p_{\kappa'}(i)$ is the sum of the mode participation function over a base of the eigenspace corresponding to the symplectic eigenvalue $\sigma_\kappa$.
From (\ref{pki general - K}) one observes that orthogonal transformations mixing eigenvectors within the same eigenspace indexed by $\kappa$ do not change $\mathcal{P}_\kappa(i)$.

Let us conclude with an observation about the uniqueness of the matrix $\hat{M}$ introduced in the proof of the Williamson's theorem found in \cite{simon-99}.
In particular, let us try to extend the analysis by considering $\hat{M}_{(a)} \equiv M^{a} J M^{a}$ for a real power $a$, which becomes $\hat{M}$ in the special case of $a=1/2$.
The matrices $\hat{M}_{(a)} $ and $J M^{2a}$ have the same spectrum. 
Since the matrix $\hat{M}_{(a)}$ is real and antisymmetric, a real orthogonal matrix $O_{(a)}$ exists such that 
\be
\label{O_a def}
O_{(a)} \hat{M}_{(a)}  O_{(a)}^{\textrm t} =
\bigg( \hspace{-.1cm} \begin{array}{cc}
 \boldsymbol{0}  &  D_{(a)} \\
 - D_{(a)}  &  \boldsymbol{0} \\
\end{array}  \hspace{-.05cm}  \bigg)\,.
\ee
By employing the orthogonality of $O_{(a)} $, it is straightforward to show that the relation 
$M = W_{(a)}^{\textrm t} \big( D_{(a)} \oplus D_{(a)} \big) W_{(a)}$ is satisfied for the matrix
$W_{(a)} \equiv (D_{(a)}^{-1/2} \oplus D_{(a)}^{-1/2}) \, O_{(a)} M^{1/2}$.
Then, by requiring that $W_{(a)} $ satisfies the symplectic condition, one finds that $O_{(a)} \hat{M}\,  O_{(a)}^{\textrm t} $ gives the r.h.s. of (\ref{O_a def}). 
This allows to conclude that $O_{(a)} \hat{M}_{(a)}  O_{(a)}^{\textrm t} = O_{(a)} \hat{M}\,  O_{(a)}^{\textrm t} $, which leads to $\hat{M}_{(a)} = \hat{M}$ and therefore $a=1/2$.


\section{On the properties of the contour function}
\label{app:properties}

In this appendix we consider the three properties of the contour function for the entanglement entropies introduced in \cite{chen-vidal} besides the constraints (\ref{E_A sum_i}) and (\ref{positiv}).
Focussing on the harmonic lattices, in the following analysis, which supports and extends the discussion made in \S\ref{sec:contour properties}, 
we show that these three further properties are satisfied by the contour function described in \S\ref{sec:contour euler} if we restrict to the canonical transformations implemented by the matrices $M \in \textrm{Sp}(\ell) \cap O(2\ell)$.

{\bf (a)} {\it Spatial symmetry.}
If $\rho_A$ is invariant under a transformation relating the sites $i$ and $j$ in the subsystem $A$, 
then $s_A^{(n)}(i)=s_A^{(n)}(j)$.


Considering a canonical transformation that maps the $i$-th site into the $j$-th site, 
for the corresponding symplectic $M$ we have that $ X^{(i)} \to X^{(j)} = M^{\textrm t}  X^{(i)}  M $.
Since $X^{(j)}$ is a projector (i.e. $(X^{(j)})^2 = X^{(j)}$), the matrix $M$ is orthogonal.
Assuming that this transformation is a symmetry of $\rho_A$, from the transformation rule (\ref{trans gamma_tilde}) we conclude that 
$ E_{\textrm{\tiny R}}^{-1} \gamma_A E_{\textrm{\tiny R}}^{-1}  =
M \big( E_{\textrm{\tiny R}}^{-1} \gamma_A E_{\textrm{\tiny R}}^{-1} \big) M^{\textrm t} $ for this particular $M$.
Thus, under this mapping we have
\bea
& & \hspace{-1.7cm}
s^{(n)}_A(i) 
\,=\,
 \frac{1}{2} \, \Tr \big[  X^{(i)}  \,s_n( E_{\textrm{\tiny R}}^{-1} \gamma_A E_{\textrm{\tiny R}}^{-1}  )\big]
 \,=\,
  \frac{1}{2} \, \Tr \big[  X^{(i)}  \,M \big( E_{\textrm{\tiny R}}^{-1} \gamma_A E_{\textrm{\tiny R}}^{-1} \big) M^{\textrm t} \big]
 \\
 \rule{0pt}{.7cm}
& & \hspace{-1.87cm}
\phantom{s^{(n)}_A(G)} 
\,=\,
 \frac{1}{2} \, \Tr \big[  M^{\textrm t}  X^{(i)}  M \big( E_{\textrm{\tiny R}}^{-1} \gamma_A E_{\textrm{\tiny R}}^{-1} \big) \big]
 \,=\,
 \frac{1}{2} \, \Tr \big[    X^{(j)}  \big( E_{\textrm{\tiny R}}^{-1} \gamma_A E_{\textrm{\tiny R}}^{-1} \big) \big]
  \,=\,
  s^{(n)}_A(j) \,.
\eea
This shows that the contour (\ref{ee s function K}) fulfils the constraint (a) about the spatial symmetry of the subsystem.

In order to formulate the remaining properties, we need to introduce  also the contour $s_A^{(n)}(G) $ of a subregion $G \subseteq A$ as follows
\be
\label{def contour region app}
s_A^{(n)}(G)   \equiv \sum_{i  \,\in\, G} s_A^{(n)}(i)\,.
\ee
In the special case of $G = A$, from (\ref{E_A sum_i}) and (\ref{def contour region app}) we find $s_A^{(n)}(A) = S_A^{(n)}$.
The contour $s_A^{(n)}(G)$ is clearly additive: for any two non intersecting spatial subsets $G \subsetneq A $ and $\tilde{G} \subsetneq A$ we have $s_A^{(n)}(G \cup \tilde{G}) = s_A^{(n)}(G) + s_A^{(n)}(\tilde{G})$.
Moreover, the contour $s_A^{(n)}(G)$ is monotonous, i.e. for $G \subseteq \tilde{G} \subseteq A $ the inequality $ s_A^{(n)}(G) \leqslant s_A^{(n)}(\tilde{G})$ holds.

{\bf (b)} {\it Invariance under local unitary transformations.}
Given a system in the state characterised by the density matrix $\rho$ and a unitary transformation $U_{G}$ acting non trivially only on $G \subseteq A$, denoting by $\rho'$ the state of the system after such transformation, the same contour $s_A^{(n)}(G)$ should be found for $\rho$ and $\rho'$.


In order to discuss the requirement (b), let us introduce the projector for the subregion $G \subseteq  A$ as follows
\be
X^{(G)} \equiv \sum_{i \,\in\,G} X^{(i)}\,.
\ee
By employing the realisation of the matrices $X^{(i)}$ described in \S\ref{sec:orthog-mat}, we have that $X^{(G)}$ is the identity on $G$, while it vanishes outside.
The unitary transformation $U_G$ corresponds to a symplectic matrix $M_{G} \in \textrm{Sp}(\ell)$ which acts non trivially only on $G$.
We recall that in our analysis we restrict to symplectic matrices that are also orthogonal, i.e. $M_{G} \in \textrm{Sp}(\ell) \cap O(2\ell)$.
Since $M_{G}$ acts non trivially only on $G$, where $X^{(G)}$ is the identity, we have that 
$ [X^{(G)} , M_{G} ] = 0 $.

From (\ref{contour s_A project - K}) and (\ref{def contour region app}) we get
\bea
\label{prop4 step1}
& & \hspace{-2.4cm}
s^{(n)}_A(G) 
\,=\,
 \frac{1}{2} \, \Tr \big[  X^{(G)}  \,s_n( E_{\textrm{\tiny R}}^{-1} \gamma_A E_{\textrm{\tiny R}}^{-1}  )\big]
 \,=\,
  \frac{1}{2} \, \Tr \big[  X^{(G)}  \,s_n( E_{\textrm{\tiny R}}^{-1} \gamma_A E_{\textrm{\tiny R}}^{-1}  )\,
 M_{G}^{\textrm t} \,M_{G} \big]
 \\
 \rule{0pt}{.7cm}
& & \hspace{-2.4cm}
\phantom{s^{(n)}_A(G)} 
\,=\,
 \frac{1}{2} \, \Tr \big[  X^{(G)}  \,
 M_{G} \,s_n( E_{\textrm{\tiny R}}^{-1} \gamma_A E_{\textrm{\tiny R}}^{-1}  )\, M_{G}^{\textrm t}  \big]
 \,=\,
  \frac{1}{2} \, \Tr \big[  X^{(G)}  \,
 s_n( M_{G} \, E_{\textrm{\tiny R}}^{-1} \gamma_A E_{\textrm{\tiny R}}^{-1}  \, M_{G}^{\textrm t}  ) \big]
   \\
 \rule{0pt}{.6cm}
& & \hspace{-2.4cm}
\phantom{s^{(n)}_A(G)} 
\,=\,
s^{(n)}_A(G) '\,,
\eea
where we have used first that $\boldsymbol{1} = M_{G}^{\textrm t} \, M_{G}$, then the cyclic property of the trace and finally the fact that
 $X^{(G)}$ and $M_{G} $ commute. 
Thus, also the property (b) is satisfied for the restricted class of unitary transformations $U_G$ associated to $M_{G} \in \textrm{Sp}(\ell) \cap O(2\ell)$.

{\bf (c)} {\it A bound.}
Given a system in the pure state $| \Psi \rangle $ and the bipartition $\mathcal{H}= \mathcal{H}_A \otimes \mathcal{H}_B$, let us assume that the further decompositions $\mathcal{H}_A = \mathcal{H}_{\Omega_A} \otimes \mathcal{H}_{\bar{\Omega}_A} $ and
$\mathcal{H}_B = \mathcal{H}_{\Omega_B} \otimes \mathcal{H}_{\bar{\Omega}_B} $ lead to the following factorisation of the state
\be
\label{factor state hyp app}
| \Psi \rangle = | \Psi_{\Omega_A \Omega_B} \rangle \otimes | \Psi_{\bar{\Omega}_A \bar{\Omega}_B} \rangle\,.
 \ee
Considering a subregion $G \subseteq A$ such that $\bigotimes_{i \in G} \mathcal{H}_i \subseteq  \mathcal{H}_{\Omega_A}$, we must have that 
\be
s^{(n)}_A(G) \leqslant S^{(n)}(\Omega_A)\,,
\ee
where $S^{(n)}(\Omega_A)$ are the entanglement entropies corresponding to the reduced density matrix $\rho_{\Omega_A}$, obtained by tracing over the degrees of freedom of $\mathcal{H}_{\bar{\Omega}_A}  \otimes \mathcal{H}_B $.

We refer the interested reader to \cite{chen-vidal} for a more detailed discussion on the motivations leading to this property.

Assuming that $G$ is made by the first $\ell_G < \ell$ sites of $A$, let us order the modes by considering the vector
$\hat{\boldsymbol{r}}_G$ given by $ \{ \hat{r}_{i,\alpha} \,|\, i\in G\} $ and 
$\hat{\boldsymbol{r}}_{\bar{G}} $ which collects the remaining ones $ \{ \hat{r}_{i,\alpha} \,|\, i \in A\setminus G\} $, 
being $\hat{r}_{i,1} \equiv \hat{q}_{i,2}$ and $\hat{r}_{i,2} \equiv \hat{p}_i $.
Similarly, we can consider the modes 
$\hat{\boldsymbol{w}}_{\Omega_A}$ and $\hat{\boldsymbol{w}}_{\bar{\Omega}_A}$ given by 
$ \{ \hat{w}_{m,\alpha} \,|\, m\in \Omega_A\} $  and
$ \{ \hat{w}_{m,\alpha} \,|\,m \in \bar{\Omega}_A \} $ respectively. 
Notice that $\ell_G \leqslant \ell_{\Omega_A}$, being  $2\ell_{\Omega_A}$ the number of elements in $\hat{\boldsymbol{w}}_{\Omega_A}$.
A consequence of the hypothesis $\bigotimes_{i \in G} \mathcal{H}_i \subseteq  \mathcal{H}_{\Omega_A}$ is that
the $\hat{\boldsymbol{r}}$ modes are related to the $\hat{\boldsymbol{w}}$  modes through a linear map such that 
all the modes $ \hat{\boldsymbol{r}}_G$ are contained into $ \hat{\boldsymbol{w}}_{\Omega_A} $, namely
\be
\label{w from r}
\bigg( \hspace{-.1cm} \begin{array}{c}
 \hat{\boldsymbol{w}}_{\Omega_A}  \\
\hat{\boldsymbol{w}}_{\bar{\Omega}_A}   
\end{array}  \hspace{-.1cm}  \bigg)
=
\bigg( \hspace{-.1cm} \begin{array}{cc}
 V_{\Omega G}  &  V_{\Omega \bar{G}}  \\
 \boldsymbol{0}   & V_{\bar{\Omega} \bar{G}}  
\end{array}  \hspace{-.05cm}  \bigg)
\bigg( \hspace{-.1cm} \begin{array}{c}
 \hat{\boldsymbol{r}}_G  \\
\hat{\boldsymbol{r}}_{\bar{G}}
\end{array}  \hspace{-.1cm}  \bigg)\,,
\ee
where $V_{\Omega G}$, $ V_{\Omega \bar{G}} $ and $V_{\bar{\Omega} \bar{G}}$ are rectangular matrices with proper sizes 
which partition the $2\ell \times 2\ell$ matrix $V$.

In order to preserve the canonical commutation relations, the matrix $V$ in (\ref{w from r}) must be symplectic, but we also impose that $V$ is orthogonal.
As for the covariance matrices $\gamma_{\Omega} $ and $\gamma_A $ corresponding to the $\hat{\boldsymbol{r}}$ modes and to the $\hat{\boldsymbol{w}}$  modes respectively, from (\ref{factor state hyp app}) we have that 
$\gamma_{\Omega}  =\gamma_{\Omega_A} \oplus \gamma_{\bar{\Omega}_A}$.
Combining this observation with (\ref{w from r}), we obtain that
\be
\gamma_{\Omega_A} \oplus \gamma_{\bar{\Omega}_A}
=
V \gamma_A V^{\textrm t}   \,.
\ee
From $s^{(n)}_A(G) $ in (\ref{prop4 step1}) and the polar decomposition (\ref{polar-dec-W}) for the symplectic matrix occurring in the Williamson's theorem for $\gamma_{\Omega} $, which provides the matrix $E_{\textrm{\tiny R},{\textrm{\tiny $\Omega$}}}$, we find that
\bea
\label{prop5 step1}
& & 
\hspace{-2.3cm}
s^{(n)}_A(G)
 \,=\,
 \frac{1}{2} \, \Tr \big[  X^{(G)}  \,s_n( E_{\textrm{\tiny R}}^{-1} \gamma_A E_{\textrm{\tiny R}}^{-1}  )\big]
\,=\,
 \frac{1}{2} \, 
 \Tr \big[  X^{(G)}  \,
 s_n( V^{\textrm t}   E_{\textrm{\tiny R},{\textrm{\tiny $\Omega$}}}^{-1} 
 (\gamma_{\Omega_A} \oplus \gamma_{\bar{\Omega}_A}) E_{\textrm{\tiny R},{\textrm{\tiny $\Omega$}}}^{-1}  V )\big]
\\
 \label{prop5 step3}
 \rule{0pt}{.7cm}
& & 
\hspace{-2.3cm}
\phantom{s^{(n)}_A(G)} 
\,=\,
 \frac{1}{2} \, 
 \Tr \big[  V X^{(G)} V^{\textrm t} \,
 s_n\big(  (E_{\textrm{\tiny R},{\textrm{\tiny $\Omega_A$}}}^{-1} \gamma_{\Omega_A}  E_{\textrm{\tiny R},{\textrm{\tiny $\Omega_A$}}}^{-1})
 \oplus  (E_{\textrm{\tiny R},{\textrm{\tiny $\bar{\Omega}_A$}}}^{-1} \gamma_{\bar{\Omega}_A} E_{\textrm{\tiny R},{\textrm{\tiny $\bar{\Omega}_A$}}}^{-1} )\big)\big] 
 \\
  \label{prop5 step4}
 \rule{0pt}{.7cm}
& & 
\hspace{-2.3cm}
\phantom{s^{(n)}_A(G)} 
\,=\,
 \frac{1}{2} \, 
 \Tr \big[  V_{\Omega G} \,V_{\Omega G}^{\textrm t}  \;
 s_n(E_{\textrm{\tiny R},{\textrm{\tiny $\Omega_A$}}}^{-1} \gamma_{\Omega_A}  E_{\textrm{\tiny R},{\textrm{\tiny $\Omega_A$}}}^{-1})
\big] 
 \\
   \label{prop5 step5}
 \rule{0pt}{.7cm}
& & 
\hspace{-2.3cm}
\phantom{s^{(n)}_A(G)} 
\,\leqslant\,
 \frac{1}{2} \, \Tr \big[  s_n( E_{\textrm{\tiny R},{\textrm{\tiny $\Omega_A$}}}^{-1} \gamma_{\Omega_A} E_{\textrm{\tiny R},{\textrm{\tiny $\Omega_A$}}}^{-1}  )\big]
 \,=\,S^{(n)}(\Omega_A)\,,
\eea
where in (\ref{prop5 step1}) we used (\ref{trans gamma_tilde}), the step (\ref{prop5 step3}) has been obtained from
$E_{\textrm{\tiny R},{\textrm{\tiny $\Omega$}}}
= E_{\textrm{\tiny R},{\textrm{\tiny $\Omega_A$}}}\oplus E_{\textrm{\tiny R},{\textrm{\tiny $\bar{\Omega}_A$}}}$,
which is a consequence of $\gamma_{\Omega}  =\gamma_{\Omega_A} \oplus \gamma_{\bar{\Omega}_A}$,
and (\ref{prop5 step4}) comes from $V X^{(G)} V^{\textrm t} = V_{\Omega G} \,V_{\Omega G}^{\textrm t} \oplus  \boldsymbol{0}$.
The inequality in (\ref{prop5 step5}) is obtained by employing the fact that $V_{\Omega G} \,V_{\Omega G}^{\textrm t} $ is a projector.
Equivalently, one can employ that from the condition $V V^{\textrm t}  = \boldsymbol{1}$ one gets
$V_{\Omega G} \,V_{\Omega G}^{\textrm t}  = \boldsymbol{1} - V_{\Omega \bar{G}} \,V_{\Omega \bar{G}}^{\textrm t}$.
Then, by first plugging the latter expression into (\ref{prop5 step4}) and then discarding the term with the positive definite matrix $V_{\Omega \bar{G}} \,V_{\Omega \bar{G}}^{\textrm t}$, we find the inequality in (\ref{prop5 step5}), which can be written by exploiting that 
$ s_n(E_{\textrm{\tiny R},{\textrm{\tiny $\Omega_A$}}}^{-1} \gamma_{\Omega_A}  E_{\textrm{\tiny R},{\textrm{\tiny $\Omega_A$}}}^{-1})$ 
is positive definite.


\section{Correlators}
\label{app:correlators}

In this appendix we collect the correlators employed in our numerical analysis.
Considering the one-dimensional harmonic chain  (\ref{HC ham}) with $L$ lattice sites, in the following we write the two-point correlators 
corresponding to either periodic (\S\ref{app:corr-per}) or Dirichlet (\S\ref{app:corr-Dir}) boundary conditions.

\subsection{Periodic chain}
\label{app:corr-per}

The periodic harmonic chain is defined by the hamiltonian (\ref{HC ham}) with boundary conditions given by $q_L = q_0$ and $p_L = p_0$.
By introducing the creation and annihilation operators in the standard way, for the vacuum state one finds the following two-point correlators 
\be
\label{corrs per T=0}
\langle \hat{q}_i \hat{q}_j  \rangle =
\frac{1}{2L} \sum_{k=0}^{L-1} \frac{1}{m \omega_k} \cos[2\pi k(i-j)/L]\,,
\,\qquad\,
\langle \hat{p}_i \hat{p}_j  \rangle =
\frac{1}{2L} \sum_{k=0}^{L-1} m \omega_k  \cos[2\pi k(i-j)/L]\,,
\ee
where the dispersion relation reads
\be
\omega_k \equiv 
\sqrt{\omega^2 +\frac{4\kappa}{m}\, \big[ \sin(\pi k/L) \big]^2}\,,
\qquad
0 \leqslant k \leqslant L-1\,.
\ee
It is important to remark that the correlator $\langle \hat{q}_i \hat{q}_j  \rangle $ in (\ref{corrs per T=0}) is not well defined for $\omega =0$.
Indeed, for the zero mode $k=0$ we have $\omega_0|_{\omega=0}=0$ and therefore the corresponding term in $\langle \hat{q}_i \hat{q}_j  \rangle $ diverges as $\omega \to 0$.
The occurrence of the zero mode is due to the translation invariance of the model with periodic boundary conditions.
In order to explore the massless regime, in our numerical analysis we keep $\omega$ non vanishing and such that $1/\omega$ is much larger
than the other scales involved in the analysis. 

We also consider the periodic harmonic chain in the thermal state with temperature $T=1/\beta$.
For this state the correlators entering in the numerical analysis read
\bea
\label{corrs per qq T}
& &
\langle \hat{q}_i \hat{q}_j  \rangle_\beta =
\frac{1}{2L} \sum_{k=0}^{L-1} \frac{1}{m \omega_k} \coth(\beta \omega_k /2) \cos[2\pi k(i-j)/L]\,,
\\
\label{corrs per pp T}
& &
\langle \hat{p}_i \hat{p}_j  \rangle_\beta =
\frac{1}{2L} \sum_{k=0}^{L-1} m \omega_k \coth(\beta \omega_k /2)  \cos[2\pi k(i-j)/L]\,.
\eea
Also in this case the correlator (\ref{corrs per qq T}) diverges as $\omega \to 0$ because of the occurrence of zero mode at $k=0$.

In the thermodynamic limit $L \to \infty$, the vacuum state correlators (\ref{corrs per T=0}) become \cite{br-04}
\bea
\label{corrs per qq thermo}
& &
\hspace{-1.6cm}
\braket{\hat q_i \hat q_j} 
\,=\, 
\frac{z^{i-j+1/2}}{2\,\sqrt{\kappa m}} \binom{i-j-1/2}{i-j}
\; {}_2F_1\big( 1/2 , i-j+1/2 ; i-j+1;z^2\big) \,,
\\
\label{corrs per pp thermo}
\rule{0pt}{.9cm}
& & \hspace{-1.6cm}
\braket{\hat p_i \hat p_j} 
\,=\, \sqrt{\kappa  m}\;\frac{z^{i-j-1/2}}{2} \binom{i-j-3/2}{i-j}
\; {}_2F_1\big(-1/2,i-j-1/2,i-j+1;z^2\big)\,, 
\eea
where $ z \equiv (\omega - \sqrt{\omega^2+4 \kappa/m}\,)^2/(4\kappa/m) $.
As for the correlators (\ref{corrs per qq T}) and (\ref{corrs per pp T}) at finite temperature, in the thermodynamic limit they become
\bea
\label{qq per T therm}
& &
\langle \hat{q}_i \hat{q}_j  \rangle_\beta = 
\frac{1}{2\pi}\int_{0}^{\pi} \frac{1}{m\omega_k} \coth(\beta \omega_k /2) \cos[k(i-j)]\, dk\,,
\\
\label{pp per T therm}
& &
\langle \hat{p}_i \hat{p}_j  \rangle_\beta = 
\frac{1}{2\pi}\int_{0}^{\pi} m\omega_k \coth(\beta \omega_k /2) \cos[k(i-j)] \, dk\,,
\eea
with $\omega_k = \sqrt{\omega^2 +(4\kappa/m)\sin^2(k/2)}$.

The above two-point functions provide the elements of the matrices $Q$ and $P$ entering in the reduced covariance matrix $\gamma_A$, whose symplectic spectrum leads to the entanglement entropies as discussed in \S\ref{sec:entHC}.

\subsection{Open chain with Dirichlet boundary conditions}
\label{app:corr-Dir}

The open harmonic chain with Dirichlet boundary conditions is defined by the hamiltonian (\ref{HC ham}) with the conditions 
$q_0 = q_{L+1} = 0$ and $p_0 = p_{L+1} = 0$ imposed at its endpoints. 
Because of these boundary conditions, the invariance under translations does not occur. 

The vacuum state correlators for this model read \cite{lievens-dirichlet}
\bea
\label{qq open}
& &
\langle \hat{q}_i \hat{q}_j  \rangle =
\frac{1}{L} \sum_{k=1}^{L-1} \frac{1}{m \tilde{\omega}_k} \, \sin(\pi k\, i/L)  \, \sin(\pi k\, j/L) \,,
\\
\rule{0pt}{.8cm}
\label{pp open}
& &
\langle \hat{p}_i \hat{p}_j  \rangle =
\frac{1}{L} \sum_{k=1}^{L-1} m \tilde{\omega}_k \, \sin(\pi k \, i/L)  \, \sin(\pi k \, j/L) \,,
\eea
where the dispersion relation is given by
\be
\tilde{\omega}_k \equiv 
\sqrt{\omega^2 +\frac{4\kappa}{m}\, \big[ \sin(\pi k/(2L)) \big]^2} \,>\,\omega\,,
\qquad
1 \leqslant k \leqslant L-1\,.
\ee
We remark that the correlators (\ref{qq open}) and (\ref{pp open}) are well defined also in the massless regime, i.e. finite quantities are obtained when $\omega = 0$.
This key feature significantly improves the comparison between the numerical results from the lattice and the corresponding ones obtained through CFT methods.

In the thermodynamic limit $L \to \infty$, the two-point functions (\ref{qq open}) and (\ref{pp open}) become
\be
\label{corr open thermo}
\braket{\hat q_i \hat q_j} = \frac{1}{\pi}\int_{0}^{\pi} \frac{1}{m \omega_k} \sin(k \,i)\sin(k \,j)  \, dk\,,
\qquad
\braket{\hat p_i \hat p_j} = \frac{1}{\pi}\int_{0}^{\pi} m \omega_k \sin(k \,i)\sin(k \, j)  \, dk\,,
\ee
where  $ \omega_k $ has been defined in the text below (\ref{pp per T therm}).
When $\omega=0$ these integrals can be performed analytically and the results read respectively 
\cite{cct-neg}
\bea
\label{corr qq dirichlet thermo}
& &
\langle \hat{q}_i \hat{q}_j  \rangle =
\frac{1}{2\pi \, \sqrt{\kappa m}} 
\Big(  \psi(1/2+i+j) -  \psi(1/2+i-j)  \Big)\,,
\\
\label{corr pp dirichlet thermo}
\rule{0pt}{.9cm}
& &
\langle \hat{p}_i \hat{p}_j  \rangle =
\frac{2\,\sqrt{\kappa m}}{\pi}  \left( \frac{1}{4(i+j)^2-1} - \frac{1}{4(i-j)^2-1} \right) ,
\eea
being $\psi(z)$ the digamma function.

\end{appendices}


\end{document}